\documentclass[a4paper,12pt]{article}

\usepackage{amsmath,amsfonts,amssymb}
\usepackage[dvips]{graphicx}

\numberwithin{equation}{section}

\def\beq#1\eeq{\begin{equation}#1\end{equation}}
\def\bes#1\ees{\begin{equation}\begin{split}#1
               \end{split}\end{equation}}
\def\bea#1\eea{\begin{align}#1\end{align}}
            
\newcommand{\abs}[1]{\lvert #1 \rvert}
\DeclareMathOperator{\trace}{Tr}
\newcommand{\Irr}[1]{\mathrm{Irr}(#1)}

\newcommand{\Z}{\mathbb{Z}}
\newcommand{\C}{\mathbb{C}}

\newcommand{\bra}[1]{\langle #1|}
\newcommand{\ket}[1]{|#1\rangle}
\newcommand{\dbra}[1]{\langle\!\langle #1|}
\newcommand{\dket}[1]{|#1 \rangle\!\rangle}

\newcommand{\tq}{\tilde{q}^{\frac{1}{2} H_c}}
\newcommand{\fusion}[3]{\mathcal{N}_{#1 #2}{}^{#3}}

\newcommand{\fusionhat}[3]{\widehat{\mathcal{N}}_{#1 #2}{}^{#3}}
\newcommand{\Ihat}{\hat{\mathcal{I}}}
\newcommand{\Shat}{\hat{S}}
\newcommand{\Ehat}{\hat{\mathcal{E}}}
\newcommand{\Vhat}{\hat{\mathcal{V}}}
\newcommand{\Nhat}{\widehat{N}}
\newcommand{\bb}{{\mathfrak b}}

\newcommand{\A}{\mathcal{A}}
\newcommand{\I}{\mathcal{I}}
\newcommand{\Sc}{\mathcal{S}}
\newcommand{\F}{\mathcal{F}}

\newcommand{\E}{\mathcal{E}}
\newcommand{\V}{\mathcal{V}}

\newcommand{\bsp}{\hspace{-0.1cm}}
\newcommand{\OO}{\mbox{\bf O}}

\begin{document} 
\baselineskip=5.3mm
\begin{titlepage}
\renewcommand{\thefootnote}{\fnsymbol{footnote}}
\nopagebreak
\vskip 5mm
\begin{flushright}
TU-754\\
AS-ITP-2005-004
\end{flushright}

\vskip 10mm
\begin{center}
\baselineskip=10mm
{\LARGE
\textbf{%
Twisted boundary states and representation of
generalized fusion algebra}}
\end{center}
\begin{center}
\vskip 10mm
Hiroshi \textsc{Ishikawa}$^1$ and Taro \textsc{Tani}$^2$~\footnote[2]{%
Address after September 2005:
Kurume National College of Technology,
1-1-1 Komorino, Kurume 830-8555, Japan.
E-mail: tani@kurume-nct.ac.jp
}
\vskip 5mm
\textsl{%
$^1$Department of Physics, Tohoku University \\
Sendai 980-8578, JAPAN\\\vspace{10pt}
$^2$Interdisciplinary Center of Theoretical Studies\\
Institute of Theoretical Physics\\
Chinese Academy of Sciences\\
Beijing 100080, CHINA\\
}
\vspace{10pt}
\texttt{%
\footnotesize
$^1$ishikawa@tuhep.phys.tohoku.ac.jp \\
$^2$tani@itp.ac.cn
}
\end{center}
\vskip 7mm 

\begin{quote}
~~~
The mutual consistency of boundary conditions twisted
by an automorphism group $G$ of the chiral algebra is studied
for general modular invariants
of rational conformal field theories.
We show that a consistent set of twisted boundary states
associated with any modular invariant
realizes a non-negative integer matrix representation (NIM-rep) 
of the generalized fusion algebra, an extension of the fusion algebra
by representations of the twisted chiral algebra associated with 
the automorphism group $G$.
We check this result for several concrete cases. 
In particular, we find
that two NIM-reps of the fusion algebra for 
$su(3)_k\,(k=3,5)$ are organized 
into a NIM-rep of the generalized fusion algebra for 
the charge-conjugation automorphism of $su(3)_k$.
We point out that the generalized fusion algebra is
non-commutative if $G$ is non-abelian and 
provide some examples for $G \cong S_3$.
Finally, we give an argument that the graph fusion algebra
associated with simple current extensions coincides with
the generalized fusion algebra for the extended chiral algebra,
and thereby explain that the graph fusion algebra contains the fusion algebra
of the extended theory as a subalgebra.

\end{quote}

\vfill
\end{titlepage}

\section{Introduction}

The classification of conformally invariant 
boundary states in two-dimensional conformal field theories (CFTs)
is an interesting subject of research,
both from the point of view of purely field theoretical problems
and applications to condensed matter physics or string theory.
In string theory, conformally invariant boundary states
correspond to D-branes, which are considered to be submanifolds embedded
in the target space of strings.
The classification of boundary states
is therefore the classification of D-branes,
which might be helpful in understanding
the nature of stringy geometry.

This classification problem of boundary states
is hard to treat in full generality,
and one would need some restriction on the problem to make it tractable.
A natural choice for such a restriction is to
consider only rational CFTs (RCFTs)~\footnote{%
For the current status of the research about non-rational cases,
see \textit{e.g.} \cite{Schomerus}.}
 and boundary states that preserve
the full chiral algebra of RCFTs.
Actually, in a RCFT, every consistent set of boundary states
should satisfy a set of simple algebraic equations,
the so-called Cardy equation~\cite{Cardy},
if the full chiral algebra is preserved in the open-string
channel.
Finding a set of Cardy states, \textit{i.e.}, the states satisfying the
Cardy equation, is equivalent to finding a non-negative 
integer matrix representation (NIM-rep) of the fusion 
algebra~\cite{BPPZ}\footnote{
It should be noted that the Cardy condition is a necessary condition 
on boundary states \cite{CL,L,Runkel}.
Actually, there is a NIM-rep which does not give 
rise to any consistent boundary states, such as the tadpole NIM-rep
of $su(2)_{2n-1}$ (see \textit{e.g.} \cite{BPPZ,Gannon}).} 
assuming that
the set of boundary states is `complete' in a sense of \cite{PSS}.
The classification of boundary states for this case 
is therefore related to
the classification of NIM-reps of the fusion algebra \cite{Gannon}.

Clearly, the boundary states preserving the full chiral algebra
constitute only a subset of conformally invariant boundary states,
since the chiral algebra in general contains the Virasoro algebra as 
a proper subalgebra.
In order to obtain, and classify, more states other 
than those preserving the full chiral algebra, 
we need some way to reduce the symmetry preserved by boundary states
in a controlled manner.
One way to accomplish this is to modify (or `twist') the boundary 
condition of the chiral currents
by an automorphism group $G$ of the chiral algebra.
The corresponding `twisted boundary states' preserve the Virasoro
algebra if $G$ keeps it fixed. 
The states preserving the full chiral algebra 
(the `untwisted' states) correspond to the identity element
of $G$ and are considered to be a particular case of the twisted states.
In this sense, 
the twisted states associated with the automorphism group $G$ 
realize a generalization of untwisted states. 

The classification problem of twisted boundary states
is systematically studied by several groups
\cite{FS,BFS,GG,Recknagel}.~\footnote{%
For the other approaches to boundary states
with less symmetry, see, \textit{e.g.}, 
\cite{FS2,MMS,QS}.}
In particular, it is pointed out in \cite{GG} 
(see also \cite{BFS,Ishikawa})
that the mutual consistency of twisted boundary states 
in the charge-conjugation modular invariant 
follows from the integrality of the structure constants
of a certain algebra
called \textit{the generalized fusion algebra}.
This algebra is defined by the fusion of 
representations of the twisted chiral algebra
and hence contains, as a subalgebra,
the ordinary fusion algebra consisting
of only representations of the (untwisted) chiral algebra. 
This relation of twisted boundary states with the generalized
fusion algebra is quite analogous to the relation of
untwisted boundary states with the ordinary fusion algebra,
and suggests that
the generalized fusion algebra also plays some role in the classification
problem of twisted boundary states.

In this paper, we give an answer to the above question concerning
the role of the generalized fusion algebra in the classification problem 
of twisted boundary states.
Namely, we show that a set of mutually consistent twisted
boundary states 
realizes \textit{a NIM-rep of the generalized fusion algebra}.
Our result is valid for any finite automorphism group 
including non-abelian ones.
We point out that the generalized fusion algebra 
for a non-abelian automorphism group is 
\textit{non-commutative} and 
provide some examples for the case of the symmetric group $S_3$.
We develop the representation theory of the generalized fusion algebras
and find their irreducible representations, which generalize
quantum dimensions for the case of the ordinary fusion algebras.
Unlike quantum dimensions, irreducible representations
of the generalized fusion algebras are possible to 
have dimension greater 
than one, reflecting the non-commutativity of the generalized
fusion algebras.
We also point out that
a NIM-rep of the generalized fusion algebra
is decomposed into NIM-reps of the ordinary fusion algebra,
since the former contains the latter as a subalgebra. 
As a check of our argument, 
we explicitly construct twisted boundary states in 
several concrete cases
and show that their overlap matrices indeed form a NIM-rep of
the corresponding generalized fusion algebra.
In particular, we find that two NIM-reps of $su(3)_k$ ($k=3,5$)
are combined into a NIM-rep of the generalized fusion
algebra of $su(3)_k$ for the automorphism group $\{1, \omega_c\}$,
where $\omega_c$ is the charge-conjugation automorphism of $su(3)_k$.
We also give an argument that some graph fusion algebra
\cite{DZ,PZ,BPPZ} can be regarded as the generalized fusion algebra.
More precisely, we show that
the graph fusion algebra associated with
boundary states in a simple current extension \cite{SY} 
coincides with the generalized fusion algebra of the extended chiral
algebra for an appropriate automorphism group.
This result naturally explains the observation \cite{DZ} that 
the graph fusion algebra contains the fusion algebra of the extended
theory as a subalgebra.

An algebraic structure of twisted boundary states is also
studied in \cite{FS,BFS} and we should clarify the difference
of our analysis from that of \cite{FS,BFS}.
In \cite{FS}, the case that a chiral algebra $\A$ is obtained
by a simple current extension is considered.
In this setting, the unextended chiral algebra $\A_0$
is characterized as the fixed point algebra of
some finite abelian automorphism group $G$ of $\A$, 
which is the dual of the simple current group used in the extension.
The authors of \cite{FS} classify boundary states
that preserve $\A_0$ in the charge-conjugation invariant of $\A$
using simple current techniques,
and show that the states preserving $\A_0$ can be regarded
as the $G$-twisted states.
They also show that the $G$-twisted boundary states correspond
to one-dimensional representations of some commutative algebra
called the classifying algebra \cite{FS3}.
In \cite{BFS}, the structure of twisted boundary states
in the charge-conjugation invariant
is investigated for an arbitrary chiral algebra, in particular
untwisted affine Lie algebras,
with the automorphism group $G \cong \Z_2$,
and the corresponding generalized fusion algebra together with
the classifying algebra are studied. 
Our stance on the problem is slightly different from these studies.
We do not intend to construct twisted boundary
states explicitly; rather we clarify a consistency condition
of twisted states assuming their existence.
We impose no restrictions on the model we consider.
The charge-conjugation modular invariant, as well as
an abelian automorphism group, is a particular case of our analysis. 
Finally, the expression \cite{BFS} for the case of
the $\Z_2$ automorphism suggests that
the classifying algebra in \cite{FS}
is the dual of the generalized fusion
algebra for abelian automorphism groups
in the sense of $C$-algebras \cite{BI,DZ,PZ},
although we have no proof for this statement.
Since this duality exchanges the irreducible representations
with the elements of the algebra, the result of \cite{FS} is
consistent with the observation \cite{GG,Ishikawa} that
the twisted boundary states in the charge-conjugation invariant
realize the generalized fusion algebra,
which is the starting point of our analysis. 

This paper is organized as follows.
In the next section, we give the definition of
the generalized fusion algebras using
the twisted boundary states in the charge-conjugation modular invariant
along the lines of \cite{GG}.
The case of the affine Lie algebra $so(8)_1$ and its triality
automorphism group is treated in detail.
In Section~\ref{sec:irrep}, we show that
the fusion coefficients of the generalized fusion
algebras can be expressed in a form similar
to the Verlinde formula even in the case of non-abelian automorphisms,
and obtain the irreducible representations
of the generalized fusion algebras.
Based on this result, 
we show in Section~\ref{sec:genNIM} that the overlap matrices of 
mutually consistent twisted boundary states 
in any modular invariant
form a NIM-rep of the generalized fusion algebra.
Some examples including the case of
non-abelian automorphisms are presented in Section~\ref{sec:example}.
In Section~\ref{sec:graph}, we argue
the relation of graph fusion algebras with the
generalized fusion algebras. 
The final section is devoted to discussions.
In the appendices, we present the details of the calculation
in our examples. 
In Appendix~\ref{sec:su3so8}, we give an explicit
realization of the algebra embedding $su(3)_3 \subset so(8)_1$
and show that the charge-conjugation automorphism group
$\{1,\omega_c\}$ of $su(3)$ has a lift to $so(8)$, which is
isomorphic to the symmetric group $S_3$.
In Appendix~\ref{sec:su3detail}, the generalized fusion
algebra of $su(3)_k\, (k=3,5)$ for the charge-conjugation 
automorphism group is determined and 
it is shown that the twisted boundary states 
associated with non-trivial modular invariants yield a NIM-rep
of the generalized fusion algebra.
The same calculation is done
in Appendix~\ref{sec:E6detail} for $su(3)_1^{\oplus 3}$
and its permutation automorphism of three factors.

\section{Generalized fusion algebra}
\label{sec:Gfusion}

In this section, we give a definition of the generalized fusion
algebra for the chiral algebra $\A$ with the automorphism
group $G$ by using the cylinder amplitude for the (twisted) 
boundary states in the charge-conjugation modular invariant of $\A$. 

\subsection{Untwisted boundary states
and ordinary fusion algebra}

We start from the construction of 
the untwisted boundary states and their relation with the ordinary
fusion algebra \cite{Cardy}.

Let $\I$ be the set of all the irreducible representations
of the chiral algebra $\A$.
Since we consider only rational cases, $\I$ is a finite set.
The vacuum representation of $\A$ is denoted by $0 \in \I$.
For each $\lambda \in \I$, the character $\chi_\lambda(q)$
of the representation $\lambda$ is defined as
\beq
  \chi_\lambda(q) = \trace_{\mathcal{H}_\lambda} q^{L_0 - \frac{c}{24}} \, ,
\eeq
where $q = e^{2 \pi i \tau}$, 
$c$ is the central charge of the theory and the trace is
taken in the irreducible
representation space $\mathcal{H}_\lambda$ with the highest-weight $\lambda$.
Under the modular transformation $\tau \mapsto -1/\tau$,
the characters $\chi_\lambda(q)$ transform as follows
\beq
  \chi_\lambda(q) = \sum_{\mu \in \I} S_{\lambda \mu} 
  \chi_\mu(\tilde{q}) \quad
  (\tilde{q} = e^{-2 \pi i/\tau} ) \, .
  \label{eq:S}
\eeq
The modular transformation matrix $S$ is unitary and symmetric.
The (ordinary) fusion algebra is an associative commutative
algebra defined by the fusion of two representations in $\I$,
\beq
  (\lambda) \times (\mu) = \sum_{\nu \in \I} 
  \fusion{\lambda}{\mu}{\nu} (\nu) \qquad
  (\lambda, \, \mu \in \I)\, ,
\eeq
where the fusion coefficients $\fusion{\lambda}{\mu}{\nu}$ 
take values in non-negative integers
and are related with the modular transformation
matrix via the Verlinde formula \cite{Verlinde},
\beq
  \fusion{\lambda}{\mu}{\nu}
  = \sum_{\sigma \in \I}
  \frac{S_{\lambda \sigma} S_{\mu \sigma} \overline{S_{\nu \sigma}}}{%
        S_{0 \sigma}} \, .
  \label{eq:Verlinde0}
\eeq

A boundary state $\ket{\alpha}$ is a coherent state in the 
closed-string channel that preserves a half of 
the (super) conformal symmetry 
on the worldsheet. 
We also require that $\ket{\alpha}$ keep the full chiral algebra $\A$. 
In terms of modes, the boundary condition can be written as
\beq
   (W_n - (-1)^h \widetilde{W}_{-n}) \ket{\alpha} =0 \, ,
\label{eq:symmetricbc}
\eeq
where $W_n$ and $\widetilde{W}_{-n}$ are respectively the modes of
the current $W(z)$ of the holomorphic chiral algebra $\A$ 
with conformal dimension $h$ and 
its counterpart
$\widetilde{W}(\bar{z})$ of the anti-holomorphic chiral algebra
$\widetilde{\A}$.
For each $\lambda \in \I$, 
we can construct a building block $\dket{(\lambda,\lambda^*)}$
of boundary states satisfying \eqref{eq:symmetricbc},
\beq
  \dket{(\lambda,\lambda^*)}
  = \frac{1}{\sqrt{S_{0\lambda}}}
  \sum_{N} \ket{\lambda; N} \otimes \overline{\ket{\lambda; N}}
  \, \in \mathcal{H}_\lambda \otimes \widetilde{\mathcal{H}}_{\lambda^*}
  \, ,
  \label{eq:Ishibashi}
\eeq
which is called the Ishibashi state \cite{Ishibashi}.
Here $\ket{\lambda; N}$ is an orthonormal basis of the representation
space $\mathcal{H}_\lambda$
and we denote by $\lambda^* \in \I$ the conjugate representation
of $\lambda \in \I$.
This notation for Ishibashi states is slightly different from that used in
other references, such as $\dket{\lambda}$. 
We adopt $\dket{(\lambda,\lambda^*)}$, instead of $\dket{\lambda}$,
to indicate explicitly that 
this state is composed of the states in 
$\mathcal{H}_\lambda \otimes \widetilde{\mathcal{H}}_{\lambda^*}$
with the highest-weight state 
$\ket{\lambda} \otimes \ket{\lambda^*}$ of $\A \times \widetilde{\A}$.
Since the boundary condition \eqref{eq:symmetricbc} is linear, 
multiplying $\dket{(\lambda,\lambda^*)}$ by some constant
still gives a solution to \eqref{eq:symmetricbc}.
The above normalization \eqref{eq:Ishibashi} is chosen
so that the following equation is satisfied,
\beq
   \dbra{(\lambda,\lambda^*)} 
   \tq \dket{(\lambda',{\lambda'}^*)}
   = \delta_{\lambda\lambda'} \frac{1}{S_{0\lambda}} \sum_N
     \bra{\lambda; N} \tilde{q}^{L_0 - \frac{c}{24}} \ket{\lambda; N}              
   = \delta_{\lambda\lambda'}\frac{1}{S_{0\lambda}}\chi_{\lambda}(\tilde{q})\, .
\label{eq:Ishibashinorm}
\eeq
Here $H_c$ is the closed string Hamiltonian
\beq
  H_c = L_0 + \tilde{L}_0 - \frac{c}{12} \, 
\eeq
and we used 
$(L_0 - \tilde{L}_0) \dket{(\lambda,\lambda^*)} = 0$ to eliminate
$\tilde{L}_0$.

In order to find solutions to \eqref{eq:symmetricbc}, 
we have to specify the spectrum of bulk fields,
or equivalently the modular invariant
\beq
  Z = \sum_{\lambda, \mu \in \I} 
  M_{\lambda \mu} \chi_\lambda \overline{\chi_\mu} \, ,
  \label{eq:Z}
\eeq
where $\chi_\lambda \overline{\chi_\mu}$ corresponds to
a bulk field $\Phi_{(\lambda,\mu^*)}$ which carries
a representation $(\lambda, \mu^*)$ of 
$\A \times \widetilde{\A}$.
In \eqref{eq:Z}, there are $M_{\lambda \mu}$ fields
with representation $(\lambda,\mu^*)$ and we should distinguish them
by putting an index like $\Phi_{(\lambda,\mu^*)_i}$
if $M_{\lambda \mu} > 1$. 
In order to keep the notation simple, however, we omit the extra index $i$
and simply use $\Phi_{(\lambda,\mu^*)}$
understanding that they appear in \eqref{eq:Z} with 
the multiplicity $M_{\lambda \mu}$.
We denote by $\mathrm{Spec} (Z)$ 
the set of labels for bulk fields 
available in \eqref{eq:Z},
\beq
  \mathrm{Spec} (Z) = \{ (\lambda, \mu^*) \, | \, 
  \Phi_{(\lambda,\mu^*)} \text{ exists in } \eqref{eq:Z} \} \, .
  \label{eq:SpecZ}
\eeq
Note that a symbol $(\lambda,\mu^*)$ appears 
$M_{\lambda \mu}$ times in $\mathrm{Spec}(Z)$
corresponding to $M_{\lambda \mu}$ independent bulk fields with 
the label $(\lambda, \mu^*)$.
In this section, we restrict ourselves to the case of 
the charge-conjugation modular invariant
\beq
  Z_c = \sum_{\lambda \in \I} \abs{\chi_\lambda}^2 \, ,
  \label{eq:Zc}
\eeq
for which the spectrum of bulk fields reads
\beq
  \mathrm{Spec}(Z_c) =
  \{(\lambda,\lambda^*) \, | \, \lambda \in \I \} \, .
\eeq
We have one Ishibashi state $\dket{(\lambda,\lambda^*)}$
for each bulk field $\Phi_{(\lambda,\lambda^*)}$.
A general boundary state $\ket{\alpha}$
satisfying \eqref{eq:symmetricbc} are therefore a linear combination
of $\dket{(\lambda,\lambda^*)}$ ($\lambda \in \I$), 
\beq
  \ket{\alpha} = \sum_{\lambda \in \I} 
  \Psi_{\alpha (\lambda,\lambda^*)} \dket{(\lambda,\lambda^*)} \, .
\eeq

A consistent set of boundary states is obtained by considering
the cylinder amplitude $Z_{\alpha \beta}$ 
with boundary conditions $\alpha$ and $\beta$.
This amplitude can be calculated in two ways:
a closed-string propagation between two boundary states,
and an open-string one-loop amplitude.
In the closed-string channel, the cylinder amplitude can be
expressed as
\beq
  Z_{\alpha \beta} = \bra{\beta} \tq \ket{\alpha} \, ,
\eeq
where $\tilde{q} = e^{-2 \pi / t}$ and $t$ is the circumference of
the cylinder. 
On the other hand, the calculation in the open-string channel 
yields
\beq
  Z_{\alpha \beta} = 
  \sum_{\lambda \in \I} (n_\lambda)_\alpha{}^\beta \,\chi_\lambda(q) \, , 
\eeq
where $q=e^{-2 \pi t}$ and $(n_\lambda)_\alpha{}^\beta$
is the multiplicity of representation $\lambda$ in the spectrum
of open string with boundary condition $[\beta, \alpha]$.
Note that this form 
of open-string spectrum,
$\bigoplus_{\lambda \in \I} (n_\lambda)_\alpha{}^\beta \,\mathcal{H}_\lambda$,
follows from the requirement that the boundary
conditions $\alpha$ and $\beta$ of open string preserve the chiral
algebra $\A$.
Comparing two expressions for $Z_{\alpha \beta}$, one obtains
\beq
  \bra{\beta} \tq \ket{\alpha} = 
  \sum_{\lambda \in \I} (n_\lambda)_\alpha{}^\beta \,\chi_\lambda(q) \, .
\eeq

The multiplicity $(n_\lambda)_\alpha{}^\beta$ takes values
in non-negative integers. In particular,
$(n_0)_\alpha{}^\alpha = 1$, since an open string with the same boundary 
condition at both ends should have the unique vacuum. 
We therefore obtain
\beq
  \bra{\alpha} \tq \ket{\alpha} = \chi_0(q) + \cdots \, ,
\eeq
for any boundary state $\ket{\alpha}$.
Clearly, the simplest situation is that the right-hand side contains
only the vacuum character $\chi_0$. 
We denote such a state by $\ket{0}$,
\beq
  \bra{0} \tq \ket{0} = \chi_0(q) \, .
\eeq
This state can be realized in the charge-conjugation modular invariant
using the Ishibashi states,
\beq
  \ket{0} = \sum_{\lambda \in \I} S_{0 \lambda} 
  \dket{(\lambda,\lambda^*)} \, .
\eeq
One can check that this state has the desired overlap with itself,
\beq
  \bra{0} \tq \ket{0} = \sum_\lambda 
  \overline{S_{0 \lambda}} S_{0 \lambda} 
  \frac{1}{S_{0\lambda}} \chi_\lambda({\tilde{q}}) 
  = \sum_\lambda 
  S_{0 \lambda} \chi_\lambda({\tilde{q}}) 
  = \chi_0 (q) \, .
\eeq
The remaining states satisfying the boundary condition
\eqref{eq:symmetricbc} are determined by 
requiring that their overlap with $\ket{0}$ contain
a single character $\chi_\lambda$,
\beq
  \bra{0} \tq \ket{\lambda} = \chi_\lambda(q) \quad
  (\lambda \in \I) \, .
  \label{eq:lambda0}
\eeq
One can solve this equation to obtain $\ket{\lambda}$ in the form
\beq
  \ket{\lambda} = \sum_{\mu \in \I} S_{\lambda \mu} 
  \dket{(\mu,\mu^*)} \, .
  \label{eq:lambda}
\eeq
Actually, 
the overlap of this state with $\ket{0}$ yields $\chi_\lambda$,
\beq
  \bra{0} \tq \ket{\lambda} = \sum_\mu 
  \overline{S_{0 \mu}} S_{\lambda \mu} 
  \frac{1}{S_{0\mu}} \chi_\mu({\tilde{q}}) 
  = \sum_\mu 
  S_{\lambda \mu} \chi_\mu({\tilde{q}}) 
  = \chi_\lambda (q) \, .
\eeq
Taking the complex conjugation of this equation, we obtain
the conjugate representation $\lambda^*$ of $\lambda \in \I$,
\beq
  \bra{\lambda} \tq \ket{0} = \chi_{\lambda^*} (q) \, .
  \label{eq:cc}
\eeq

\begin{figure}
\begin{center}
\includegraphics[height=.2\textheight]{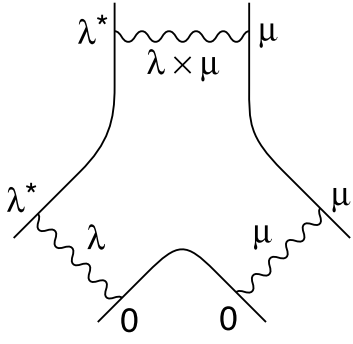}
\end{center}
\caption{Joining two open strings with boundary condition
$[\lambda^*, 0]$ and $[0, \mu]$ yields a string with
$[\lambda^*, \mu]$.}
\label{fig:three}
\end{figure}

One can relate the overlap of two boundary states with the 
fusion rule coefficients.
Suppose that we have two open strings, one of which has
the boundary condition $[\lambda^*, 0]$ and the other has
$[0, \mu]$ (see Fig.\ref{fig:three}). 
The spectrum of these two can be calculated as
\beq
  \bra{\lambda^*} \tq \ket{0} = \chi_{\lambda} (q) \, , \quad
  \bra{0} \tq \ket{\mu} = \chi_\mu (q) \, .
\eeq
Since the boundary condition `$0$' is common to both strings,
one can join two strings at the boundary `$0$' to
yield a string with the boundary condition $[\lambda^*, \mu]$. 
The spectrum of this string can be determined by the fusion of 
the spectra of the initial strings. 
In the present case, this is nothing but the fusion of 
$\lambda$ and $\mu$. 
Hence we obtain
\beq
  \bra{\lambda^*} \tq \ket{\mu} 
  = \sum_{\nu \in \I} \fusion{\lambda}{\mu}{\nu} \chi_\nu(q) \, .
\eeq
On the other hand, we can explicitly calculate the left-hand side using
the Ishibashi states,
\bes
  \bra{\lambda^*} \tq \ket{\mu} 
  &= \sum_{\sigma \in \I} \overline{S_{\lambda^* \sigma}} S_{\mu \sigma}
    \frac{1}{S_{0 \sigma}} \chi_\sigma(\tilde{q})  \\
  &= \sum_{\sigma \in \I} \overline{S_{\lambda^* \sigma}} S_{\mu \sigma}
    \frac{1}{S_{0 \sigma}} \sum_{\nu \in \I} 
    (S^{-1})_{\sigma \nu} \chi_\nu(q) \\
  &= \sum_{\nu, \sigma \in \I} 
     \frac{S_{\lambda \sigma} S_{\mu \sigma} \overline{S_{\nu \sigma}}}{%
     S_{0 \sigma}}
     \chi_\nu(q) \, ,
\ees
where we used the unitarity of $S$ and the property
\beq
  \overline{S_{\lambda \mu}} = S_{\lambda^* \mu} \, . 
  \label{eq:Scc}
\eeq
Comparison of these two expressions for $\bra{\lambda^*} \tq \ket{\mu}$
yields \cite{Cardy} the Verlinde formula \eqref{eq:Verlinde0}.

\subsection{Twisted boundary states
and generalized fusion algebra}
\label{sec:twisted}

We next turn to the case of twisted boundary states
and define the generalized fusion algebra
\cite{BFS,GG,Ishikawa}.

Let $G$ be a group of automorphisms of the chiral algebra $\A$,
which is in general a subgroup of the full automorphism group
$\mathrm{Aut}(\A)$. 
We restrict ourselves to the case that
$G$ is finite (but not necessarily abelian). 
We also require that $G$ fixes the Virasoro algebra in $\A$.
Then we can modify the boundary condition of the chiral algebra
using $\omega \in G$ as follows,
\beq
  (\omega(W_n) - (-1)^h \widetilde{W}_{-n}) \ket{\tilde{\alpha}} = 0 \, ,
  \label{eq:twistbc}
\eeq
which we call twisted boundary conditions 
while the condition \eqref{eq:symmetricbc}
the untwisted one.
This boundary condition preserves
the conformal symmetry since $G$ fixes the Virasoro algebra by assumption.

An automorphism $\omega \in G$ determines a unitary operator
$R(\omega)$ on the representation space 
$\bigoplus_{\lambda \in \I} \mathcal{H}_\lambda$ of $\A$
through the relation
\beq
  \omega(W_n) = R(\omega) W_n R(\omega)^{-1} \, .
\eeq
The operator $R(\omega)$ in general maps the representation space
$\mathcal{H}_\lambda$ to the different one,
which we denote by $\mathcal{H}_{\omega(\lambda)}$,
\beq
  R(\omega) : \mathcal{H}_\lambda \rightarrow \mathcal{H}_{\omega(\lambda)} \, .
\eeq
Using this $R(\omega)$,
the basis of solutions to \eqref{eq:twistbc} can be constructed
from the Ishibashi states \eqref{eq:Ishibashi}
for the untwisted condition \eqref{eq:symmetricbc}
in the following manner,
\bes
  \dket{(\lambda,\mu^*); \omega} 
  &= R(\omega) \dket{(\mu,\mu^*)} \\
  &= \frac{1}{\sqrt{S_{0\mu}}}
  \sum_{N} R(\omega) \ket{\mu; N} \otimes \overline{\ket{\mu; N}}
  \, \in \mathcal{H}_\lambda \otimes \widetilde{\mathcal{H}}_{\mu^*}
  \quad (\lambda = \omega(\mu) )
  \, ,
  \label{eq:twistIshibashi}
\ees
where $R(\omega)$ acts only on the holomorphic sector and
$\lambda = \omega(\mu)$. 
One can verify that this state satisfies
the twisted boundary condition \eqref{eq:twistbc}
as follows
\bes
  (-1)^h \widetilde{W}_{-n} \dket{(\lambda,\mu^*); \omega}
  &= R(\omega) (-1)^h \widetilde{W}_{-n} 
     \dket{(\mu,\mu^*)} \\
  &= R(\omega) W_n \dket{(\mu,\mu^*)} \\
  &= R(\omega) W_n R(\omega)^{-1} R(\omega)
     \dket{(\mu,\mu^*)} \\
  &= \omega(W_n) \dket{(\lambda,\mu^*); \omega} \, .
  \label{eq:checkbc}
\ees
We call these states 
$\dket{(\lambda,\mu^*); \omega}$ twisted Ishibashi states.
The untwisted one \eqref{eq:Ishibashi}
can be considered as the particular case 
$\omega=1$ of twisted states. 
Note that the $\omega$-twisted Ishibashi state 
$\dket{(\lambda,\mu^*); \omega}$ exists if and only if
$\lambda = \omega(\mu)$.

The overlap of twisted Ishibashi states can be calculated
in the same way as the untwisted case,
\bes
   \dbra{(\lambda,\mu^*); \omega} 
   \tq \dket{(\lambda',{\mu'}^*); \omega'} 
   &= \dbra{(\mu,\mu^*)} R(\omega)^\dagger
   \tq R(\omega') \dket{(\mu',{\mu'}^*)} \\
   &= \delta_{\mu \mu'} \frac{1}{S_{0\mu}} \sum_N
      \bra{\mu; N} R(\omega^{-1} \omega') 
      \tilde{q}^{L_0 - \frac{c}{24}} \ket{\mu; N} \, .
\ees
Clearly, this expression vanishes unless 
$\omega^{-1} \omega' (\mu) = \mu$, or equivalently
$\omega^{-1} \omega' \in \Sc(\mu)$, where $\Sc(\mu) \subset G$ is
the stabilizer of $\mu \in \I$,
\beq
  \Sc(\mu) = \{ \omega \in G \, | \, \omega(\mu) = \mu \} \, .
  \label{eq:stab}
\eeq
Using the relations
$\lambda = \omega(\mu)$ and 
$\lambda' = \omega'(\mu') = \omega'(\mu)$,
we can rewrite the condition $\omega^{-1} \omega' \in \Sc(\mu)$
into that for $\lambda$ and $\lambda'$,
\beq
  \lambda' = \omega'(\mu) 
  = \omega \omega^{-1} \omega' (\mu)
  = \omega (\mu)
  = \lambda \, .
\eeq
Conversely, $\lambda = \lambda'$ implies $\omega^{-1} \omega' \in \Sc(\mu)$,
\beq
  \omega^{-1} \omega'(\mu)
  = \omega^{-1} (\lambda')
  = \omega^{-1} (\lambda)
  = \mu \, .
\eeq
Putting these things together,
we can express the overlap of twisted Ishibashi states
in the following form
\beq
   \dbra{(\lambda,\mu^*); \omega} 
   \tq \dket{(\lambda',{\mu'}^*); \omega'} 
   = \delta_{(\lambda,\mu^*) (\lambda',{\mu'}^*)} 
     \frac{1}{S_{0\mu}} \chi_\mu^{\omega^{-1} \omega'}(\tilde{q}) \quad
   (\lambda = \omega(\mu) , \, \lambda' = \omega'(\mu') ) \, ,
\label{eq:Ishibashinormtwisted}
\eeq
where $\chi^\omega_\mu \,(\omega \in \Sc(\mu))$
is the twining character \cite{FSS1} of $\A$,
\beq
  \chi^\omega_\mu(q) = \mathrm{Tr}_{\mathcal{H}_\mu} R(\omega) 
  q^{L_0 - \frac{c}{24}} \, .
  \label{eq:twin}
\eeq
As we have noted above, $\chi^\omega_\mu$ for a given $\mu \in \I$
vanishes unless $\omega \in \Sc(\mu)$. 
Equivalently, 
for a given $\omega \in G$, $\chi^\omega_\mu$ exists if and only if
$\omega(\mu) = \mu$.
To express the condition for non-vanishing $\chi^\omega_\lambda$
for a given $\omega \in G$, 
we introduce a set 
$\I(\omega)$ which consists of representations of $\A$
fixed by $\omega$
\beq
  \I(\omega) = \{\lambda \in \I \, | \, \omega(\lambda) = \lambda \} \, .
  \label{eq:Iomega}
\eeq
It is useful to note that the following two conditions are equivalent,
\beq
  \lambda \in \I(\omega) \,\, \Leftrightarrow \,\, 
  \omega \in \Sc(\lambda) \, .
  \label{eq:ISrel}
\eeq

\begin{figure}
\begin{center}
\includegraphics[height=.2\textheight]{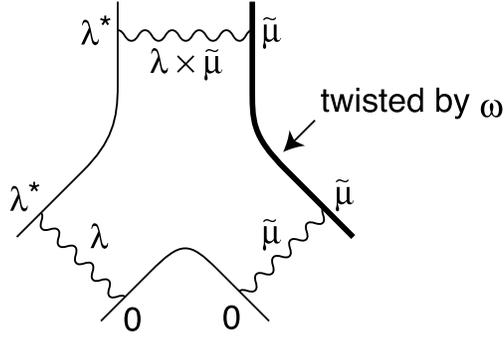}
\end{center}
\caption{Joining two open strings with boundary condition
$[\lambda^*, 0]$ and $[0, \tilde{\mu}]$ yields a string with
$[\lambda^*, \tilde{\mu}]$.
The thick line stands for a twist by the automorphism
$\omega \in G$.}
\label{fig:threetwist}
\end{figure}

Since the twisted Ishibashi state $\dket{(\lambda,\mu^*);\omega}$
composed from the representation $(\lambda,\mu^*)$
exists only when $\lambda = \omega(\mu)$,
the set of $\omega$-twisted Ishibashi states available
in the charge-conjugation invariant \eqref{eq:Zc} 
have the form
\beq
  \{\dket{(\lambda,\lambda^*);\omega} \, | \,
    \lambda \in \I(\omega) \}  \, . 
\eeq
A general boundary state $\ket{\tilde{\alpha}}$
satisfying the boundary condition
\eqref{eq:twistbc} in the charge-conjugation
modular invariant is therefore expressed as follows,
\beq
  \ket{\tilde{\alpha}} = \sum_{\lambda \in \I(\omega)}
  \Psi^\omega_{\tilde{\alpha} (\lambda, \lambda^*)}
  \dket{(\lambda,\lambda^*);\omega} \, .
\eeq
In order to determine the coefficients 
$\Psi^\omega_{\tilde{\alpha} (\lambda, \lambda^*)}$,
we consider the cylinder amplitude 
$\bra{0} \tq \ket{\tilde{\alpha}}$
with boundary conditions $0 \in \I$ and $\tilde{\alpha}$.
Since the boundary condition of the corresponding
open string is twisted by $\omega$
at only one end (see Fig.\ref{fig:threetwist}),
the chiral algebra in the open-string channel is
the twisted chiral algebra $\A^\omega$ of $\A$
associated with $\omega \in G$,
which is generated by the currents of $\A$ with the 
twisted boundary condition,
\beq
  W(e^{2 \pi i} z) = \omega(W(z)) \, .
\eeq
The cylinder amplitude $\bra{0} \tq \ket{\tilde{\alpha}}$
is therefore 
expanded into the characters of $\A^\omega$.

We denote by $\I^\omega$ the set of all the irreducible representations
of the twisted chiral algebra $\A^\omega$
and by $\chi_{\tilde{\lambda}}$ 
the character of $\tilde{\lambda} \in \I^\omega$.
The modular transformation of $\chi_{\tilde{\lambda}}$ can be expanded
into the twining characters \eqref{eq:twin} of $\A$.
Actually, evaluating the cylinder amplitude 
$\bra{0} \tq \ket{\tilde{\alpha}}$ in the closed-string channel
yields a linear combination of the overlap
$\dbra{(\lambda,\lambda^*)} \tq 
\dket{(\lambda,\lambda^*);\omega}$, 
which in turn provides the twining character 
$\chi^\omega_\lambda(\tilde{q})$ as we have shown in 
\eqref{eq:Ishibashinormtwisted}.~\footnote{%
This argument gives an explanation for the fact \cite{FSS1}
that the twining character 
for an affine Lie algebra $\mathfrak{g}^{(1)}$ with
the diagram automorphism of its horizontal Lie algebra 
as an automorphism $\omega$ of order $r$
can be expressed by characters
of a twisted affine Lie algebra $\mathfrak{g}'{}^{(r)}$, 
since the modular transformation
of characters of a twisted affine Lie algebra 
$\mathfrak{g}^{(r)}$ is expanded
into characters of a (generally different) 
twisted affine Lie algebra $\mathfrak{g}'{}^{(r)}$ \cite{Kac}.
}
Accordingly, we can express the modular transformation of the character
$\chi_{\tilde{\lambda}}$ in the following form,
\beq
  \chi_{\tilde{\lambda}}(q) = 
  \sum_{\mu \in \I(\omega)} 
  S^\omega_{\tilde{\lambda} \mu} \chi^\omega_\mu(\tilde{q}) 
  \quad (\tilde{\lambda} \in \I^\omega )
  \, .
  \label{eq:Somega}
\eeq
The matrix $S^\omega$ relates the representations of $\A^\omega$
with those of $\A$ fixed by $\omega$. 
We assume that $S^\omega$ is unitary.
In particular, two sets, $\I^\omega$ and $\I(\omega)$, have the same
order, $\abs{\I^\omega} = \abs{\I(\omega)}$. 
This holds for many examples including the case of 
an affine Lie algebra of untwisted type
with the diagram automorphism of its horizontal Lie algebra
as an automorphism $\omega$.~\footnote{%
A detailed study of the matrix $S^\omega$ for the case of
$\omega^2 = 1$ is given in \cite{BFS}. 
}

We can determine the coefficients of twisted Ishibashi states
in a way parallel to the untwisted one,
namely we require the condition
\beq
  \bra{0} \tq \ket{\tilde{\lambda}} 
  = \chi_{\tilde{\lambda}}(q) \quad (\tilde{\lambda} \in \I^\omega)
  \, .
  \label{eq:lambdatilde0}
\eeq
In the same way as the untwisted case,
twisted boundary states satisfying the above condition
can be constructed using the modular transformation
matrix of the chiral algebra in the open-string channel,
this time $\A^\omega$,
\beq
  \ket{\tilde{\lambda}} 
  = \sum_{\mu \in \I(\omega)} 
  S^\omega_{\tilde{\lambda} \mu} \dket{(\mu,\mu^*); \omega} \, .
  \label{eq:lt}
\eeq
We can check this expression yields the desired overlap 
\eqref{eq:lambdatilde0} with $\ket{0}$,
\beq
  \bra{0} \tq \ket{\tilde{\lambda}}
  = \sum_{\mu \in \I(\omega)} 
  \overline{S_{0 \mu}} S^\omega_{\tilde{\lambda} \mu}
  \frac{1}{S_{0 \mu}} \chi^\omega_\mu (\tilde{q})
  = \sum_{\mu \in \I(\omega)} 
    S^\omega_{\tilde{\lambda} \mu} \chi^\omega_\mu (\tilde{q})
  = \chi_{\tilde{\lambda}}(q) \, .
\eeq

Having obtained the explicit form of the twisted states,
one can calculate 
the overlap of the twisted state 
$\ket{\tilde{\mu}} \, (\tilde{\mu} \in \I^\omega)$
with generic untwisted
states,
\bes
  \bra{\lambda^*} \tq \ket{\tilde{\mu}}
  &= \sum_{\sigma \in \I(\omega)} \overline{S_{\lambda^* \sigma}} 
    S^\omega_{\tilde{\mu} \sigma}
  \frac{1}{S_{0 \sigma}} \chi^\omega_\sigma (\tilde{q}) \\
  &= \sum_{\sigma \in \I(\omega)} \overline{S_{\lambda^* \sigma}} 
    S^\omega_{\tilde{\mu} \sigma}
  \frac{1}{S_{0 \sigma}} 
  \sum_{\tilde{\nu} \in \I^\omega} 
  (S^\omega)^{-1}_{\sigma \tilde{\nu}} \chi_{\tilde{\nu}}(q) \\
  &= \sum_{\tilde{\nu} \in \I^\omega, \, \sigma \in \I(\omega)}
  \frac{%
    S_{\lambda \sigma} 
    S^\omega_{\tilde{\mu} \sigma}
    \overline{S^\omega_{\tilde{\nu} \sigma}}
    }{S_{0 \sigma}}
    \chi_{\tilde{\nu}}(q) \, . 
  \label{eq:lmt}
\ees
In the same way as the case of untwisted states,
one can regard this as the fusion of two open strings,
$[\lambda^*, 0]$ and $[0, \tilde{\mu}]$
(see Fig.\ref{fig:threetwist}). 
The spectrum of these two strings contains
only one representation,
$\lambda$ and $\tilde{\mu}$, respectively. 
Therefore the spectrum of the joined string $[\lambda^*, \tilde{\mu}]$
can be considered to express
the fusion $(\lambda) \times (\tilde{\mu})$ of two representations,
$\lambda \in \I$ and $\tilde{\mu} \in \I^\omega$. 
Since the string $[\lambda^*, \tilde{\mu}]$ has a twist by $\omega$
at only one end $\tilde{\mu}$, its spectrum 
consists of representations of the twisted chiral algebra $\A^\omega$. 
Hence one can conclude that
the product $(\lambda) \times (\tilde{\mu})$ is
expanded into representations of $\A^\omega$,
\beq
  (\lambda) \times (\tilde{\mu}) = \sum_{\tilde{\nu} \in \I^\omega}
  \fusion{\lambda}{\tilde{\mu}}{\tilde{\nu}} (\tilde{\nu}) \qquad
  (\lambda \in \I, \, \tilde{\mu} \in \I^\omega) \, .
\eeq
The coefficients $\fusion{\lambda}{\tilde{\mu}}{\tilde{\nu}}$
express the multiplicity of representation $(\tilde{\nu})$
in the product $(\lambda) \times (\tilde{\mu})$
and take values in non-negative integers. 
In terms of these coefficients, the overlap of
untwisted states with twisted states can be written in the 
following form,
\beq
  \bra{\lambda^*} \tq \ket{\tilde{\mu}}
  = \sum_{\tilde{\nu} \in \I^\omega}
    \fusion{\lambda}{\tilde{\mu}}{\tilde{\nu}}
    \chi_{\tilde{\nu}}(q) \, .
\eeq
Comparing this with the closed-string channel calculation
\eqref{eq:lmt}, one can obtain
a formula for $\fusion{\lambda}{\tilde{\mu}}{\tilde{\nu}}$,
\beq
  \fusion{\lambda}{\tilde{\mu}}{\tilde{\nu}}
  = \sum_{\rho \in \I(\omega)}
  \frac{%
    S_{\lambda \rho} 
    S^\omega_{\tilde{\mu} \rho}
    \overline{S^\omega_{\tilde{\nu} \rho}}
    }{S_{0 \rho}} \qquad
    (\lambda \in \I ; \, \tilde{\mu}, \tilde{\nu} \in \I^\omega)
    \, .
\eeq

The above result can be readily generalized to 
the fusion of two generic representations.
To state the result in a concise form,
we treat the untwisted states as the particular
case $\omega = 1$ of the twisted states,
and introduce a set $\Ihat$ of all the representations
of twisted chiral algebras $\A^\omega$ for $\omega \in G$,
\beq
  \Ihat = \coprod_{\omega \in G} \I^\omega \, ,
\eeq
which contains $\I = \I^{\omega = 1}$ as a subset.
We use a capital letter
such as $L$ for expressing an element of $\Ihat$ 
and denote the automorphism type of $L \in \Ihat$ by 
putting a subscript like $\omega_L \in G$.
In this notation, 
the modular transformation, 
eqs.\eqref{eq:S} and \eqref{eq:Somega},
of characters of $\A^\omega$
can be written as follows,
\beq
  \chi_L(q) = \sum_{\lambda \in \I(\omega_L)} 
  S^{\omega_L}_{L \lambda} \, \chi^{\omega_L}_\lambda(\tilde{q}) 
  \quad 
  (L \in \Ihat)
  \, .
  \label{eq:S_gen}
\eeq
If $L$ is a representation of $\A$, $L \in \I$, 
the modular transformation matrix $S^{\omega_L}$ is the ordinary
$S$-matrix and the twining character $\chi^{\omega_L}_\lambda$ is
nothing but the character $\chi_\lambda$ of $\A$.
The boundary states are labeled by the element of $\Ihat$ and
take the form given in eqs.\eqref{eq:lambda} and \eqref{eq:lt},
\beq
  \ket{L} 
  = \sum_{\lambda \in \I(\omega_L)} 
  S^{\omega_L}_{L \lambda} \dket{(\lambda,\lambda^*); \omega_L} 
  \quad (L \in \Ihat)
  \, .
  \label{eq:L}
\eeq
The overlap of $\ket{L}$ with the untwisted state $\ket{0}$
yields the character $\chi_L$ in the open-string channel,
\beq
  \bra{0} \tq \ket{L} = \chi_L(q) 
  \quad (L \in \Ihat) 
  \, ,
\eeq
which means that the open string $[0, L]$ has a spectrum consisting of
only one representation $L \in \Ihat$.

Taking the complex conjugation of this equation,
one obtains
the spectrum of the string $[L, 0]$.
Since the string $[L,0]$ has an orientation
opposite to that of $[0, L]$,
its chiral algebra has a boundary condition twisted by
$\omega_L^{-1}$ instead of $\omega_L$.
The spectrum of the string $[L,0]$ is therefore expanded 
into characters of $\A^{\omega_L^{-1}}$.
Since the string $[0,L]$ has only one representation, 
the spectrum of the string $[L,0]$ also consists of only one representation.
We call this the conjugate representation of $L$ and
denote it by $L^* \in \I^{\omega_L^{-1}} \subset \Ihat$,
\beq
  \bra{L} \tq \ket{0} = \chi_{L^*}(q) 
  \quad (L \in \Ihat) \, .
  \label{eq:cc_gen}
\eeq
This generalizes the complex conjugation of 
ordinary representations (see \eqref{eq:cc}).
Calculating the left-hand side of eq.\eqref{eq:cc_gen}
in terms of the Ishibashi states yields
\beq
  \chi_{L^*}(q) = \sum_{\lambda \in \I(\omega_L)}
  \overline{S^{\omega_L}_{L\lambda}} \,
  \chi^{\omega_L^{-1}}_\lambda(\tilde{q}) \, .
\eeq
Comparing this with eq.\eqref{eq:S_gen}, we obtain the following formula
\beq
  \overline{S^{\omega_L}_{L \lambda}} = 
  S^{\omega_L^{-1}}_{L^* \lambda}
  \quad (L \in \Ihat) \, ,
  \label{eq:Sbar}
\eeq
which reduces to eq.\eqref{eq:Scc} for the case of $\omega_L = 1$. 

Since the spectrum of two open strings, $[L^*, 0]$ and $[0, M]$, 
consists of only one representation, $L$ and $M$, respectively,
the spectrum of the joined string $[L^*, M]$ is given by
the fusion of two representations $L$ and $M$.
From the expression given in eq.\eqref{eq:L}, 
one can calculate the overlap of $\ket{L^*}$ and $\ket{M}$ 
as follows, 
\bes
  \bra{L^*} \tq \ket{M} 
  &= \sum_{\lambda \in \I(\omega_L) \cap \I(\omega_M)}
  \overline{S^{\omega_L^{-1}}_{L^* \lambda}} S^{\omega_M}_{M \lambda} \,
  \dbra{(\lambda,\lambda^*);\omega_L^{-1}} \tq 
  \dket{(\lambda,\lambda^*);\omega_M} \\
  &= \sum_{\lambda \in \I(\omega_L) \cap \I(\omega_M)}
  S^{\omega_L}_{L \lambda} S^{\omega_M}_{M \lambda} 
  \frac{1}{S_{0\lambda}} \chi^{\omega_L \omega_M}_\lambda(\tilde{q}) \\
  &= \sum_{\lambda \in \I(\omega_L) \cap \I(\omega_M)}
  S^{\omega_L}_{L \lambda} S^{\omega_M}_{M \lambda}
  \frac{1}{S_{0\lambda}} 
  \sum_{N \in \I^{\omega_L \omega_M}} 
  (S^{\omega_L \omega_M})^{-1}_{\lambda N} \, \chi_N(q) \\
  &= \sum_{N \in \I^{\omega_L \omega_M}} 
     \sum_{\lambda \in \I(\omega_L) \cap \I(\omega_M)}
  \frac{%
    S^{\omega_L}_{L \lambda} S^{\omega_M}_{M \lambda}
    \overline{S^{\omega_L \omega_M}_{N \lambda}}
    }{S_{0 \lambda}}
    \chi_{N}(q) \, ,
    \label{eq:LM}
\ees
where we used the formula \eqref{eq:Ishibashinormtwisted}
and the unitarity of $S^{\omega_L \omega_M}$.
This equation shows that the fusion of two representations $L$ and $M$
($L, M \in \Ihat$) can be expanded into representations
of the automorphism type $\omega_L \omega_M \in G$,
which are also the elements of $\Ihat$.
Therefore the set $\Ihat$ is closed under the fusion of representations
and forms an algebra
\beq
  (L) \times (M) = \sum_{N \in \Ihat}  \fusionhat{L}{M}{N} (N) \, ,
  \label{eq:gen_fusion}
\eeq
which generalizes the fusion algebra of ordinary untwisted representations. 
We call this the generalized fusion algebra and denote
it by $\F(\A; G)$.~\footnote{$\F$ stands for fusion.}
The coefficients $\fusionhat{L}{M}{N}$ take values in non-negative
integers and we call them the generalized fusion coefficients. 
As is shown in the overlap \eqref{eq:LM},
the generalized fusion coefficients are given by the
following formula
\beq
  \fusionhat{L}{M}{N} = 
  \sum_{\lambda \in
   \I(\omega_L) \cap \I(\omega_M)}
  \frac{%
    S^{\omega_L}_{L \lambda} 
    S^{\omega_M}_{M \lambda}
    \overline{S^{\omega_N}_{N \lambda}}
    }{S_{0 \lambda}} \times 
    \delta_{\omega_L \omega_M,  \, \omega_N} \, ,
    \label{eq:gen_Verlinde}
\eeq
which generalizes the Verlinde formula \eqref{eq:Verlinde0}
for the ordinary fusion coefficients.

As we have argued above,
the fusion of a representation
$L$ with $M$ has the automorphism type $\omega_L \omega_M$ and
is given by a linear combination of the elements in
$\I^{\omega_L \omega_M}$. 
An immediate consequence of this fact is 
that $(L) \times (M) \neq (M) \times (L)$ in general,
since two sets $\I^{\omega_L \omega_M}$ and $\I^{\omega_M \omega_L}$
are distinct from each other
if $\omega_L \omega_M \neq \omega_M \omega_L$. 
Therefore the generalized fusion algebra $\F(\A; G)$
is \textit{non-commutative}
if the automorphism group $G$ is non-abelian. 
In contrast to this,
there seems to be no special reason for
a generalized fusion algebra to be non-commutative
for an abelian $G$.
Actually, it follows from the definition \eqref{eq:gen_Verlinde} 
of the generalized fusion coefficients that
$(L) \times (M) = (M) \times (L)$ if
$\omega_L \omega_M = \omega_M \omega_L$. 
In particular, $\F(\A; G)$ is commutative if $G$ is abelian.
The ordinary fusion algebra $\F(\A)$ corresponds to the trivial case 
$\F(\A; \{1\})$
and is hence commutative. 

\subsection{An example: generalized fusion algebras for $so(8)_1$}
\label{sec:so8}

As an illustration of our argument,
we give the explicit form of
the generalized fusion algebra for the affine Lie algebra
$so(8)_1$ and its triality automorphism group,
which is isomorphic to the symmetric group $S_3$.
Since $S_3$ is non-abelian,
the generalized fusion algebra $\F(so(8)_1; S_3)$~\footnote{%
For brevity, we denote the automorphism group itself by $S_3$. 
} 
is non-commutative.

There are four irreducible representations for $so(8)_1$,
\beq
  \I = \{O = \Lambda_0, V = \Lambda_1, S = \Lambda_3, C = \Lambda_4 \} ,
  \label{eq:so8I}
\eeq
where $\Lambda_i$ is the fundamental weight of $so(8)_1$
and each symbol stands for the vacuum, vector, spinor and
conjugate spinor representation, respectively.
The modular transformation matrix reads
\beq
  S = \frac{1}{2} \begin{pmatrix}
   1 &  1 &  1 &  1 \\
   1 &  1 & -1 & -1 \\
   1 & -1 &  1 & -1 \\
   1 & -1 & -1 &  1 
   \end{pmatrix} \, ,
\label{eq:so8S}
\eeq
where the rows and the columns are ordered as \eqref{eq:so8I}.

The chiral algebra $so(8)_1$ has an automorphism group 
consisting of all the permutations of three legs of the Dynkin diagram
for the horizontal subalgebra $so(8) \subset so(8)_1$,
which is isomorphic to the symmetric group $S_3$.
The automorphism group $S_3$ is generated by two elements
$\pi$ and $\sigma$, which act on the elements of $\I$ as
\bes
  \pi    &: V \mapsto S \mapsto C 
  \mapsto V \, , \,\,
  O \mapsto O \, , \\
  \sigma &: 
  O \mapsto O \, , \,\,
  V \mapsto V \, , \,\,
  S \mapsto C \mapsto S \, .
  \label{eq:so8pisigma}
\ees
In terms of $\pi$ and $\sigma$, the elements of 
$S_3$ can be expressed as follows, 
\beq
  S_3 = \langle \pi, \sigma \rangle = 
  \{1, \pi, \pi^2, \sigma, \pi \sigma, \pi^2 \sigma \} \, . 
\eeq

In order to obtain the generalized fusion algebra,
we need to know twisted representations and their modular
transformation for each element of $S_3$. 
For $\pi$ and $\pi^2$, the fixed point set of the automorphism
consists of a single element
\beq
  \I(\pi) = \I(\pi^2) = \{ O \} \, .
\eeq
Correspondingly, there is only one representation
for each twisted chiral algebra, which is isomorphic to
the twisted affine Lie algebra $D_4^{(3)}$ at level 1.
(The irreducible representations and their modular transformation
matrix for the affine Lie algebra $D_4^{(r)} \, (r=1,2,3)$ 
are presented, \textit{e.g.}, in \cite{IY}.)
We denote the twisted representations for $\pi$ and $\pi^2$ 
as follows,
\beq
  \I^\pi = \{ (0)_\pi \} \, , \quad
  \I^{\pi^2} = \{ (0)_{\pi^2} \} \, .
\eeq
The modular transformation matrix in this case is simply a number
\beq
  S^\pi = (1) \, , \quad S^{\pi^2} = (1) \, .
\eeq
There are two representations fixed by $\sigma \in S_3$,
\beq
  \I(\sigma) = \{ O, \, V \} \, .
\eeq
The twisted chiral algebra for this case is isomorphic to 
$D_4^{(2)}$ at level 1. 
There are two representations of the twisted chiral algebra
for $\sigma$,
which we denote as
\beq
  \I^\sigma = \{ (0)_\sigma, \, (1)_\sigma \} \, .
\eeq
The modular transformation matrix reads
\beq
  S^\sigma = \frac{1}{\sqrt{2}}
  \begin{pmatrix} 1 & 1 \\ 1 & -1 \end{pmatrix} \, .
  \label{eq:so8sigma}
\eeq
For the other elements $\pi \sigma$ and $\pi^2 \sigma$,
one can proceed in the same way as $\sigma$ to obtain
\bea
  \qquad \I(\pi \sigma) &= \{ O, \, C \} \, , & 
  \I^{\pi \sigma} &= \{(0)_{\pi\sigma}, \, (1)_{\pi\sigma} \} \, , \qquad \\
  \I(\pi^2 \sigma) &= \{ O, \, S \} \, , & 
  \I^{\pi^2 \sigma} &= \{(0)_{\pi^2\sigma}, \, (1)_{\pi^2\sigma} \} \, .
\eea
The modular transformation matrix is the same as that for $\sigma$,
\beq
  S^{\pi\sigma} = S^{\pi^2 \sigma} = S^\sigma \, .
\eeq
Consequently, the set $\Ihat$ of all the twisted representations
consists of 12 elements,
\beq
  \abs{\Ihat} = \abs{\I} + \abs{\I^\pi} + \abs{\I^{\pi^2}}
  + \abs{\I^\sigma} + \abs{\I^{\pi \sigma}} + \abs{\I^{\pi^2 \sigma}}
  = 4 + 1 + 1 + 2 + 2 + 2 = 12 \, .
\eeq
Since all the modular transformation matrices are real, 
the relation \eqref{eq:Sbar} implies that the representations
in $\I, \I^{\sigma}, \I^{\pi\sigma}$ and $\I^{\pi^2 \sigma}$
are self-conjugate.
On the other hand,
for the representations in $\I^\pi$ and $\I^{\pi^2}$,
the conjugation changes the automorphism type,
$(0)_\pi^* = (0)_{\pi^2}$, since $\pi^{-1} = \pi^2$.
\begin{table}
{\footnotesize 
\bes
  &\renewcommand{\arraystretch}{1.5}
  \begin{array}{c | cccccc}
    & O & V & S & C & (0)_\pi & (0)_{\pi^2} \\
    \hline
  O     & O & V & S & C & (0)_\pi & (0)_{\pi^2} \\
  V     & V & O & C & S & (0)_\pi & (0)_{\pi^2} \\
  S     & S & C & O & V & (0)_\pi & (0)_{\pi^2} \\
  C     & C & S & V & O & (0)_\pi & (0)_{\pi^2} \\
  (0)_\pi & (0)_\pi & (0)_\pi & (0)_\pi & (0)_\pi & 
  2 \,(0)_{\pi^2} & O + V + S + C \\
  (0)_{\pi^2} & (0)_{\pi^2} & (0)_{\pi^2} & (0)_{\pi^2} & (0)_{\pi^2} & 
  O + V + S + C & 2 \,(0)_\pi \\
  (0)_\sigma & (0)_\sigma & (0)_\sigma & (1)_\sigma & (1)_\sigma &
  (0)_{\pi^2 \sigma} + (1)_{\pi^2 \sigma} &
  (0)_{\pi \sigma} + (1)_{\pi \sigma} \\
  (1)_\sigma & (1)_\sigma & (1)_\sigma & (0)_\sigma & (0)_\sigma &
  (0)_{\pi^2 \sigma} + (1)_{\pi^2 \sigma} &
  (0)_{\pi \sigma} + (1)_{\pi \sigma} \\
  (0)_{\pi \sigma} & (0)_{\pi \sigma} & (1)_{\pi \sigma} & 
  (1)_{\pi \sigma} & (0)_{\pi \sigma} &
  (0)_{\sigma} + (1)_{\sigma} &
  (0)_{\pi^2 \sigma} + (1)_{\pi^2 \sigma} \\
  (1)_{\pi \sigma} & (1)_{\pi \sigma} & (0)_{\pi \sigma} & 
  (0)_{\pi \sigma} & (1)_{\pi \sigma} &
  (0)_{\sigma} + (1)_{\sigma} &
  (0)_{\pi^2 \sigma} + (1)_{\pi^2 \sigma} \\
  (0)_{\pi^2 \sigma} & (0)_{\pi^2 \sigma} & (1)_{\pi^2 \sigma} & 
  (0)_{\pi^2 \sigma} & (1)_{\pi^2 \sigma} &
  (0)_{\pi \sigma} + (1)_{\pi \sigma} &
  (0)_{\sigma} + (1)_{\sigma} \\
  (1)_{\pi^2 \sigma} & (1)_{\pi^2 \sigma} & (0)_{\pi^2 \sigma} & 
  (1)_{\pi^2 \sigma} & (0)_{\pi^2 \sigma} &
  (0)_{\pi \sigma} + (1)_{\pi \sigma} &
  (0)_{\sigma} + (1)_{\sigma} 
  \end{array}\\[5\jot]
  &\renewcommand{\arraystretch}{1.5}
  \begin{array}{c | cccccc}
    & (0)_\sigma & (1)_\sigma & (0)_{\pi \sigma} & (1)_{\pi \sigma} &
      (0)_{\pi^2 \sigma} & (1)_{\pi^2 \sigma} \\
    \hline
  O     & (0)_\sigma & (1)_\sigma & (0)_{\pi \sigma} & (1)_{\pi \sigma} &
      (0)_{\pi^2 \sigma} & (1)_{\pi^2 \sigma} \\
  V     & (0)_\sigma & (1)_\sigma & (1)_{\pi \sigma} & (0)_{\pi \sigma} &
      (1)_{\pi^2 \sigma} & (0)_{\pi^2 \sigma} \\
  S     & (1)_\sigma & (0)_\sigma & (1)_{\pi \sigma} & (0)_{\pi \sigma} &
      (0)_{\pi^2 \sigma} & (1)_{\pi^2 \sigma} \\
  C     & (1)_\sigma & (0)_\sigma & (0)_{\pi \sigma} & (1)_{\pi \sigma} &
      (1)_{\pi^2 \sigma} & (0)_{\pi^2 \sigma} \\
  (0)_\pi & 
  (0)_{\pi \sigma} + (1)_{\pi \sigma} & 
  (0)_{\pi \sigma} + (1)_{\pi \sigma} & 
  (0)_{\pi^2 \sigma} + (1)_{\pi^2 \sigma} & 
  (0)_{\pi^2 \sigma} + (1)_{\pi^2 \sigma} & 
  (0)_{\sigma} + (1)_{\sigma} & 
  (0)_{\sigma} + (1)_{\sigma} \\
  (0)_{\pi^2} & 
  (0)_{\pi^2 \sigma} + (1)_{\pi^2 \sigma} & 
  (0)_{\pi^2 \sigma} + (1)_{\pi^2 \sigma} & 
  (0)_{\sigma} + (1)_{\sigma} & 
  (0)_{\sigma} + (1)_{\sigma} & 
  (0)_{\pi\sigma} + (1)_{\pi\sigma} & 
  (0)_{\pi\sigma} + (1)_{\pi\sigma} \\
  (0)_\sigma &
  O+V & S+C & (0)_{\pi^2} & (0)_{\pi^2} & (0)_{\pi} & (0)_{\pi} \\
  (1)_\sigma &
  S+C & O+V & (0)_{\pi^2} & (0)_{\pi^2} & (0)_{\pi} & (0)_{\pi} \\
  (0)_{\pi\sigma} &
  (0)_{\pi} & (0)_{\pi} & O+C & V+S & (0)_{\pi^2} & (0)_{\pi^2} \\
  (1)_{\pi\sigma} &
  (0)_{\pi} & (0)_{\pi} & V+S & O+C & (0)_{\pi^2} & (0)_{\pi^2} \\
  (0)_{\pi^2\sigma} &
  (0)_{\pi^2} & (0)_{\pi^2} & (0)_{\pi} & (0)_{\pi} & O+S & V+C \\
  (1)_{\pi^2\sigma} &
  (0)_{\pi^2} & (0)_{\pi^2} & (0)_{\pi} & (0)_{\pi} & V+C & O+S
  \end{array}
\notag
\ees
}
\caption{Multiplication table of the generalized fusion algebra
$\F(so(8)_1; S_3)$.
The subscripts stand for the automorphism type of twisted representations.
Since $\sigma \pi = \pi^2 \sigma \neq \pi \sigma$, this algebra
is non-commutative. 
}
\label{tab:so8S3}
\end{table}

Having obtained twisted representations and their modular transformation
matrices, it is straightforward to calculate the generalized fusion 
coefficients of $\F(so(8)_1; S_3)$
using the formula \eqref{eq:gen_Verlinde}. 
For example, the coefficient 
$\fusionhat{(0)_{\pi}}{(0)_{\sigma}}{(0)_{\pi \sigma}}$
can be obtained as follows,
\beq
  \fusionhat{(0)_{\pi}}{(0)_{\sigma}}{(0)_{\pi \sigma}}
  =   \sum_{\lambda \in
  \I(\pi) \cap \I(\sigma)}
  \frac{%
    S^{\pi}_{(0)_{\pi}\, \lambda} \,
    S^{\sigma}_{(0)_{\sigma}\, \lambda} \,
    \overline{S^{\pi \sigma}_{(0)_{\pi \sigma}\, \lambda}}
    }{S_{O \lambda}} 
  = \frac{%
    S^{\pi}_{(0)_{\pi}\, O} \,
    S^{\sigma}_{(0)_{\sigma}\, O} \,
    \overline{S^{\pi \sigma}_{(0)_{\pi \sigma}\, O}}
    }{S_{O O}}  
  = 1 \, .
\eeq
The other cases can be calculated in the same manner;
we give the result in Table~\ref{tab:so8S3}.
As this table shows explicitly, 
the generalized fusion algebra $\F(so(8)_1; S_3)$
is non-commutative.
For instance, the fusion of $(0)_\pi$ with $(0)_\sigma$
yields 
\bes
  (0)_\pi \times (0)_\sigma &= 
  (0)_{\pi \sigma} + (1)_{\pi \sigma} \, , \\
  (0)_\sigma \times (0)_\pi &= 
  (0)_{\pi^2 \sigma} + (1)_{\pi^2 \sigma} \, .
\ees
Since $\pi^2 \sigma \neq \pi \sigma$,
these two expressions are distinct from each other, namely
$(0)_\sigma \times (0)_\pi \neq (0)_\pi \times (0)_\sigma$,
which shows the non-commutativity of this algebra. 
One can also check the associativity of this algebra, 
for example,
\bes
  (0)_{\pi} \times \bigl((0)_\pi \times (0)_\sigma \bigr)
  &= (0)_{\pi} \times (0)_{\pi \sigma} + 
    (0)_{\pi} \times (1)_{\pi \sigma}
  = 2\, (0)_{\pi^2 \sigma} + 2\, (1)_{\pi^2 \sigma} \, , \\
  \bigl( (0)_{\pi} \times (0)_\pi \bigr) \times (0)_\sigma
  &= 2\, (0)_{\pi^2} \times (0)_\sigma
  = 2\, (0)_{\pi^2 \sigma} + 2\, (1)_{\pi^2 \sigma} \, .
\ees
A proof for the associativity of the generalized fusion
algebras is given in the next section.

The generalized fusion algebra $\F(so(8)_1; S_3)$ contains
non-trivial subalgebras other than the ordinary fusion algebra
$\F(so(8)_1)$.
Actually, a subalgebra of $\F(so(8)_1; S_3)$ exists 
for each subgroup of $S_3$,
\beq
  \{1\} , \,
  \langle \sigma \rangle , \,
  \langle \pi \sigma \rangle , \, 
  \langle \pi^2 \sigma \rangle ,  \,
  \langle \pi \rangle  = A_3 \cong \Z_3 \, .
\eeq
For example, 
corresponding to the three elements $\{1, \pi, \pi^2 \}$ of 
the alternating group $A_3 \subset S_3$, 
the representations in $\I, \I^\pi$ and $\I^{\pi^2}$
form a subalgebra which we denote as $\F(so(8)_1; A_3)$.
The generalized fusion algebra $\F(so(8)_1;S_3)$ is then
decomposed into representations of $\F(so(8)_1;A_3)$.
Indeed, from Table~\ref{tab:so8S3}, 
one can check that the set
$\I^\sigma\, \amalg\, \I^{\pi \sigma}\, \amalg\, \I^{\pi^2 \sigma}$
is invariant under the action of $\F(so(8)_1;A_3)$
and hence gives a representation of it.
Together with the regular representation of $\F(so(8)_1;A_3)$
on $\I \, \amalg \, \I^\pi \, \amalg \, \I^{\pi^2}$,
we eventually obtain two representations of $\F(so(8)_1;A_3)$ 
from the decomposition
of the regular representation of $\F(so(8)_1;S_3)$.

\section{Irreducible representations}
\label{sec:irrep}

For ordinary fusion algebras, the formula
\eqref{eq:gen_Verlinde} for the generalized fusion coefficients
reduces to the ordinary Verlinde formula
\beq
  \fusion{L}{M}{N} = 
  \sum_{\lambda \in \I}
  \frac{S_{L \lambda} S_{M \lambda} \overline{S_{N \lambda}}}{%
        S_{0 \lambda}} \, ,
  \label{eq:Verlinde}
\eeq
from which one can show that the generalized quantum dimension
\beq
  b_\lambda(L) = \frac{S_{L \lambda}}{S_{0 \lambda}}
  \quad (\lambda \in \I)
  \label{eq:qd}
\eeq
realizes a one-dimensional representation of the fusion algebra $\F(\A)$, 
\beq
  b_\lambda(L) b_\lambda(M) = \sum_{N \in \I}
  \fusion{L}{M}{N} \, b_\lambda(N) \, .
\eeq
The aim of this section is to extend this result to the case
of the generalized fusion algebra $\F(\A; G)$
for a general finite group $G$. 
As we shall show below, we have irreducible representations with
dimension greater than $1$ due to the non-commutativity of
$\F(\A; G)$ if $G$ is non-abelian.

We start our analysis by introducing a new basis for 
twining characters. 
Namely, we define the following linear combination
\beq
  \chi_{(\lambda; \rho)}^{ab}(q) = 
  \sqrt{\frac{\dim \rho}{\abs{\Sc(\lambda)}}}
  \sum_{\omega \in \Sc(\lambda)} 
  \overline{\rho^{ab}(\omega)} \, \chi^\omega_\lambda(q)
  = \sqrt{\frac{\dim \rho}{\abs{\Sc(\lambda)}}} 
    \trace_\lambda \Bigl(
     \sum_{\omega \in \Sc(\lambda)}
     R(\overline{\rho^{ab}(\omega)} \, \omega) \, 
     q^{L_0 - \frac{c}{24}} \Bigr) \, .
     \label{eq:twin2}
\eeq
Here $\Sc(\lambda) \subset G$ is the stabilizer of $\lambda$
defined in \eqref{eq:stab}, and
$\rho \in \Irr{\Sc(\lambda)}$ is a unitary
irreducible representation of
$\Sc(\lambda)$,
\beq
  \rho(\omega) = 
  \bigl(\rho^{ab}(\omega)\bigr)_{1 \le \, a, \, b \, \le n}
  \in U(n ; \C ) \quad (n = \dim \rho) \, . 
\eeq
The set 
$\{\chi_{(\lambda; \rho)}^{ab} \, | \, 
\rho \in \Irr{\Sc(\lambda)} ; \, 1 \le a, b \le \dim \rho\}$ 
forms a basis for the twining
characters of $\lambda \in \I$. 
Actually, one can express $\chi^\omega_\lambda$ in terms of 
$\chi_{(\lambda; \rho)}^{ab}$,
\beq
  \chi^\omega_\lambda(q) = \frac{1}{\sqrt{\abs{\Sc(\lambda)}}} 
  \sum_{\substack{%
  \rho \in \Irr{\Sc(\lambda)}\\
  1 \le a, b \le \dim \rho}} 
  \sqrt{\dim \rho} \, 
  \rho^{ab}(\omega) \, \chi_{(\lambda;\rho)}^{ab}(q)  \, .
\eeq
Here we used the orthogonality relations for matrix elements of
irreducible representations
\begin{subequations}
\bea
  \frac{1}{\abs{H}} \sum_{h \in H}
  \rho_i^{ab}(h) \, \overline{\rho_j^{cd}(h)} 
  = \frac{\delta_{ij} \delta^{ac} \delta^{bd}}{\dim \rho_i}  \quad
  (\rho_i, \rho_j \in \Irr{H})&  , 
  \label{eq:ortho1} \\
  \frac{1}{\abs{H}} 
  \sum_{\substack{%
  \rho \in \Irr{H} \\ 
  1 \le a, b \le \dim \rho}} 
  (\dim \rho) \, \rho^{ab}(h)\, \overline{\rho^{ab}(h')} = \delta_{h h'} \quad
  (h, h' \in H)&  \, ,
  \label{eq:ortho2}
\eea
\end{subequations}
which hold for any finite group $H$.

In terms of $\chi_{(\lambda;\rho)}^{ab}$, 
the modular transformation of the character
$\chi_L \, (L \in \Ihat)$
can be expressed as follows,
\bes
  \chi_L(q) &= \sum_{\lambda \in \I(\omega_L)} 
  S^{\omega_L}_{L \lambda} \chi^{\omega_L}_\lambda(\tilde{q})\\
  &= \sum_{\lambda \in \I(\omega_L)} S^{\omega_L}_{L \lambda} 
  \frac{1}{\sqrt{\abs{\Sc(\lambda)}}} 
  \sum_{\substack{%
  \rho \in \Irr{\Sc(\lambda)}\\
  1 \le a, b \le \dim \rho}} 
  \sqrt{\dim \rho} \, 
  \rho^{ab}(\omega_L) \, \chi_{(\lambda;\rho)}^{ab}(\tilde{q}) \\
    &= \sum_{\lambda \in \I(\omega_L)}
     \sum_{\substack{%
     \rho \in \Irr{\Sc(\lambda)}\\
     1 \le a, b \le \dim \rho}}
     S^{\omega_L}_{L \lambda}
     \sqrt{\frac{\dim \rho}{\abs{\Sc(\lambda)}}}
     \rho^{ab}(\omega_L) \, \chi_{(\lambda;\rho)}^{ab}(\tilde{q}) \, .
\ees
We write this in the following form,
\beq
  \chi_L(q) = \sum_{(\lambda;\rho) \in \Ihat^*}
  \sum_{1 \le a, b \le \dim \rho} 
  \Shat_{L (\lambda; \rho)}^{ab}
  \chi_{(\lambda; \rho)}^{ab}(\tilde{q}) \, . 
\eeq
Here the set $\Ihat^*$ and the matrix $\Shat$ are defined as follows,
\bea
  \Ihat^* &= \{ (\lambda; \rho) \, | \, \lambda \in \I; \, 
  \rho \in \Irr{\Sc(\lambda)} \} \, , 
  \label{eq:Ihatstar} \\
  \Shat_{L (\lambda; \rho)}^{ab} 
  &= \begin{cases}
    S^{\omega_L}_{L \lambda} 
    \sqrt{\frac{\dim \rho}{\abs{\Sc(\lambda)}}} \rho^{ab}(\omega_L) \quad
    &(\lambda \in \I(\omega_L) , \, \rho \in \Irr{\Sc(\lambda)}) \, ,\\
    0 \quad &(\lambda \notin \I(\omega_L)).
    \end{cases}
    \label{eq:Shat} 
\eea
We call $\Shat$ the generalized $S$-matrix of the chiral algebra $\A$. 
This matrix $\Shat$ is square since 
\beq
  \abs{\Ihat} = \sum_{\omega \in G} \abs{\I^\omega}
  = \sum_{\omega \in G} \abs{\I(\omega)} 
  = \sum_{\lambda \in \I} \abs{\Sc(\lambda)} 
  = \sum_{\lambda \in \I} \sum_{\rho \in \Irr{\Sc(\lambda)}} (\dim \rho)^2 
  = \sum_{(\lambda; \rho) \in \Ihat^*} (\dim \rho)^2 \, .
\eeq
Moreover, $\Shat$ is unitary, 
\bes
  \sum_{(\lambda; \rho) \in \Ihat^*; a, b} 
  \Shat_{L (\lambda;\rho)}^{ab} \,
  \overline{\Shat_{M (\lambda; \rho)}^{ab}} 
  &= \sum_{\lambda \in \I(\omega_L) \cap \I(\omega_M)}
     S^{\omega_L}_{L \lambda} \, \overline{S^{\omega_M}_{M \lambda}}
     \sum_{\rho \in \Irr{\Sc(\lambda)}}
     \sum_{a, b}
     \frac{\dim \rho}{\abs{\Sc(\lambda)}}
     \rho^{ab}(\omega_L) \overline{\rho^{ab}(\omega_M)} \\
  &= \sum_{\lambda \in \I(\omega_L)}
     S^{\omega_L}_{L \lambda} \, \overline{S^{\omega_L}_{M \lambda}}
     \times \delta_{\omega_L \omega_M} \\
  &= \delta_{LM} \, ,
\ees
where we used the orthogonality relation \eqref{eq:ortho2}.
The same matrix for the case of $G \cong \Z_2$
is found in \cite{BFS,GG}.

One can express the generalized fusion coefficients 
\eqref{eq:gen_Verlinde} in a form completely 
parallel to the ordinary Verlinde formula \eqref{eq:Verlinde}
using the generalized $S$-matrix \eqref{eq:Shat}
in place of the ordinary $S$-matrix. 
Namely, we can prove the following formula
\beq
  \fusionhat{L}{M}{N} =
  \sum_{(\lambda; \rho) \in \Ihat^*; a, b, c}
  \frac{%
  \Shat_{L (\lambda; \rho)}^{ab}
  \Shat_{M (\lambda; \rho)}^{bc}
  \overline{\Shat_{N (\lambda; \rho)}^{ac}}}{%
  \Shat_{0 (\lambda; \rho)}^{11}} \, . 
  \label{eq:gen_Verlinde2}
\eeq
This can be readily checked by a straightforward calculation, 
\bes
  \lefteqn{\text{r.h.s. of \eqref{eq:gen_Verlinde2}}} \\
  &= \sum_{\lambda \in \I(\omega_L) \cap \I(\omega_M) \cap \I(\omega_N)}
  \frac{%
  S^{\omega_L}_{L \lambda}\, S^{\omega_M}_{M \lambda} \,
  \overline{S^{\omega_N}_{N \lambda}}}{S_{0 \lambda}}
  \sum_{\rho \in \Irr{\Sc(\lambda)}} \sum_{a, b, c}
  \frac{\dim \rho}{\abs{\Sc(\lambda)}}
  \rho^{ab}(\omega_L) \rho^{bc}(\omega_M) 
  \overline{\rho^{ac}(\omega_N)} \\
  &= \sum_{\lambda \in \I(\omega_L) \cap \I(\omega_M) \cap \I(\omega_N)}
  \frac{%
  S^{\omega_L}_{L \lambda}\, S^{\omega_M}_{M \lambda} \,
  \overline{S^{\omega_N}_{N \lambda}}}{S_{0 \lambda}}
  \sum_{\rho \in \Irr{\Sc(\lambda)}}
  \sum_a \frac{\dim \rho}{\abs{\Sc(\lambda)}}
  \rho^{aa}(\omega_L \omega_M \omega_N^{-1}) \\
  &= \sum_{\lambda \in \I(\omega_L) \cap \I(\omega_M)}
  \frac{%
  S^{\omega_L}_{L \lambda}\, S^{\omega_M}_{M \lambda} \,
  \overline{S^{\omega_N}_{N \lambda}}}{S_{0 \lambda}}
  \times \delta_{\omega_L \omega_M, \omega_N} \\
  &= \fusionhat{L}{M}{N} \, ,
\ees
where we used the orthogonality relation \eqref{eq:ortho2}
for $h'=1$. 
We call eq.\eqref{eq:gen_Verlinde2} the generalized Verlinde
formula. 

The similarity of eq.\eqref{eq:gen_Verlinde2} to the ordinary Verlinde
formula \eqref{eq:Verlinde} suggests that  
the quantum dimension \eqref{eq:qd} is also generalized to 
the case of generalized fusion algebras. 
Actually, one can prove the following
\beq
  \bb_{(\lambda;\rho)}(L) \,\bb_{(\lambda;\rho)}(M)
  = \sum_{N \in \Ihat} \fusionhat{L}{M}{N}\, \bb_{(\lambda;\rho)}(N) \, ,
  \label{eq:bbalgebra}
\eeq
where $\bb_{(\lambda;\rho)}(L)$ is a matrix defined as
\beq
  \bb_{(\lambda;\rho)}(L) 
  = \Biggl(
  \frac{\Shat_{L (\lambda;\rho)}^{ab}}{\Shat_{0 (\lambda;\rho)}^{11}}
  \Biggr)_{1 \le a, b \le n} 
  \in M_n(\C) 
  \quad
  (n = \dim \rho, (\lambda; \rho) \in \Ihat^* )
  \, .
  \label{eq:bb}
\eeq
The proof of eq.\eqref{eq:bbalgebra} is straightforward,
\bes
  \sum_{N \in \Ihat} \fusionhat{L}{M}{N}\, \bb_{(\lambda;\rho)}^{ab}(N)
  &= \sum_{N \in \Ihat}
  \sum_{(\lambda'; \rho') \in \Ihat^*; a', b', c'}
  \frac{%
  \Shat_{L (\lambda'; \rho')}^{a'b'}
  \Shat_{M (\lambda'; \rho')}^{b'c'}
  \overline{\Shat_{N (\lambda'; \rho')}^{a'c'}}}{%
  \Shat_{0 (\lambda'; \rho')}^{11}} \,
  \frac{\Shat_{N (\lambda;\rho)}^{ab}}{\Shat_{0 (\lambda;\rho)}^{11}} \\
  &= \sum_{(\lambda'; \rho') \in \Ihat^*; a', b', c'}
  \frac{%
  \Shat_{L (\lambda'; \rho')}^{a'b'}
  \Shat_{M (\lambda'; \rho')}^{b'c'}}{%
  \Shat_{0 (\lambda'; \rho')}^{11} \Shat_{0 (\lambda; \rho)}^{11}} \, 
  \delta_{\lambda' \lambda} \delta_{\rho' \rho} 
  \delta^{a' a} \delta^{c' b} \\
  &= \sum_{b'}
  \frac{%
  \Shat_{L (\lambda; \rho)}^{a b'}}{%
  \Shat_{0 (\lambda; \rho)}^{11}} \, 
  \frac{%
  \Shat_{M (\lambda; \rho)}^{b' b}}{%
  \Shat_{0 (\lambda; \rho)}^{11}} \\
  &= \sum_{b'} \bb_{(\lambda;\rho)}^{ab'}(L) \, 
     \bb_{(\lambda;\rho)}^{b'b}(M) \, .
\ees
The above equation \eqref{eq:bbalgebra} means that 
$\bb_{(\lambda;\rho)}$ defined in \eqref{eq:bb} realizes 
a representation of the generalized fusion algebra $\F(\A; G)$
by linearly extending its action on the basis $(L)$ to 
the entire algebra.  
For the case of ordinary fusion algebras,
$\bb_{(\lambda; \rho)}$ reduces
to the quantum dimension $b_\lambda$ since the generalized
$S$-matrix $\Shat$ coincides with the ordinary $S$-matrix
for ordinary fusion algebras. 
Therefore $\bb_{(\lambda;\rho)}$ actually gives a generalization
of the one-dimensional representations $b_\lambda$ to the case of 
generalized fusion algebras. 

Since we define the generalized fusion algebra by products of
the boundary vertex operators, we can expect that the generalized
fusion algebra is associative.
However the associativity is not manifest from the definition
\eqref{eq:gen_Verlinde} of the generalized fusion coefficients
and its proof is desirable.
Let $\Nhat_L$ be a matrix $(\Nhat_L)_M{}^N = \fusionhat{M}{L}{N}$.  
The associativity of the generalized fusion algebra is equivalent
with the following condition for $\Nhat_L$, 
\beq
  \Nhat_{L} \Nhat_{M} = \sum_{N \in \Ihat} \fusionhat{L}{M}{N} \Nhat_{N} \, .
  \label{eq:Nhatalgebra}
\eeq
One can easily see that this follows from the generalized Verlinde formula
\eqref{eq:gen_Verlinde2}.
First note that
\bes
  (\Nhat_L)_M{}^N =
  \fusionhat{M}{L}{N} &=
  \sum_{(\lambda; \rho); a, b}
  \sum_{(\lambda'; \rho'); a', b'}
    \Shat_{M (\lambda; \rho)}^{ab} \,
    \delta_{\lambda \lambda'} \delta_{\rho \rho'} \delta^{a a'}
  \frac{\Shat_{L (\lambda; \rho)}^{b b'}}{\Shat_{0 (\lambda; \rho)}^{11}} \,
    \overline{\Shat_{N (\lambda'; \rho')}^{a'b'}} \\
  &=
  \sum_{(\lambda; \rho); a, b}
  \sum_{(\lambda'; \rho'); a', b'}
    \Shat_{M (\lambda; \rho)}^{ab} \,
    \delta_{\lambda \lambda'} \delta_{\rho \rho'} \delta^{a a'}
    \bb_{(\lambda;\rho)}^{b b'}(L)  \,
    \overline{\Shat_{N (\lambda'; \rho')}^{a'b'}} \\  
  &= \Biggl[\Shat \,\,
  \bigoplus_{(\lambda;\rho)} \bigl(\,
  \underbrace{%
  \bb_{(\lambda; \rho)}(L) \oplus \dots \oplus \bb_{(\lambda; \rho)}(L)
  }_{\text{$(\dim \rho)$ terms}} \, \bigr) \,\,
  \Shat^\dagger \Biggr]_M^{\,\, N} \, ,
  \label{eq:Shat_decomposition}
\ees
which shows that the matrix $\Nhat_L$ is similar to a direct sum of 
$\bb_{(\lambda;\rho)}(L)$. 
The equation \eqref{eq:Nhatalgebra} is an immediate consequence
of this fact
since $\bb_{(\lambda;\rho)}(L)$ satisfies the generalized fusion algebra
\eqref{eq:bbalgebra} for all $(\lambda;\rho) \in \Ihat^*$. 
The generalized fusion algebra is therefore associative and
the matrix $\Nhat_L$ realizes the (right) regular representation
of the generalized fusion algebra. 

We next show the mutual independence, in particular irreducibility,
of $\bb_{(\lambda;\rho)}$ for $(\lambda;\rho) \in \Ihat^*$. 
To see this, we use the explicit form of $\bb_{(\lambda;\rho)}(L)$,
\beq
  \bb_{(\lambda; \rho)}(L) =
  \begin{cases}
    \frac{S^{\omega_L}_{L \lambda}}{S_{0 \lambda}} \rho(\omega_L)
    \quad &(\omega_L \in \Sc(\lambda)) \, ,\\
    0     &(\text{otherwise}) \, ,
  \end{cases}
  \label{eq:bb2}
\eeq
which follows from eq.\eqref{eq:Shat}. 
Suppose that $\bb_{(\lambda;\rho)}$ is reducible.
Then there exists a matrix $X$ such that
\beq
  X \bb_{(\lambda; \rho)}(L) X^{-1} =
  \begin{pmatrix} * & * \\ 0 & * \end{pmatrix}
  \quad \text{for all $L \in \Ihat$} \, .
\eeq
However this implies that $\rho \in \Irr{\Sc(\lambda)}$ is also reducible,
since there exists at least one representation $L \in \I^{\omega}$
for each $\omega \in \Sc(\lambda)$
such that $S^{\omega}_{L \lambda} \neq 0$
and 
$\rho(\omega) = \frac{S_{0 \lambda}}{S^\omega_{L \lambda}}
\bb_{(\lambda;\rho)}(L)$.
This contradiction proves that $\bb_{(\lambda;\rho)}$ is irreducible for all
$(\lambda;\rho) \in \Ihat^*$. 

In order to show the mutual independence of $\bb_{(\lambda;\rho)}$, 
we introduce the character of $\bb_{(\lambda;\rho)}$,
\beq
  \hat{b}_{(\lambda;\rho)}(L) = \sum_a \bb_{(\lambda;\rho)}^{aa}(L)
  = \begin{cases}
    \frac{S^{\omega_L}_{L \lambda}}{S_{0 \lambda}} \psi(\omega_L)
    \quad &(\omega_L \in \Sc(\lambda)) \, ,\\
    0     &(\text{otherwise}) \, ,
  \end{cases}
\eeq
where $\psi$ is the group character of the representation
$\rho \in \Irr{\Sc(\lambda)}$. 
These characters satisfy the following orthogonality
relation
\beq
  \sum_{L \in \Ihat}
  \hat{b}_{(\lambda; \rho)}(L) \,
  \overline{\hat{b}_{(\lambda';\rho')}(L)} = 
  \sum_{a, b} \sum_L
  \frac{\Shat_{L (\lambda;\rho)}^{aa}}{\Shat_{0 (\lambda;\rho)}^{11}} \,
  \frac{\overline{\Shat_{L (\lambda';\rho')}^{bb}}}{%
  \Shat_{0 (\lambda';\rho')}^{11}} 
  =
  \sum_a 
  \frac{%
  \delta_{\lambda \lambda'} \delta_{\rho \rho'} \delta^{aa}}{%
  (S_{0 \lambda})^2 \frac{\dim \rho}{\abs{\Sc(\lambda)}}} 
  = 
  \frac{%
  \delta_{\lambda \lambda'} \delta_{\rho \rho'}}{%
  \frac{1}{\abs{\Sc(\lambda)}}(S_{0 \lambda})^2 } \, .
  \label{eq:bhatortho}
\eeq
Suppose that two representations $\bb_{(\lambda;\rho)}$ and
$\bb_{(\lambda';\rho')}$ are similar, \textit{i.e.},
there exists a matrix $Y$ such that
\beq
  \bb_{(\lambda;\rho)}(L) = Y \bb_{(\lambda';\rho')}(L) Y^{-1}
  \quad \text{for all $L \in \Ihat$} \, .
\eeq
Then the corresponding characters are the same,
$\hat{b}_{(\lambda;\rho)}(L) = \hat{b}_{(\lambda';\rho')}(L)$.
However this implies that $(\lambda; \rho) = (\lambda'; \rho')$
since
\beq
  \sum_{L \in \Ihat}
  \hat{b}_{(\lambda; \rho)}(L) \,
  \overline{\hat{b}_{(\lambda';\rho')}(L)} = 
  \sum_L \abs{\hat{b}_{(\lambda; \rho)}(L)}^2 > 0 
\eeq
which is not possible for $(\lambda; \rho) \neq (\lambda'; \rho')$
from the orthogonality relation \eqref{eq:bhatortho}.
Therefore two representations $\bb_{(\lambda;\rho)}$ and
$\bb_{(\lambda';\rho')}$ are distinct if 
$(\lambda; \rho) \neq (\lambda'; \rho')$.
To summarize, we have shown that
$\{\bb_{(\lambda;\rho)} | (\lambda; \rho) \in \Ihat^* \}$
is a set of mutually independent irreducible representations of
the generalized fusion algebra. 
\bigskip

\begin{table}
{
\beq
  \begin{array}{c | cccccc}
    & 1 & \pi & \pi^2 & \sigma & \pi \sigma & \pi^2 \sigma \\
    \hline \vphantom{\tilde{\Ihat}}
  \rho_0 & 1 & 1 & 1 & 1 & 1 & 1 \\[2\jot]
  \rho_1 & 1 & 1 & 1 & -1 & -1 & -1 \\[2\jot]
  \rho_2 
  & \begin{pmatrix} 1 & 0 \\ 0 & 1 \end{pmatrix}
  & \begin{pmatrix} \kappa & 0 \\ 0 & \bar{\kappa} \end{pmatrix}
  & \begin{pmatrix} \bar{\kappa} & 0 \\ 0 & \kappa \end{pmatrix}
  & \begin{pmatrix} 0 & 1 \\ 1 & 0 \end{pmatrix}
  & \begin{pmatrix} 0 & \kappa \\ \bar{\kappa} & 0 \end{pmatrix}
  & \begin{pmatrix} 0 & \bar{\kappa} \\ \kappa & 0 \end{pmatrix}
  \end{array}\notag
\eeq
}
\caption{Irreducible representations 
$\rho_0, \rho_1, \rho_2$
of the symmetric group $S_3 = \langle \pi, \sigma \rangle$.
$\kappa$ is a cube root of $1$, 
$\kappa = e^{\frac{2\pi i}{3}}$.}
\label{tab:S3irreps}
\end{table}
We illustrate our results by the generalized fusion
algebra $\F(so(8)_1;S_3)$ presented in Section~\ref{sec:so8}.
The stabilizer \eqref{eq:stab} 
for the representations \eqref{eq:so8I} of $so(8)_1$
is determined from the action \eqref{eq:so8pisigma} of $S_3$
on the representations,
\beq
  \Sc(O) = S_3 \, , \quad
  \Sc(V) = \{1, \sigma \} \, , \quad
  \Sc(S) = \{1, \pi^2 \sigma \} \, , \quad
  \Sc(C) = \{1, \pi \sigma \} \, .
\eeq
The stabilizers for $V, S, C$ are isomorphic to $\Z_2$,
for which there are two irreducible representations.
We denote them as $\rho_\pm \in \Irr{\Z_2}$, which are
defined for the case of $\Sc(V)$ as
\beq
  \rho_\pm(1) = 1 \, , \quad
  \rho_\pm(\sigma) = \pm 1 \, .
  \label{eq:Z2irreps}
\eeq
For the symmetric group $S_3$, there are three irreducible 
representations, $\rho_0, \rho_1$ and $\rho_2$, 
two of which are one-dimensional
while the remaining one is two-dimensional
(see Table~\ref{tab:S3irreps}).
Accordingly, the set $\Ihat^*$ in \eqref{eq:Ihatstar}
takes the form
\bes
  \Ihat^* &= \{ (\lambda; \rho) \, | \, \lambda \in \I; \, 
  \rho \in \Irr{\Sc(\lambda)} \} \\
  &= \{(O;0) , (O;1), (O;2), (V;+), (V;-), (S;+), (S;-), (C;+), (C;-) \} \, ,
\ees
where we write $\rho_a \,(a=0,1,2;\pm)$ as $a$ for simplicity. 
The generalized $S$-matrix \eqref{eq:Shat}, 
which is a $12 \times 12$ matrix in the present case,
is then obtained in the following form,
\bes
  &\Shat = \frac{1}{2\sqrt{6}} \times \\
  &{\scriptsize
  \renewcommand{\arraystretch}{2}
  \begin{array}{c | cccccccccccc}
  & (O;0) & (O;1) & (O;2)_{11} & (O;2)_{22} & (O;2)_{12} & (O;2)_{21}
  & (V;+) & (V;-) & (S;+) & (S;-) & (C;+) & (C;-) \\
  \hline
  O &
  1 & 1 & \sqrt{2} & \sqrt{2} & 0 & 0 & 
  \sqrt{3} & \sqrt{3} & \sqrt{3} & \sqrt{3} & \sqrt{3} & \sqrt{3} \\
  V &
  1 & 1 & \sqrt{2} & \sqrt{2} & 0 & 0 &  
  \sqrt{3} & \sqrt{3} & -\sqrt{3} & -\sqrt{3} & -\sqrt{3} & -\sqrt{3} \\
  S &
  1 & 1 & \sqrt{2} & \sqrt{2} & 0 & 0 & 
  -\sqrt{3} & -\sqrt{3} & \sqrt{3} & \sqrt{3} & -\sqrt{3} & -\sqrt{3} \\
  C &
  1 & 1 & \sqrt{2} & \sqrt{2} & 0 & 0 & 
  -\sqrt{3} & -\sqrt{3} & -\sqrt{3} & -\sqrt{3} & \sqrt{3} & \sqrt{3} \\
  (0)_\pi &
  2 & 2 & \sqrt{8} \kappa & \sqrt{8} \bar{\kappa} & 0 & 0 &
  0 & 0 & 0 & 0 & 0 & 0 \\
  (0)_{\pi^2} &
  2 & 2 & \sqrt{8} \bar{\kappa} & \sqrt{8} \kappa & 0 & 0 &
  0 & 0 & 0 & 0 & 0 & 0 \\
  (0)_\sigma &
  \sqrt{2} & -\sqrt{2} & 0 & 0 & 2 & 2 & 
  \sqrt{6} & -\sqrt{6} & 0 & 0 & 0 & 0 \\
  (1)_\sigma &
  \sqrt{2} & -\sqrt{2} & 0 & 0 & 2 & 2 & 
  -\sqrt{6} & \sqrt{6} & 0 & 0 & 0 & 0 \\
  (0)_{\pi \sigma} &
  \sqrt{2} & -\sqrt{2} & 0 & 0 & 2\kappa & 2\bar{\kappa} & 
  0 & 0 & 0 & 0 & \sqrt{6} & -\sqrt{6} \\
  (1)_{\pi \sigma} &
  \sqrt{2} & -\sqrt{2} & 0 & 0 & 2\kappa & 2\bar{\kappa} & 
  0 & 0 & 0 & 0 & -\sqrt{6} & \sqrt{6} \\
  (0)_{\pi^2 \sigma} &
  \sqrt{2} & -\sqrt{2} & 0 & 0 & 2\bar{\kappa} & 2\kappa & 
  0 & 0 & \sqrt{6} & -\sqrt{6} & 0 & 0 \\
  (1)_{\pi^2 \sigma} &
  \sqrt{2} & -\sqrt{2} & 0 & 0 & 2\bar{\kappa} & 2\kappa & 
  0 & 0 & -\sqrt{6} & \sqrt{6} & 0 & 0 
  \end{array} } \, ,
\ees
where $\kappa = e^{\frac{2 \pi i}{3}}$.
One can readily check that this matrix is unitary and that
the generalized Verlinde formula \eqref{eq:gen_Verlinde2}
reproduces the generalized fusion coefficients 
of $\F(so(8)_1;S_3)$ given in Table~\ref{tab:so8S3}.
From this matrix $\Shat$, one can construct
the irreducible representations \eqref{eq:bb} of $\F(so(8)_1;S_3)$.
We give the result in Table~\ref{tab:so8bb}.
From this table,
one can check that $\bb_{(\lambda;\rho)}(L) \,((\lambda;\rho) \in \Ihat^*)$
indeed satisfies the generalized fusion algebra in Table~\ref{tab:so8S3}
and realizes a representation of $\F(so(8)_1;S_3)$.
\bigskip

Before concluding this section, we comment on the case that the automorphism
group $G$ is abelian. 
Since all the irreducible representations are one-dimensional for 
an abelian $G$, the representation matrix $\rho(\omega)$ has
only one component $\rho^{11}(\omega)$. 
Therefore
we can omit the suffix and write $\rho(\omega)$
instead of $\rho^{11}(\omega)$. 
This enables us to express the formulas we have given above
in a simple form.
First, the basis \eqref{eq:twin2} of twining characters
and its modular transformation 
can be written as follows,
\bea
  \chi_{(\lambda; \rho)}(q) &= 
  \frac{1}{\sqrt{\abs{\Sc(\lambda)}}}
  \sum_{\omega \in \Sc(\lambda)} 
  \overline{\rho(\omega)} \, \chi^\omega_\lambda(q) \, , 
     \label{eq:twin2abel} \\
  \chi_L(q) &= \sum_{(\lambda;\rho) \in \Ihat^*}
  \Shat_{L (\lambda; \rho)}
  \chi_{(\lambda; \rho)}(\tilde{q}) \, ,
\eea
where the generalized $S$-matrix is defined as
\beq
  \Shat_{L (\lambda; \rho)} 
  = \begin{cases}
    S^{\omega_L}_{L \lambda} 
    \frac{1}{\sqrt{\abs{\Sc(\lambda)}}} \rho(\omega_L) \quad
    &(\lambda \in \I(\omega_L) , \, \rho \in \Irr{\Sc(\lambda)}) \, ,\\
    0 \quad &(\lambda \notin I(\omega_L)).
    \end{cases}
    \label{eq:Shatabel} 
\eeq
The generalized Verlinde formula \eqref{eq:gen_Verlinde2}
then takes the form
\beq
  \fusionhat{L}{M}{N} =
  \sum_{(\lambda; \rho) \in \Ihat^*}
  \frac{%
  \Shat_{L (\lambda; \rho)}
  \Shat_{M (\lambda; \rho)}
  \overline{\Shat_{N (\lambda; \rho)}}}{%
  \Shat_{0 (\lambda; \rho)}} \, . 
  \label{eq:gen_Verlinde2abel}
\eeq
For $G \cong \Z_2$, 
the generalized $S$-matrix \eqref{eq:Shatabel} 
and the generalized Verlinde formula \eqref{eq:gen_Verlinde2abel} 
reproduce the result given in \cite{BFS,GG}.  
The matrix $\Shat$ is denoted as $\tilde{S}$ in \cite{BFS},
which is considered to be a particular case of the same matrix in \cite{FS}
for a general finite abelian $G$. 
This suggests that two matrices $\Shat$ in \eqref{eq:Shatabel}
and $\tilde{S}$ in \cite{FS} are the same,
although we have no proof for this statement.
If this is the case, the classifying algebra in \cite{FS} is the dual
of the generalized fusion algebra \eqref{eq:gen_Verlinde2abel}
in the sense of $C$-algebras \cite{BI,DZ,PZ} with the structure constants
\beq
  M_{(\lambda;\rho)(\lambda';\rho')}{}^{(\lambda'';\rho'')} =
  \sum_{L \in \Ihat}
  \frac{%
  \Shat_{L (\lambda; \rho)}
  \Shat_{L (\lambda'; \rho')}
  \overline{\Shat_{L (\lambda''; \rho'')}}}{%
  \Shat_{L (0; \rho_0)}} \, ,
\eeq
where $\rho_0$ is the identity representation 
$\rho_0(\omega) = 1 \, (\forall \omega \in G)$ of $G$.
Clearly, 
the irreducible representation of this algebra takes the form
$(\lambda;\rho) \mapsto 
\frac{\Shat_{L (\lambda; \rho)}}{\Shat_{L (0; \rho_0)}}$
and is labeled by $L \in \Ihat$.
\begin{table}
{\footnotesize
\beq
\hspace*{-1cm}
  \begin{array}{c | cccccccccccc}
    & O & V & S & C & (0)_\pi & (0)_{\pi^2} 
    & (0)_\sigma & (1)_\sigma 
    & (0)_{\pi\sigma} & (1)_{\pi\sigma} 
    & (0)_{\pi^2\sigma} & (1)_{\pi^2\sigma} \\
    \hline \vphantom{\tilde{\Ihat}}
  \bb_{(O;0)} 
  & 1 & 1 & 1 & 1 & 2 & 2 
  & \sqrt{2} & \sqrt{2} & \sqrt{2} 
  & \sqrt{2} & \sqrt{2} & \sqrt{2} \\[2\jot]
  \bb_{(O;1)} & 1 & 1 & 1 & 1 & 2 & 2 
  & -\sqrt{2} & -\sqrt{2} & -\sqrt{2} 
  & -\sqrt{2} & -\sqrt{2} & -\sqrt{2} \\[2\jot]
  \bb_{(O;2)} 
  & \bigl(\begin{smallmatrix} 1 & 0 \\ 0 & 1 \end{smallmatrix}\bigr)
  & \bigl(\begin{smallmatrix} 1 & 0 \\ 0 & 1 \end{smallmatrix}\bigr)
  & \bigl(\begin{smallmatrix} 1 & 0 \\ 0 & 1 \end{smallmatrix}\bigr)
  & \bigl(\begin{smallmatrix} 1 & 0 \\ 0 & 1 \end{smallmatrix}\bigr)
  & 2\bigl(\begin{smallmatrix} 
  \kappa & 0 \\ 0 & \bar{\kappa} \end{smallmatrix}\bigr)
  & 2\bigl(\begin{smallmatrix} 
  \bar{\kappa} & 0 \\ 0 & \kappa \end{smallmatrix}\bigr) 
  & \sqrt{2} \bigl(\begin{smallmatrix}
  0 & 1 \\ 1 & 0 \end{smallmatrix} \bigr)
  & \sqrt{2} \bigl(\begin{smallmatrix}
  0 & 1 \\ 1 & 0 \end{smallmatrix} \bigr)
  & \sqrt{2} \bigl(\begin{smallmatrix}
  0 & \kappa \\ \bar{\kappa} & 0 \end{smallmatrix} \bigr)
  & \sqrt{2} \bigl(\begin{smallmatrix}
  0 & \kappa \\ \bar{\kappa} & 0 \end{smallmatrix} \bigr)
  & \sqrt{2} \bigl(\begin{smallmatrix}
  0 & \bar{\kappa} \\ \kappa & 0 \end{smallmatrix} \bigr)
  & \sqrt{2} \bigl(\begin{smallmatrix}
  0 & \bar{\kappa} \\ \kappa & 0 \end{smallmatrix} \bigr) \\[2\jot]
  \bb_{(V;+)} 
  & 1 & 1 & -1 & -1 & 0 & 0 
  & \sqrt{2} & -\sqrt{2} & 0 & 0 & 0 & 0 \\[2\jot]
  \bb_{(V;-)} 
  & 1 & 1 & -1 & -1 & 0 & 0 
  & -\sqrt{2} & \sqrt{2} & 0 & 0 & 0 & 0 \\[2\jot]
  \bb_{(S;+)} 
  & 1 & -1 & 1 & -1 & 0 & 0 
  & 0 & 0 & 0 & 0 & \sqrt{2} & -\sqrt{2} \\[2\jot]
  \bb_{(S;-)} 
  & 1 & -1 & 1 & -1 & 0 & 0 
  & 0 & 0 & 0 & 0 & -\sqrt{2} & \sqrt{2} \\[2\jot]
  \bb_{(C;+)} 
  & 1 & -1 & -1 & 1 & 0 & 0 
  & 0 & 0 & \sqrt{2} & -\sqrt{2} & 0 & 0 \\[2\jot]
  \bb_{(C;-)} 
  & 1 & -1 & -1 & 1 & 0 & 0 
  & 0 & 0 & -\sqrt{2} & \sqrt{2} & 0 & 0   
  \end{array}\notag
\eeq
}
\caption{Irreducible representations of the generalized fusion
algebra $\F(so(8)_1;S_3)$.
$\kappa$ is a cube root of 1, $\kappa = e^{\frac{2\pi i }{3}}$.
Unlike ordinary fusion algebras,
there is an irreducible representation of dimension 2
due to the non-commutativity of $\F(so(8)_1;S_3)$.}
\label{tab:so8bb}
\end{table}

\section{Boundary states as a NIM-rep of generalized fusion algebras}
\label{sec:genNIM}

It is now well understood that boundary states preserving a chiral 
algebra $\A$ form a non-negative integer matrix 
representation (NIM-rep) of the fusion algebra $\F(\A)$ \cite{BPPZ}. 
Since the generalized fusion algebra $\F(\A; G)$ is defined
through the twisted boundary states associated with the charge-conjugation
modular invariant,
it is natural to expect
a close relationship between $\F(\A; G)$
and twisted boundary states for general modular invariants, such as
simple current invariants \cite{SY}
or exceptional ones. 
As we shall show below, this expectation turns out to be true;
a set of mutually consistent boundary states
associated with any modular invariant
realizes a NIM-rep of the generalized fusion algebra.

\subsection{Charge-conjugation invariants}

We begin our analysis by rewriting the twisted boundary states
for the charge-conjugation modular invariant in the form that
the relation to the generalized fusion algebra is more apparent.
We call them the regular states since their mutual overlap is expressed
by the generalized fusion coefficients, which realize the regular
representation of the generalized fusion algebra. 

Corresponding to the character $\chi_{(\lambda;\rho)}^{ab}$ 
defined in eq.\eqref{eq:twin2},
we introduce a new basis for the twisted Ishibashi states, 
\beq
  \dket{((\lambda,\lambda^*); \rho)_{ab}} 
  = \sqrt{\frac{\dim \rho}{\abs{\Sc(\lambda)}}}
  \sum_{\omega \in \Sc(\lambda)} \overline{\rho^{ab}(\omega)} \,
  \dket{(\lambda,\lambda^*); \omega} \quad
  (\rho \in \Irr{\Sc(\lambda)})  \, .
  \label{eq:Ishibashinew}
\eeq
One can express the original basis in terms of
$\dket{((\lambda,\lambda^*); \rho)_{ab}}$ as
\beq
  \dket{(\lambda,\lambda^*);\omega} 
  = \frac{1}{\sqrt{\abs{\Sc(\lambda)}}}
  \sum_{\substack{%
  \rho \in \Irr{\Sc(\lambda)}\\
  1 \le a, b \le \dim \rho}} \sqrt{\dim \rho} \, 
  \rho^{ab}(\omega) \, \dket{((\lambda,\lambda^*); \rho)_{ab}} \, ,
  \label{eq:Ishibashiorg}
\eeq
where we used the orthogonality relation \eqref{eq:ortho2} for 
the matrix elements of $\rho$.
We denote the set of labels for the new basis 
$\dket{((\lambda,\lambda^*); \rho)_{ab}}$ of twisted Ishibashi states
by $\Ehat_0$,
\beq
  \Ehat_0 = \{ ((\lambda,\lambda^*);\rho) \, | \, 
  \lambda \in \I ; \, \rho \in \Irr{\Sc(\lambda)} \} \, .
\eeq
This is essentially the same set as $\Ihat^*$ defined in \eqref{eq:Ihatstar}.
We have introduced the new symbol $\Ehat_0$ for the labels of
Ishibashi states in order to reserve $\Ihat^*$ for 
expressing the chiral quantity.
The overlap of $\dket{((\lambda,\lambda^*); \rho)_{ab}}$ with
$\dket{((\lambda,\lambda^*); \rho')_{a'b'}}$
yields $\chi_{(\lambda;\rho)}^{b'b}$, 
\bea
  \dbra{((\lambda,\lambda^*);\rho')_{a'b'}} 
  \tq \dket{((\lambda,\lambda^*);\rho)_{ab}}
  &= \frac{\sqrt{\dim \rho'} \sqrt{\dim \rho}}{\abs{\Sc(\lambda)}}
   \sum_{\omega, \omega' \in \Sc(\lambda)}
    \rho'^{a'b'}(\omega') \overline{\rho^{ab}(\omega)} 
    \frac{1}{S_{0\lambda}} \chi^{\omega'^{-1} \omega}_\lambda(\tilde{q}) 
    \notag \\
  &= \frac{\sqrt{\dim \rho'} \sqrt{\dim \rho}}{\abs{\Sc(\lambda)}} 
   \sum_{\omega, \omega' \in \Sc(\lambda)}
    \rho'^{a'b'}(\omega') \overline{\rho^{ab}(\omega' \omega)} 
    \frac{1}{S_{0\lambda}} \chi^{\omega}_\lambda(\tilde{q}) 
    \notag \\
  &= \delta_{\rho \rho'} \delta^{a a'} 
     \sum_{\omega \in \Sc(\lambda)} 
     \overline{\rho^{b' b}(\omega)} 
     \frac{1}{S_{0\lambda}} \chi^\omega_\lambda(\tilde{q}) \\
  &= \delta_{\rho \rho'} \delta^{a a'} \frac{1}{S_{0\lambda}} 
     \sqrt{\frac{\abs{\Sc(\lambda)}}{\dim \rho}} \, 
     \chi_{(\lambda;\rho)}^{b'b}(\tilde{q}) \notag \\
  &= \delta_{\rho \rho'} \delta^{a a'} 
     \frac{1}{\Shat_{0 (\lambda; \rho)}^{11}} 
     \chi_{(\lambda;\rho)}^{b'b}(\tilde{q}) \, . \notag
\eea
Using $\dket{((\lambda,\lambda^*);\rho)_{ab}}$,
we can rewrite the regular state $\ket{L} \, (L \in \Ihat)$ as follows
\bes
  \ket{L} &= \sum_{\lambda \in \I(\omega_L)} 
  S^{\omega_L}_{L \lambda} \dket{(\lambda,\lambda^*); \omega_L} \\
  &= \sum_{\lambda \in \I(\omega_L)} S^{\omega_L}_{L \lambda}
  \frac{1}{\sqrt{\abs{\Sc(\lambda)}}}
  \sum_{\substack{%
  \rho \in \Irr{\Sc(\lambda)}\\
  1 \le a, b \le \dim \rho}} \sqrt{\dim \rho} \, 
  \rho^{ab}(\omega_L) \dket{((\lambda,\lambda^*); \rho)_{ab}} \\
  &= \sum_{((\lambda,\lambda^*);\rho) \in \Ehat_0}
     \sum_{a, b} 
     \Shat_{L (\lambda;\rho)}^{ab} 
     \dket{((\lambda,\lambda^*); \rho)_{ab}} \, .
  \label{eq:regular}
\ees
The boundary state coefficients of the regular state
$\ket{L}$ are therefore given by
the generalized $S$-matrix $\hat{S}$.
It is instructive to calculate the overlaps of $\ket{L}$ starting from
the form given above,
\bes
  \bra{M} \tq \ket{L} 
  &= \sum_{(\lambda;\rho) \in \Ihat^*} \sum_{a,b} \sum_{a',b'}
  \overline{\Shat_{M (\lambda;\rho)}^{a'b'}}
  \Shat_{L (\lambda;\rho)}^{ab} 
  \delta^{a a'} \frac{1}{\Shat_{0 (\lambda;\rho)}^{11}} 
  \chi_{(\lambda;\rho)}^{b'b}(\tilde{q}) \\
  &= \sum_{(\lambda;\rho)} \sum_{a,b} \sum_{b'}
  \overline{\Shat_{M (\lambda;\rho)}^{a b'}}
  \Shat_{L (\lambda;\rho)}^{ab} 
  \frac{1}{\Shat_{0 (\lambda;\rho)}^{11}} 
  \sum_{N} \overline{\Shat_{N (\lambda;\rho)}^{b'b}}\, \chi_{N}(q) \\
  &= \sum_N \sum_{(\lambda;\rho); a,b,b'}
  \frac{%
  \Shat_{L (\lambda;\rho)}^{ab}
  \Shat_{N^* (\lambda;\rho)}^{bb'}\,
  \overline{\Shat_{M (\lambda;\rho)}^{a b'}}
  }{\Shat_{0 (\lambda;\rho)}^{11}} \, \chi_N(q) \\
  &= \sum_N \fusionhat{L}{N}{M}
   \chi_{N^*}(q) \\
  &= \sum_N (\Nhat_{N})_L{}^{M}
   \chi_{N^*}(q)  \, ,
  \label{eq:LMoverlap}
\ees
where we used the formula
\beq
  \overline{\Shat_{L (\lambda;\rho)}^{ab}} 
  = \Shat_{L^* (\lambda; \rho)}^{ba} 
\eeq
which follows from eq.\eqref{eq:Sbar}. 
In this way, we reproduce the regular representation $\Nhat_L$
of the generalized fusion algebra as the overlap matrices
of the regular states. 

\subsection{General cases}
\label{sec:general}

We next turn to the case of twisted boundary states associated with
a general modular invariant $Z$ (see \eqref{eq:Z})
of the chiral algebra $\A$.
In order to express twisted boundary states,
we have to determine what kinds of twisted Ishibashi states
are available in the modular invariant \eqref{eq:Z}. 
As we have argued in Section~\ref{sec:twisted},
the twisted Ishibashi state 
$\dket{(\lambda,\mu^*);\omega}$ exists if and only if 
$\lambda = \omega(\mu)$.
Accordingly, the set of labels for
$\omega$-twisted Ishibashi states,
which we denote by $\E(\omega)$, takes the following form
\beq
  \E(\omega) = \{ (\lambda,\mu^*) \, | \,
      (\lambda,\mu^*) \in \mathrm{Spec}(Z); \,
      \lambda = \omega(\mu) \} \, , 
  \label{eq:Eomega}
\eeq
where $\mathrm{Spec}(Z)$ is the spectrum of bulk fields in $Z$
(see \eqref{eq:SpecZ}).
A general boundary state $\ket{A}$
satisfying the boundary condition \eqref{eq:twistbc} 
is then written as 
\beq
  \ket{A} = \sum_{(\lambda,\mu^*) \in \E(\omega)} 
  \Psi^{\omega}_{A (\lambda,\mu^*)} 
  \dket{(\lambda,\mu^*); \omega} \quad
  (A \in \V^{\omega} \subset \Vhat)
  \, .
  \label{eq:A}
\eeq
Here $\V^\omega$ 
is the set of labels for the $\omega$-twisted boundary states.
We have also introduced the set
$\Vhat$ of all the labels of twisted 
boundary states,
\beq
  \Vhat = \coprod_{\omega \in G} \V^\omega \, .
  \label{eq:Vhat}
\eeq
As in the case of the regular states,
we assume that the number of twisted boundary states
is equal to that of twisted Ishibashi states \cite{PSS},
\beq
  \abs{\V^\omega} = \abs{\E(\omega)} \quad (\omega \in G) \, .
  \label{eq:completeness}
\eeq
This condition together with the mutual consistency of
boundary states
implies that the boundary states labeled by
$\V^\omega$ form a NIM-rep of the (ordinary) fusion algebra
$\F(\A)$ of $\A$ \cite{BPPZ}.
In particular, the matrix
$(\Psi^\omega)_{A\, (\lambda,\mu^*)} = 
\Psi^{\omega}_{A (\lambda,\mu^*)}$ is unitary.

In the same way as the regular states \eqref{eq:LM},
the overlap of two twisted boundary states can be expanded into
a sum of the characters for representations of twisted chiral algebras,
\bes
  \bra{B} \tq \ket{A} 
  &= \sum_{(\lambda,\mu^*) \in \E(\omega_A) \cap \E(\omega_B)}
  \Psi^{\omega_A}_{A (\lambda,\mu^*)} 
  \overline{\Psi^{\omega_B}_{B (\lambda,\mu^*)}}
  \frac{1}{S_{0\mu}} \chi_\mu^{\omega_B^{-1} \omega_A}(\tilde{q}) \\
  &= \sum_{(\lambda,\mu^*) \in \E(\omega_A) \cap \E(\omega_B)}
    \Psi^{\omega_A}_{A (\lambda,\mu^*)} 
    \overline{\Psi^{\omega_B}_{B (\lambda,\mu^*)}} 
    \frac{1}{S_{0 \mu}}
    \sum_{N \in \I^{\omega_B^{-1} \omega_A}} 
    \overline{S^{\omega_B^{-1} \omega_A}_{N \mu}}
    \chi_{N}(q) \\
  &= \sum_{N \in \I^{\omega_A^{-1} \omega_B}} 
     \sum_{(\lambda,\mu^*) \in \E(\omega_A) \cap \E(\omega_B)}
    \Psi^{\omega_A}_{A (\lambda,\mu^*)} 
    \frac{S^{\omega_A^{-1} \omega_B}_{N \mu}}{S_{0 \mu}}
    \overline{\Psi^{\omega_B}_{B (\lambda,\mu^*)}} \,
    \chi_{N^*}(q) \\
  &= \sum_{N \in \I^{\omega_A^{-1} \omega_B}} 
     (\hat{n}_N)_A{}^B \, \chi_{N^*}(q) \, ,
    \label{eq:BA}
\ees
where $\omega_A$ is the automorphism type of $\ket{A}$ and
we used the formula \eqref{eq:Ishibashinormtwisted}.
A $\abs{\Vhat} \times \abs{\Vhat}$ matrix $\hat{n}_N$ is 
defined for each $N \in \Ihat$ as
\beq
  (\hat{n}_N)_A{}^B =
     \sum_{(\lambda,\mu^*) \in \E(\omega_A) \cap \E(\omega_B)}
    \Psi^{\omega_A}_{A (\lambda,\mu^*)} 
    \frac{S^{\omega_N}_{N \mu}}{S_{0 \mu}}
    \overline{\Psi^{\omega_B}_{B (\lambda,\mu^*)}} 
    \times \delta_{\omega_A \omega_N, \omega_B} \, .
    \label{eq:nhat}
\eeq
The entry of $\hat{n}_N$ should be non-negative integer
for the mutual consistency of boundary states,
since it represents the multiplicity of representation $N^* \in \Ihat$
in the open-string spectrum.
In the rest of this section, we show that 
these matrices $\{ \hat{n}_N | N \in \Ihat \}$ satisfy
the generalized fusion algebra $\F(\A; G)$.
\bigskip

We first rewrite the boundary states \eqref{eq:A} in a form similar
to the regular case \eqref{eq:regular}. 
To do so, we have to generalize the basis 
\eqref{eq:Ishibashinew} of Ishibashi states for
$(\lambda, \lambda^*) \in \mathrm{Spec}(Z_c)$
to that for $(\lambda, \mu^*) \in \mathrm{Spec}(Z)$.
In contrast with the case of the charge-conjugation invariant $Z_c$,
the label $(\lambda,\mu^*)$ of bulk fields
is in general not symmetric;
there exists some 
$(\lambda,\mu^*) \in \mathrm{Spec}(Z)$ with $\lambda \neq \mu$.
Although the untwisted Ishibashi state can not be constructed
for $(\lambda, \mu^*)$ with $\lambda \neq \mu$,
it is possible to obtain twisted states for $(\lambda,\mu^*)$
if $\lambda = \omega(\mu)$ for some $\omega \in G$. 
In other words, twisted Ishibashi states are available
for $(\lambda, \mu^*)$ if $\lambda$ belongs to 
the $G$-orbit $G \mu$ of $\mu$, 
\beq
   G \mu = \{\omega(\mu) \, |\, \omega \in G \} \subset \I \, .
\eeq
In the following, we consider only the case of $\lambda \in G \mu$,
since the label $(\lambda,\mu^*)$ with $\lambda \notin G \mu$
does not appear in the twisted boundary states for 
the automorphism group $G$.

Let $\omega_{\lambda \mu}$ be an element of $G$ that relates
$\lambda \in G \mu$ with $\mu$, namely
$\lambda = \omega_{\lambda \mu}(\mu)$. 
In general, this element $\omega_{\lambda \mu}$ can not be determined
uniquely due to the stabilizer $\Sc(\mu)$ of $\mu$. 
Suppose that we have another element $\omega_{\lambda \mu}' \in G$
satisfying $\lambda = \omega_{\lambda \mu}'(\mu)$. 
Then $\omega_{\lambda \mu}^{-1} \omega_{\lambda \mu}'$
is an element of $\Sc(\mu)$ since
$\omega_{\lambda \mu}^{-1} \omega_{\lambda \mu}' (\mu)
= \omega_{\lambda \mu}^{-1} (\lambda) = \mu$. 
In other words, $\omega_{\lambda \mu}'$ belongs to the coset
\beq
  \omega_{\lambda \mu} \Sc(\mu) = \{\omega_{\lambda \mu} \omega \, | \,
  \omega \in \Sc(\mu) \} \subset G \, .  
\eeq
This result shows that, for a given $(\lambda,\mu^*) \in \mathrm{Spec}(Z)$,  
the twisted Ishibashi state $\dket{(\lambda,\mu^*);\omega}$
exists if and only if $\omega \in \omega_{\lambda \mu} \Sc(\mu)$. 

Based on this fact, one can generalize
the basis \eqref{eq:Ishibashinew} of twisted Ishibashi states
in the following manner, 
\beq
  \dket{((\lambda,\mu^*);\rho)_{ab}} = 
  \sqrt{\frac{\dim \rho}{\abs{\Sc(\mu)}}}
  \sum_{\omega \in \Sc(\mu)} 
  \overline{\rho^{ab}(\omega)} \,
  \dket{(\lambda,\mu^*); \omega_{\lambda \mu} \omega} 
  \quad 
  (\lambda = \omega_{\lambda \mu} (\mu); \, \rho \in \Irr{\Sc(\mu)} )
  \, .
  \label{eq:Ishibashinewtilde}
\eeq
Similarly to eq.\eqref{eq:Ishibashiorg},
one can express the original basis 
$\dket{(\lambda,\mu^*); \omega}$ in terms of
$\dket{((\lambda,\mu^*); \rho)_{ab}}$, 
\beq
  \dket{(\lambda,\mu^*);\omega} =
  \frac{1}{\sqrt{\abs{\Sc(\mu)}}}
  \sum_{\substack{%
  \rho \in \Irr{\Sc(\mu)}\\
  1 \le a, b \le \dim \rho}} \sqrt{\dim \rho} \, 
  \rho^{ab}(\omega_{\lambda \mu}^{-1} \omega) 
  \dket{((\lambda,\mu^*); \rho)_{ab}} \quad
  (\omega \in \omega_{\lambda \mu} \Sc(\mu)) \, .
  \label{eq:Ishibashinewtildereverse}
\eeq
The overlap of $\dket{((\lambda,\mu^*); \rho)_{ab}}$ with
$\dket{((\lambda,\mu^*); \rho')_{a'b'}}$
can be calculated in exactly the same way as 
$\dket{((\lambda,\lambda^*);\rho)_{ab}}$, 
\bes
  \lefteqn{\dbra{((\lambda,\mu^*);\rho')_{a'b'}} 
  \tq \dket{((\lambda,\mu^*);\rho)_{ab}} } \qquad & \\
  &= \frac{\sqrt{\dim \rho'} \sqrt{\dim \rho}}{\abs{\Sc(\mu)}}
   \sum_{\omega, \omega' \in \Sc(\mu)}
    \rho'^{a'b'}(\omega') \overline{\rho^{ab}(\omega)} 
    \dbra{(\lambda,\mu^*);\omega_{\lambda \mu} \omega'} \tq
    \dket{(\lambda,\mu^*);\omega_{\lambda \mu} \omega}
    \\
  &= \frac{\sqrt{\dim \rho'} \sqrt{\dim \rho}}{\abs{\Sc(\mu)}}
   \sum_{\omega, \omega' \in \Sc(\mu)}
    \rho'^{a'b'}(\omega') \overline{\rho^{ab}(\omega)} 
    \frac{1}{S_{0\mu}} \chi^{\omega'^{-1} \omega}_\mu(\tilde{q}) 
     \\
  &= \frac{\sqrt{\dim \rho'} \sqrt{\dim \rho}}{\abs{\Sc(\mu)}} 
   \sum_{\omega, \omega' \in \Sc(\mu)}
    \rho'^{a'b'}(\omega') \overline{\rho^{ab}(\omega' \omega)} 
    \frac{1}{S_{0\mu}} \chi^{\omega}_\mu(\tilde{q}) 
     \\
  &= \delta_{\rho \rho'} \delta^{a a'} 
     \sum_{\omega \in \Sc(\mu)} 
     \overline{\rho^{b' b}(\omega)} 
     \frac{1}{S_{0\mu}} \chi^\omega_\mu(\tilde{q}) \\
  &= \delta_{\rho \rho'} \delta^{a a'} 
     \frac{1}{\Shat_{0 (\mu; \rho)}^{11}} 
     \chi_{(\mu;\rho)}^{b'b}(\tilde{q}) \, . 
  \label{eq:Ishibashinorm2}
\ees
The overlap between two states with different labels,
$(\lambda',{\mu'}^*) \neq (\lambda,\mu^*)$, vanishes
from the formula \eqref{eq:Ishibashinormtwisted}.

Substituting \eqref{eq:Ishibashinewtildereverse} in \eqref{eq:A},
the boundary state $\ket{A}$ can be brought to the following form,
\bes
  \ket{A} &= \sum_{(\lambda,\mu^*) \in \E(\omega)} 
  \Psi^{\omega_A}_{A (\lambda,\mu^*)} 
  \dket{(\lambda,\mu^*); \omega_A} \\
  &= \sum_{(\lambda,\mu^*) \in \E(\omega)} 
  \Psi^{\omega_A}_{A (\lambda,\mu^*)}
    \frac{1}{\sqrt{\abs{\Sc(\mu)}}}
  \sum_{\substack{%
  \rho \in \Irr{\Sc(\mu)}\\
  1 \le a, b \le \dim \rho}} \sqrt{\dim \rho} \, 
  \rho^{ab}(\omega_{\lambda \mu}^{-1} \omega_A) 
  \dket{((\lambda,\mu^*); \rho)_{ab}} \\  
  &= \sum_{((\lambda,\mu^*);\rho) \in \hat{\E}} 
  \sum_{a,b}
  \hat{\Psi}_{A ((\lambda,\mu^*);\rho)}^{ab} 
  \dket{((\lambda,\mu^*);\rho)_{ab}} \, ,
  \label{eq:A2}
\ees
where we denote by $\hat{\E}$ the set of labels for the basis 
$\dket{((\lambda,\mu^*);\rho)_{ab}}$
available in the modular invariant \eqref{eq:Z},
\beq
  \hat{\E} = \biggl\lbrace ((\lambda,\mu^*); \rho) \, \bigg\vert \, 
  (\lambda, \mu^*) \in \bigcup_{\omega \in G} \E(\omega) ; \, 
  \rho \in \Irr{\Sc(\mu)} \biggr\rbrace \, .
  \label{eq:Ehat}
\eeq
It should be noted that some $(\lambda,\mu^*)$ may appear
more than once in $\Ehat$ corresponding to the multiple occurrence
of bulk fields with representation $(\lambda,\mu^*)$
in the modular invariant \eqref{eq:Z}.
The boundary state coefficient 
$\hat{\Psi}_{A ((\lambda,\mu^*);\rho)}^{ab}$
is related to the original coefficient
$\Psi^{\omega_A}_{A (\lambda,\mu^*)}$ as follows
\beq
  \hat{\Psi}_{A ((\lambda,\mu^*);\rho)}^{ab} =
  \begin{cases}
  \Psi^{\omega_A}_{A (\lambda,\mu^*)} 
  \sqrt{\frac{\dim \rho}{\abs{\Sc(\mu)}}}
  \rho^{ab}(\omega_{\lambda \mu}^{-1} \omega_A) \quad &
  ((\lambda,\mu^*) \in \E(\omega_A), \, 
    \rho \in \Irr{\Sc(\mu)} ) \, , \\
  0 & ((\lambda,\mu^*) \notin \E(\omega_A)) \, .
  \end{cases} 
  \label{eq:psihat}
\eeq
This form
is completely parallel to the generalized $S$-matrix \eqref{eq:Shat}.
Actually, from the unitarity of $\Psi^\omega$ for all $\omega \in G$,
one can prove that the matrix $\hat{\Psi}$ is also unitary,
\beq
  \abs{\Vhat} = \sum_{\omega \in G} \abs{\V^\omega}
  = \sum_{\omega \in G} \abs{\E(\omega)} 
  = \sum_{(\lambda,\mu^*) \in \bigcup_{\omega \in G} \E(\omega)} 
  \abs{\Sc(\mu)} 
  = \sum_{((\lambda,\mu^*); \rho) \in \Ehat} (\dim \rho)^2 \, ,
\eeq
\bes
  \lefteqn{\sum_{((\lambda,\mu^*); \rho) \in \Ehat; a, b} 
  \hat{\Psi}_{A ((\lambda,\mu^*); \rho)}^{ab} \,
  \overline{\hat{\Psi}_{B ((\lambda,\mu^*); \rho)}^{ab}} } \qquad & \\
  &= \sum_{(\lambda,\mu^*) \in \E(\omega_A) \cap \E(\omega_B)}
     \Psi^{\omega_A}_{A (\lambda,\mu^*)} \, 
     \overline{\Psi^{\omega_B}_{B (\lambda,\mu^*)}}
     \sum_{\rho \in \Irr{\Sc(\mu)}}
     \sum_{a, b}
     \frac{\dim \rho}{\abs{\Sc(\mu)}}
     \rho^{ab}(\omega_{\lambda \mu}^{-1} \omega_A)
     \overline{\rho^{ab}(\omega_{\lambda \mu}^{-1} \omega_B)} \\
  &= \sum_{(\lambda,\mu^*) \in \E(\omega_A)}
     \Psi^{\omega_A}_{A (\lambda,\mu^*)} \, 
     \overline{\Psi^{\omega_A}_{B (\lambda,\mu^*)}}
     \times \delta_{\omega_A \omega_B} \\
  &= \delta_{AB} \, ,
\ees
where we used the fact that 
$\omega_{\lambda \mu}^{-1} \omega_A = 
\omega_{\lambda \mu}^{-1} \omega_B$ in $\Sc(\mu) \subset G$
implies $\omega_A = \omega_B$ in $G$.

Using the form given in \eqref{eq:A2}
together with the formula \eqref{eq:Ishibashinorm2}, 
one can calculate 
the overlap of two boundary states $\ket{A}$ and $\ket{B}$ 
in the following way, 
\bes
  \bra{B} \tq \ket{A} 
  &= \sum_{((\lambda,\mu^*);\rho) \in \Ehat} \sum_{a,b} \sum_{a',b'}
  \overline{\hat{\Psi}_{B ((\lambda,\mu^*);\rho)}^{a'b'}}
  \hat{\Psi}_{A ((\lambda,\mu^*);\rho)}^{ab} \,
  \delta^{a a'}
     \frac{1}{\Shat_{0 (\mu;\rho)}^{11}}
     \chi_{(\mu;\rho)}^{b'b}(\tilde{q}) \\
  &= \sum_{((\lambda,\mu^*);\rho) \in \Ehat} \sum_{a,b} \sum_{b'}
  \overline{\hat{\Psi}_{B ((\lambda,\mu^*);\rho)}^{a b'}}
  \hat{\Psi}_{A ((\lambda,\mu^*);\rho)}^{ab} \, 
     \frac{1}{\Shat_{0 (\mu;\rho)}^{11}}
     \sum_{N \in \Ihat}
     \overline{\Shat_{N (\mu; \rho)}^{b'b}} \,\chi_N(q) \\
  &= \sum_{((\lambda,\mu^*);\rho) \in \Ehat} \sum_{a,b} \sum_{b'}
  \overline{\hat{\Psi}_{B ((\lambda,\mu^*);\rho)}^{a b'}}
  \hat{\Psi}_{A ((\lambda,\mu^*);\rho)}^{ab} \,
  \sum_{N \in \Ihat} 
     \frac{\Shat_{N^* (\mu; \rho)}^{b b'}}{%
     \Shat_{0 (\mu;\rho)}^{11}} \, \chi_N(q)  \\
  &=   \sum_{N \in \Ihat}
  \sum_{((\lambda,\mu^*);\rho) \in \Ehat} \sum_{a,b} \sum_{a' b'}
  \hat{\Psi}_{A ((\lambda,\mu^*);\rho)}^{ab} \,
     \delta^{a a'}
       \bb_{(\mu;\rho)}^{bb'}(N^*)
  \overline{\hat{\Psi}_{B ((\lambda,\mu^*);\rho)}^{a' b'}} \,
       \chi_N(q)  \, .
  \label{eq:BA2}
\ees
Comparing this with eq.\eqref{eq:BA}, 
we obtain the following expression for
the overlap matrix $\hat{n}_N$,
\beq
  \hat{n}_N 
  = \hat{\Psi} \Biggl[\,
    \bigoplus_{((\lambda,\mu^*);\rho) \in \Ehat} \bigl(\,
    \underbrace{%
  \bb_{(\mu; \rho)}(N) \oplus \dots \oplus \bb_{(\mu; \rho)}(N)
  }_{\text{$(\dim \rho)$ terms}} \, \bigr) \Biggr]
  \hat{\Psi}^\dagger \, .
  \label{eq:nhat2}
\eeq
This has exactly the same structure as eq.\eqref{eq:Shat_decomposition} for
the regular representation matrix $\Nhat_N$, 
from which we have shown $\Nhat_N$ satisfies the generalized
fusion algebra. 
We can repeat the same thing here since the matrix $\hat{\Psi}$ is
unitary from the assumption \eqref{eq:completeness} of completeness.
For a unitary $\hat{\Psi}$, the above expression \eqref{eq:nhat2} 
for the overlap matrix $\hat{n}_N$ means that
$\hat{n}_N$ is similar to
a direct sum of $\bb_{(\mu;\rho)}(N)$. 
Since $\bb_{(\mu;\rho)}(N)$ satisfies the generalized fusion algebra
for all $(\mu;\rho) \in \Ihat^*$, 
the matrix $\hat{n}_N$ also satisfies the generalized
fusion algebra. 
The mutual consistency of twisted boundary states requires that 
the coefficients $(\hat{n}_N)_A{}^B$ of $\chi_{N*}$ in the cylinder
amplitude take values in non-negative integers for all $N \in \Ihat$.
Hence we are led to the conclusion that the overlap matrix $\hat{n}_N$
for a set of mutually consistent boundary states
realizes a NIM-rep of the generalized fusion algebra $\F(\A; G)$
if the condition \eqref{eq:completeness} is satisfied.

\section{Examples}
\label{sec:example}

As we have shown in the previous section,
a consistent set of
twisted boundary states in any modular invariant
form a NIM-rep of the generalized fusion algebra. 
In this section, we check this for three concrete chiral algebras,
$u(1)_k, su(3)_k$ and 
$su(3)_1^{\oplus 3}$, 
by explicitly constructing twisted boundary states for each case.

\subsection{$u(1)_k$}
\label{sec:u1}

The simplest chiral algebra with non-trivial automorphisms is
$u(1)_k$, where the level $k$ is a positive integer.
There are $2k$ irreducible representations labeled by 
positive integer $n \mod 2k$,
\beq
  \I = \{ (n) \, | \, n = 0, 1, \dots, 2k-1 \} \, ,
\eeq
for which the modular transformation matrix takes the form
\beq
  S_{mn} = \frac{1}{\sqrt{2k}} e^{-\frac{\pi i}{k} mn} \, . 
\eeq
Therefore, for the charge-conjugation modular invariant,
we have $2k$ untwisted boundary states,
\beq
  \ket{m} = \sum_{n \in \I} S_{mn} \dket{(n,n^*)} \quad
  (m \in \I) \, . 
  \label{eq:u1bs}
\eeq

The charge-conjugation $\omega_c$ is an automorphism of
$u(1)_k$, which acts on $\I$ as 
\beq
  \omega_c \, : \, (n) \mapsto (n^*) = (-n) = (2k -n) \quad
  (n \in \I) \, . 
\eeq
There are two representations fixed by $\omega_c$,
\beq
  \I(\omega_c) = \{ (0) , \, (k)  \} \, . 
\eeq
Corresponding to this, we have two twisted boundary states,
which we denote by $\ket{\pm}$,
\beq
  \ket{\pm} = 
  \frac{1}{\sqrt{2}} (\dket{(0,0); \omega_c} \pm \dket{(k,k); \omega_c}) \, .
\eeq
The overlap of $\ket{\pm}$ with the untwisted state $\ket{0}$
yields the character $\chi_\pm$ of the twisted chiral algebra
$u(1)_k^{\omega_c}$
(see eq.\eqref{eq:lambdatilde0}),
\beq
  \chi_\pm(q) = \bra{0} \tq \ket{\pm} 
  = \frac{1}{\sqrt{2}} (\chi^{\omega_c}_0(\tilde{q}) \pm 
    \chi^{\omega_c}_k(\tilde{q})) \, , 
\eeq
where $\chi^{\omega_c}_n$ is the twining character of $(n)$
for $\omega_c$. 
From this equation, one obtains the modular transformation
matrix $S^{\omega_c}$
\beq
  S^{\omega_c} = \frac{1}{\sqrt{2}} 
  \begin{pmatrix} 1 & 1 \\ 1 & -1 \end{pmatrix} \, , 
\eeq
where the rows and the columns are ordered as $\I^{\omega_c} = \{+, -\}$
and $\I(\omega_c) = \{0, k\}$, respectively. 
Since $\omega_c^{-1} = \omega_c$ and $S^{\omega_c}$ is real,
the relation \eqref{eq:Sbar} implies that
the twisted representations $(\pm) \in \I^{\omega_c}$ are self-conjugate,
$(\pm^{*}) = (\pm)$.

Using these data and the formula
(see eq.\eqref{eq:LM})
\beq
  \bra{L^*} \tq \ket{M} = (L) \times (M) = 
  \sum_{N \in \Ihat} \fusionhat{L}{M}{N} \chi_N(q) 
  \quad (L, M \in \Ihat)
  \, , 
\eeq
one can determine 
the generalized fusion coefficients $\fusionhat{L}{M}{N}$
for the chiral algebra $u(1)_k$ 
and the automorphism group $G_c = \{1, \omega_c\} \cong \Z_2$.
The result is as follows,
\bes
  (l) \times (m) &= (l+m) \, , \\
  (l) \times (\pm) &= (\pm) \times (l) = 
    \begin{cases} (\pm) & (l = 0 \mod 2)\\ 
                  (\mp) & (l = 1 \mod 2)\end{cases}  \, , \\
  (+) \times (+) &= (-) \times (-) =
  (0) + (2) + \dots + (2k-2) \, , \\
  (+) \times (-) &= (-) \times (+) =
  (1) + (3) + \dots + (2k-1) \, ,
  \label{eq:u1Z2}
\ees
where $(l), (m) \in \I$ and $(\pm) \in \I^{\omega_c}$. 
This defines the generalized fusion algebra $\F(u(1)_k; G_c)$ for
$G_c = \{1, \omega_c\} \cong \Z_2$.
As is easily seen, this algebra is generated by $(1)$ and $(+)$.
Hence $\F(u(1)_k; G_c)$ is an extension of the ordinary fusion algebra
$\F(u(1)_k)$, which is the group algebra of $\Z_{2k}$, 
by the twisted representation $(+)$. 
Note that $\F(u(1)_k; G_c)$ is commutative, 
$(L) \times (M) = (M) \times (L)$, reflecting the fact that
the automorphism group $G_c \cong \Z_2$ is abelian. 
\bigskip

We next turn to the case of modular invariants
other than the charge-conjugation one 
and check that the associated boundary states form a NIM-rep 
of the generalized fusion algebra \eqref{eq:u1Z2}. 
We consider two cases:
a simple current extension of $k=4$ and 
a non-diagonal invariant of $k=6$. 
\bigskip

\noindent
\underline{$k=4$}
\medskip

For $k=4$, we have the following block-diagonal invariant 
\beq
  Z = \abs{\chi_0 + \chi_4}^2 + \abs{\chi_2 + \chi_6}^2 \, ,
   \label{eq:u1ex}
\eeq
which is a simple current extension \cite{SY}
of $u(1)_4$ by $(4) \in \I$.
The chiral algebra of the extended theory is $u(1)_1$
and each block corresponds to an irreducible representation
of $u(1)_1$.
Since there are two irreducible representations for $u(1)_1$,
we have two untwisted boundary states that keep $u(1)_1$
(see eq.\eqref{eq:u1bs})
\beq
  \frac{1}{\sqrt{2}} (\dket{(0',0')} \pm \dket{(1',1')}) \, ,
  \label{eq:k=1}
\eeq
where we distinguish representations of $u(1)_1$ 
from those of $u(1)_4$ by putting a prime. 
Since $u(1)_1$ contains $u(1)_4$ as a subalgebra,
we can regard these states as untwisted boundary states
of $u(1)_4$ associated with the modular invariant \eqref{eq:u1ex}. 
From the branching rule
\beq
  (0') = (0) \oplus (4) \, , \quad
  (1') = (2) \oplus (6) \, , 
\eeq
and our normalization \eqref{eq:Ishibashinorm} for Ishibashi states,
we can express Ishibashi states of $u(1)_1$ by
those of $u(1)_4$,
\beq
   \dket{(0',0')} = 
   \frac{1}{\sqrt{2}} (\dket{(0,0)} + \dket{(4,4)}) \, , \quad
   \dket{(1',1')} = 
   \frac{1}{\sqrt{2}} (\dket{(2,2^*)} + \dket{(6,6^*)}) \, .
\eeq
Substituting this into eq.\eqref{eq:k=1}, we obtain two 
untwisted boundary states of $u(1)_4$,
\bes
  \ket{\tilde{0}} &= \frac{1}{2} (\dket{(0,0)} + \dket{(4,4)}) +
             \frac{1}{2} (\dket{(2,2^*)} + \dket{(6,6^*)}) \, , \\
  \ket{\tilde{2}} &= \frac{1}{2} (\dket{(0,0)} + \dket{(4,4)}) -
             \frac{1}{2} (\dket{(2,2^*)} + \dket{(6,6^*)}) \, . 
\ees
Since there are four untwisted Ishibashi states
in \eqref{eq:u1ex},
\beq
  \E(1) = \{(0,0), (2,2^*), (4,4), (6,6^*) \} \, , 
  \label{eq:k4E}
\eeq
we need two more boundary states for completeness. 
The remaining two can be constructed by, \textit{e.g.}, 
the fusion with representations of the unextended chiral algebra
\cite{AOS,BPPZ,IT}. 
(See \cite{IT} for a detailed exposition of this procedure.) 
The result is as follows,
\bes
  \ket{\tilde{1}} &= \frac{1}{2} (\dket{(0,0)} - \dket{(4,4)}) -
             \frac{i}{2} (\dket{(2,2^*)} - \dket{(6,6^*)}) \, , \\
  \ket{\tilde{3}} &= \frac{1}{2} (\dket{(0,0)} - \dket{(4,4)}) +
             \frac{i}{2} (\dket{(2,2^*)} - \dket{(6,6^*)}) \, . 
\ees
Consequently,
the boundary state coefficients $\Psi$ for
untwisted boundary states associated with
\eqref{eq:u1ex} take the following form
\beq
  \Psi = \frac{1}{2}
  \begin{pmatrix} 
    1 &  1 &  1 &  1 \\
    1 & -i & -1 &  i \\
    1 & -1 &  1 & -1 \\
    1 &  i & -1 & -i 
  \end{pmatrix} \, ,
  \label{eq:k4psi}
\eeq
where the column is ordered as in \eqref{eq:k4E}.

The restriction of
the charge-conjugation of $u(1)_1$ to the subalgebra $u(1)_4$ is 
clearly the charge-conjugation of $u(1)_4$.
In other words, 
the charge-conjugation of $u(1)_4$ has a lift to the extended
chiral algebra $u(1)_1$.
(The lifting of automorphisms for RCFTs is studied in \cite{FSW}.)
This enables us to construct twisted boundary states of $u(1)_4$
starting from those of $u(1)_1$. Since two representations of $u(1)_1$
are fixed by the charge-conjugation,
we have two twisted boundary states of $u(1)_1$,
\beq
  \frac{1}{\sqrt{2}} 
  (\dket{(0',0'); \omega_c} \pm \dket{(1',1'); \omega_c}) \, . 
\eeq
This is completely the same form as the untwisted states.
Hence one can proceed in the same way as the untwisted states
to obtain four twisted states of $u(1)_4$.
The boundary state coefficients $\Psi^{\omega_c}$ of twisted states
coincide with $\Psi$ given in \eqref{eq:k4psi} with the spectrum
\beq
  \E(\omega_c) = \{(0,0), (6,2^*), (4,4), (2,6^*) \}  
\eeq
instead of $\E(1)$.

From these data, one can calculate the mutual overlap of 
boundary states and the associated open string spectrum 
$\hat{n}$ defined in \eqref{eq:nhat}. 
Since we have $4 +  4 = 8$ boundary states, $\hat{n}$ is 
a $8 \times 8$ matrix and takes the form
\beq
    \hat{n}_l = \begin{pmatrix}
                (n^{(4)}_1)^l &    O     \\
                O        & (n^{(4)}_1)^l 
                \end{pmatrix} \,
     (l = 0,1,\dots,7) , \quad 
    \hat{n}_{\pm} = \begin{pmatrix}
                    O            &  n_{\pm}  \\
                    n_{\pm} &       O
                    \end{pmatrix} \, , 
  \label{eq:u14NIM}                    
\eeq
where 
$O$ is a $4 \times 4$ matrix with all entries being $0$ and
$n_{1,\pm}$ are defined as
\beq
    n^{(4)}_1 = \begin{pmatrix}
               0 & 1 & 0 & 0 \\
               0 & 0 & 1 & 0 \\
               0 & 0 & 0 & 1 \\
               1 & 0 & 0 & 0 
               \end{pmatrix}  , \,\,
    n_+ 
             = \begin{pmatrix}
               1 & 0 & 1 & 0 \\
               0 & 1 & 0 & 1 \\
               1 & 0 & 1 & 0 \\
               0 & 1 & 0 & 1 
               \end{pmatrix} , \,\,
    n_- 
             = \begin{pmatrix}
               0 & 1 & 0 & 1 \\
               1 & 0 & 1 & 0 \\
               0 & 1 & 0 & 1 \\
               1 & 0 & 1 & 0 
               \end{pmatrix} \, .
    \label{eq:k4NIM}
\eeq
One can confirm that these 10 matrices actually
satisfy the generalized fusion 
algebra \eqref{eq:u1Z2} for $k=4$, namely, 
\beq
  (\hat{n}_1)^8 = 1 \, , \quad
  \hat{n}_1 \hat{n}_+ = \hat{n}_- \, , \quad
  \hat{n}_+ \hat{n}_+ = \hat{n}_0 + \hat{n}_2 + \hat{n}_4 + \hat{n}_6 \, .
\eeq
The boundary states associated with the simple current
extension \eqref{eq:u1Z2} therefore realize a NIM-rep of the generalized 
fusion algebra $\F(u(1)_4; G_c)$.
Note that 
this 8-dimensional NIM-rep is distinct from the regular one,
which is $8 + 2 = 10$ dimensional. 
Hence we have obtained two NIM-reps of $\F(u(1)_4; G_c)$;
one of them corresponds to the charge-conjugation invariant
whereas the other originates from the block diagonal invariant
\eqref{eq:u1Z2}. 

Since the generalized fusion algebra contains the ordinary fusion
algebra as a subalgebra, one can decompose a NIM-rep of the former
into NIM-reps of the latter. 
Conversely, several NIM-reps (more precisely, $\abs{G}$ NIM-reps)
of the ordinary fusion
algebra may be combined to yield a NIM-rep of the generalized fusion algebra.
For $u(1)_4$, 
there are four NIM-reps of the ordinary fusion algebra $\F(u(1)_4)$
corresponding to four subgroups of the simple current group $\Z_8$
\cite{Gannon}.
Two of them, the regular NIM-rep and the two-dimensional one,
form the regular NIM-rep of the generalized fusion algebra.
The four-dimensional one \eqref{eq:k4psi} is
paired with itself to form a 8-dimensional NIM-rep of the generalized
fusion algebra, as we have seen above.
The remaining one, which is one-dimensional
and unphysical, 
is also paired with itself to yield a two-dimensional NIM-rep of
the generalized fusion algebra.
Consequently, there are three NIM-reps for the generalized 
fusion algebra $\F(u(1)_4; G_c)$; two of them are physical and correspond to 
the charge-conjugation modular invariant and the simple current extension
\eqref{eq:u1Z2}, respectively, while the remaining one is unphysical
and has no associated modular invariant. 
\bigskip

\noindent
\underline{$k=6$}
\medskip

For $k=6$, we have the following non-diagonal invariant,
\bes
  Z &= \abs{\chi_0}^2 + \abs{\chi_2}^2 + \abs{\chi_4}^2
    + \abs{\chi_6}^2 + \abs{\chi_8}^2 + \abs{\chi_{10}}^2 \\
    & \qquad
    + \chi_1 \overline{\chi_7}
    + \chi_3 \overline{\chi_9} 
    + \chi_5 \overline{\chi_{11}} 
    + \chi_7 \overline{\chi_1}
    + \chi_9 \overline{\chi_3} 
    + \chi_{11} \overline{\chi_5} \, , 
  \label{eq:u1k6}
\ees
which is the $\Z_2$-orbifold of $u(1)_6$.
Let $\ket{m}$ ($m \in \I =  \{0,1,2,\cdots,11 \}$) be the untwisted
boundary state
in the charge-conjugation modular invariant of $u(1)_6$ defined in 
\eqref{eq:u1bs}. 
The orbifold action on $\ket{m}$ reads
$\ket{m} \mapsto \ket{m + 6}$,
and the untwisted state in \eqref{eq:u1k6} can be obtained
by averaging $\ket{m}$ over the $\Z_2$-orbit,
\beq
  \ket{m}_{\Z_2} = \frac{1}{\sqrt{2}} (\ket{m} + \ket{m+6}) 
  = \frac{1}{\sqrt{6}} \sum_{n = 0,1,\dots,5}
  e^{-\frac{\pi i}{3} m n} \dket{(2n,2n^*)}
  \, .
\eeq
The $\omega_c$-twisted boundary states in \eqref{eq:u1k6}
can be constructed as the untwisted states in the $\Z_3$-orbifold
of $u(1)_6$, since the $\Z_3$-orbifold is the T-dual
of \eqref{eq:u1k6}.
The untwisted states in the $\Z_3$-orbifold are obtained
by averaging $\ket{m}$ over the $\Z_3$-orbit,
\beq
  \ket{m}_{\Z_3} = 
  \frac{1}{\sqrt{3}} (\ket{m} + \ket{m+4} + \ket{m+8})
  = \frac{1}{2} \sum_{n=0,1,2,3} 
  e^{-\frac{\pi i}{2} m n} \dket{(3n,3n^*)} \, .
\eeq
The $\omega_c$-twisted states in the $\Z_2$-orbifold \eqref{eq:u1k6},
which we denote by $\ket{m; \omega_c}_{\Z_2}$,
are constructed by acting $\omega_c$ on the holomorphic sector of 
$\ket{m}_{\Z_3}$,
\beq
  \ket{m; \omega_c}_{\Z_2}
  = R(\omega_c) \ket{m}_{\Z_3}
  = \frac{1}{2} \sum_{n=0,1,2,3} 
  e^{-\frac{\pi i}{2} m n} \dket{(12-3n,3n^*);\omega_c} \, .
\eeq
In this way, we obtain 6 untwisted states and 4 twisted states
in \eqref{eq:u1k6}.
The overlap matrix $\hat{n}$ is a $10 \times 10$ matrix and
takes the form,
\beq
  \hat{n}_l = \begin{pmatrix} 
  (n^{(6)}_1)^l & O^{(6,4)} \\ O^{(4,6)} & (n^{(4)}_1)^l \end{pmatrix} \,
  (l = 0,1,\dots,11) , \quad
  \hat{n}_\pm = \begin{pmatrix}
  O^{(6,6)} & n_\pm \\ n_\pm^T & O^{(4,4)} \end{pmatrix} \, ,
\eeq
where $n^{(4)}_1$ is defined in \eqref{eq:k4NIM},
$O^{(m,n)}$ is a $m \times n$ zero matrix
and the other matrices are defined as
\beq
  n^{(6)}_1 =
  \begin{pmatrix}
  0 & 1 & 0 & 0 & 0 & 0 \\
  0 & 0 & 1 & 0 & 0 & 0 \\
  0 & 0 & 0 & 1 & 0 & 0 \\
  0 & 0 & 0 & 0 & 1 & 0 \\
  0 & 0 & 0 & 0 & 0 & 1 \\
  1 & 0 & 0 & 0 & 0 & 0 
  \end{pmatrix} , \,
  n_+ = 
  \begin{pmatrix}
  1 & 0 & 1 & 0 \\
  0 & 1 & 0 & 1 \\
  1 & 0 & 1 & 0 \\
  0 & 1 & 0 & 1 \\
  1 & 0 & 1 & 0 \\
  0 & 1 & 0 & 1 
  \end{pmatrix} , \,
  n_- = 
  \begin{pmatrix}
  0 & 1 & 0 & 1 \\
  1 & 0 & 1 & 0 \\
  0 & 1 & 0 & 1 \\
  1 & 0 & 1 & 0 \\
  0 & 1 & 0 & 1 \\
  1 & 0 & 1 & 0 
  \end{pmatrix} . \,
\eeq
One can easily check that these matrices satisfy the generalized
fusion algebra \eqref{eq:u1Z2} for $k=6$.

The simple current group of $u(1)_6$ is isomorphic to
$\Z_{12}$. 
Hence we have 6 NIM-reps of the ordinary fusion algebra
$\F(u(1)_6)$ corresponding to
6 divisors $\{1,2,3,4,6,12\}$ of 12 \cite{Gannon}.
In the same way as the case of $k=4$,
two of them, the regular and the two-dimensional ones, form the regular
NIM-rep of the generalized fusion algebra $\F(u(1)_6; G_c)$.
NIM-reps of dimension 6 and 4 are combined
to yield another NIM-rep of $\F(u(1)_6; G_c)$ associated with
the $\Z_2$-orbifold (or equivalently, the $\Z_3$-orbifold) of $u(1)_6$,
as we have shown above.
In contrast with the case of $k=4$, however, the remaining two NIM-reps
of dimension 1 and 3 can not form a NIM-rep of $\F(u(1)_6; G_c)$.
Actually, one can show that the overlap matrices $\hat{n}$ 
containing the one-dimensional NIM-rep of $\F(u(1)_6)$ as a factor
and satisfying the generalized fusion algebra $\F(u(1)_6; G_c)$
necessarily have non-integral entries and can not be a NIM-rep.
The case of the three-dimensional NIM-rep is shown in the same way.
Consequently, we have two NIM-reps of $\F(u(1)_6; G_c)$,
one of which corresponds to the charge-conjugation
(or the diagonal) invariant while the other is associated
with the $\Z_2$-orbifold (or its T-dual) of $u(1)_6$.

\subsection{$su(3)_k$}

Our next example is $su(3)_k$ with the charge-conjugation $\omega_c$
as an automorphism of the chiral algebra. 
Since the automorphism group $G_c = \{1, \omega_c\}$ consists of
two elements, we can expect that two NIM-reps of the ordinary fusion algebra
$\F(su(3)_k)$ are combined to yield a NIM-rep of the generalized
fusion algebra $\F(su(3)_k; G_c)$. 
We check this for the case of $k=3$ and $k=5$,
in which there are respectively four and six 
independent NIM-reps of $\F(su(3)_k)$.
We present here only the outline of our analysis
and give the details in Appendix~\ref{sec:su3detail}. 
\bigskip

\noindent
\underline{$k=3$}
\medskip

For $k=3$, there are four NIM-reps
($A^{(6)}, A^{(6)*}, D^{(6)}$ and $D^{(6)*}$
\footnote{%
The superscript stands for the sum $k + h^\vee = k + 3$,
where $h^\vee$ is the dual Coxeter number of $su(3)$.})
of the ordinary fusion algebra $\F(su(3)_3)$ \cite{DZ,BPPZ}.
Two of them, $A^{(6)}$ and $A^{(6)*}$, correspond to respectively
the untwisted and the twisted boundary states
in the charge-conjugation modular invariant
and constitute the regular NIM-rep of the generalized fusion 
algebra $\F(su(3)_3; G_c)$.  
The remaining two, $D^{(6)}$ and $D^{(6)*}$, express respectively
the untwisted and the twisted boundary states corresponding to 
the following block diagonal modular invariant, 
\beq
  Z_{D^{(6)}} = 
  \abs{\chi_{(0,0)} + \chi_{(3,0)} + \chi_{(0,3)}}^2 
  + 3 \abs{\chi_{(1,1)}}^2 \, ,
  \label{eq:D6}
\eeq
which is a simple current extension of $su(3)_3$ by $(3,0)$.
Here the subscripts of characters denote the Dynkin label of $su(3)$. 
The extended chiral algebra of this invariant is $so(8)_1$
and one can construct the boundary states associated with \eqref{eq:D6}
using the data of $so(8)_1$.
Actually, as is shown in Appendix~\ref{sec:su3so8},
the automorphism group $G_c$ has a lift $\tilde{G}_c$ to $so(8)$,
which is isomorphic to the symmetric group $S_3$.
Accordingly,
the boundary states for $D^{(6)}$ and $D^{(6)*}$ are obtained
by the $\tilde{G}_c$-twisted states in $so(8)$.
By an explicit calculation of the overlap of the twisted states
with the untwisted ones in $su(3)$, we can show that two NIM-reps,
$D^{(6)}$ and $D^{(6)*}$, of $\F(su(3)_3)$
form a NIM-rep of $\F(su(3)_3; G_c)$. 
In this way, four NIM-reps of the ordinary
fusion algebra $\F(su(3)_3)$ are organized into two NIM-reps
of the generalized fusion algebra $\F(su(3)_3; G_c)$;
one corresponds to the charge-conjugation modular invariant
while the other is originated from the simple current extension
\eqref{eq:D6}.
\bigskip

\noindent
\underline{$k=5$}
\medskip

For $k=5$, there are six NIM-reps 
($A^{(8)}, A^{(8)*}, D^{(8)}, D^{(8)*}, E^{(8)}$ and $E^{(8)*}$)
of the ordinary fusion algebra $\F(su(3)_5)$ \cite{DZ,BPPZ}. 
Similarly to the case of $k=3$, 
$A^{(8)}$ and $A^{(8)*}$ express the boundary states
in the charge-conjugation invariant and form
the regular NIM-rep of the generalized fusion algebra
$\F(su(3)_5; G_c)$. 
The remaining four correspond to other modular invariants.
Two of them, $D^{(8)}$ and $D^{(8)*}$, represent the boundary states
in the simple current invariant
\bes
  Z_{D^{(8)}} &= 
  \abs{\chi_{(0,0)}}^2 + \abs{\chi_{(1,1)}}^2 + \abs{\chi_{(2,2)}}^2
  + \abs{\chi_{(3,0)}}^2 + \abs{\chi_{(0,3)}}^2
  + \abs{\chi_{(4,1)}}^2 + \abs{\chi_{(1,4)}}^2 \\
  & \quad + \bigl( 
        \chi_{(5,0)} \overline{\chi_{(0,5)}} 
      + \chi_{(3,1)} \overline{\chi_{(1,3)}}
      + \chi_{(1,2)} \overline{\chi_{(2,1)}} \\
  & \qquad + 
        \chi_{(2,3)} \overline{\chi_{(0,2)}}
      + \chi_{(2,0)} \overline{\chi_{(3,2)}}
      + \chi_{(0,4)} \overline{\chi_{(1,0)}}
      + \chi_{(0,1)} \overline{\chi_{(4,0)}}
      + \,\mathrm{c.c }\, \bigr) \, ,
\label{eq:D8}
\ees
while $E^{(8)}$ and $E^{(8)*}$ correspond to the exceptional 
invariant
\bes
   Z_{E^{(8)}} & = \abs{\chi_{(0,0)}+\chi_{(2,2)}}^2 
              + \abs{\chi_{(0,2)}+\chi_{(3,2)}}^2 
              + \abs{\chi_{(1,2)}+\chi_{(5,0)}}^2 \\
       & \phantom{=}
         + \abs{\chi_{(3,0)}+\chi_{(0,3)}}^2 
         + \abs{\chi_{(2,1)}+\chi_{(0,5)}}^2 
         + \abs{\chi_{(2,0)}+\chi_{(2,3)}}^2 \, ,    
\label{eq:E8}
\ees
which originates from the conformal embedding $su(3)_5 \subset su(6)_1$. 

The invariant \eqref{eq:D8} is the $\Z_3$-orbifold of $su(3)_5$
and we can obtain untwisted states by averaging the $\Z_3$-orbit
of untwisted states in the charge-conjugation invariant.
On the other hand,
the twisted states in the charge-conjugation invariant
are fixed points of the orbifold action and we need an appropriate
resolution of them by the Ishibashi states from the twisted sector
of the orbifold to obtain twisted states in \eqref{eq:D8}.
For the invariant \eqref{eq:E8}, one can construct 
twisted boundary states from the data of $su(6)$,
since the charge-conjugation of $su(3)$ is lifted to that of $su(6)$.
One obtains six untwisted states together with two twisted states
using the data of $su(6)$.
The remaining states of $E^{(8)}$ and $E^{(8)*}$
can be obtained by, \textit{e.g.},
the fusion with representations of $su(3)$.

From the explicit form of boundary states in
\eqref{eq:D8} and \eqref{eq:E8},
we can calculate the overlap of twisted states with untwisted ones
to show that two NIM reps of $\F(su(3)_5)$ corresponding to
the same modular invariant form a NIM-rep of $\F(su(3)_5; G_c)$. 
Consequently, we obtain three NIM-reps of
the generalized fusion algebra $\F(su(3)_5; G_c)$
corresponding to three modular invariants available for $k=5$. 

\subsection{$su(3)_1^{\oplus 3}$}

The final example we consider is
$su(3)_1^{\oplus 3} = su(3)_1 \oplus su(3)_1 \oplus su(3)_1$.
The details are given in Appendix~\ref{sec:E6detail}.
This chiral algebra has the automorphism group
consisting of all the permutations of three factors, 
which is isomorphic to $S_3$, and
we obtain a non-commutative generalized fusion algebra
$\F(su(3)_1^{\oplus 3}; S_3)$.
For the chiral algebra $su(3)_1^{\oplus 3}$, 
we have a non-trivial
block-diagonal modular invariant
\bes
  Z_{E_6} &= 
  \abs{\chi_{(0,0,0)} + \chi_{(1,1,1)} + \chi_{(2,2,2)}}^2 \\
  & \qquad
  + \abs{\chi_{(0,2,1)} + \chi_{(1,0,2)} + \chi_{(2,1,0)}}^2
  + \abs{\chi_{(0,1,2)} + \chi_{(2,0,1)} + \chi_{(1,2,0)}}^2 \, ,
  \label{eq:su3E6main}
\ees
where the subscripts stands for the label of representations of 
$su(3)_1^{\oplus 3}$ and the numbers $0, 1, 2$ correspond respectively to
the fundamental weights $\Lambda_0, \Lambda_1, \Lambda_2$ of $su(3)_1$.
This invariant originates from
the conformal embedding $su(3)_1^{\oplus 3} \subset E_{6,1}$.~\footnote{%
$E_{6, 1}$ means $E_6$ at level 1.
} 
Since the automorphism group $S_3$ of $su(3)_1^{\oplus 3}$
has a lift to $E_{6,1}$, we can construct twisted boundary states
for $S_3$ \cite{Recknagel,GSN} in this invariant.
In the same way as the case of $su(3)_k$, we can calculate the overlap
of twisted states to check that the overlap matrices 
indeed form a NIM-rep
of the generalized fusion algebra $\F(su(3)_1^{\oplus 3}; S_3)$.

\section{Graph fusion algebras}
\label{sec:graph}

Given a boundary state coefficient matrix 
$\Psi_{\alpha \lambda} (\alpha \in \V, \lambda \in \E)$~\footnote{%
In this section, we omit the label for the anti-holomorphic 
sector of Ishibashi states for simplicity.
}
and $0 \in \V$ such that $\Psi_{0 \lambda} \neq 0$,
one can define the graph fusion algebra \cite{DZ,PZ,BPPZ}
\beq
  (\alpha) \times (\beta) = \sum_{\gamma \in \V}
  G_{\alpha \beta}{}^{\gamma} \, (\gamma) \qquad
  (\alpha, \beta \in \V)
\eeq
with the structure constant
\beq
  G_{\alpha \beta}{}^\gamma = 
  \sum_{\lambda \in \E} 
  \frac{\Psi_{\alpha \lambda} \Psi_{\beta \lambda} 
        \overline{\Psi_{\gamma \lambda}}}{\Psi_{0 \lambda}} \, .
  \label{eq:Gfusion}
\eeq
Since this has the same form as the Verlinde formula
\eqref{eq:Verlinde},
one can consider the graph fusion algebra as
a generalization of ordinary fusion algebras. 
Actually, for the case of boundary states 
associated with a block diagonal invariant, 
it is observed \cite{DZ} that the graph fusion
algebra contains the fusion algebra of the extended theory 
as a subalgebra.

For the case of simple current extensions,
this relation of graph fusion
algebras and the fusion algebra of the extended theory can be explained
from the point of view of the generalized fusion algebra.
As an illustration,
we consider the $D$-type modular invariant of $su(2)_4$,
\beq
  Z = \abs{\chi_0 + \chi_4}^2 + 2 \abs{\chi_2}^2 \, ,
  \label{eq:su2D}
\eeq
where the subscript stands for the Dynkin label of $su(2)$. 
This invariant is obtained from a simple current extension
of $su(2)_4$ by the simple current group 
$\Gamma = \{(0),(4)\} \cong \Z_2$.
The extended chiral algebra is $su(3)_1$ and one can regard
\eqref{eq:su2D} as the charge-conjugation invariant of $su(3)_1$
\beq
  Z = \abs{\chi_{(0,0)}}^2 + \abs{\chi_{(1,0)}}^2
  + \abs{\chi_{(0,1)}}^2 \, ,
  \label{eq:su31}
\eeq
where representations are again denoted by their Dynkin labels.
One obtains \eqref{eq:su2D} from \eqref{eq:su31} using
the branching rule for the representations of $su(3)_1$
\beq
  (0,0) = (0) \oplus (4) \, , \quad
  (1,0) = (2) \, , \quad
  (0,1) = (2) \, .
  \label{eq:su3su2}
\eeq

The unextended chiral algebra in simple current extensions 
can be characterized by the fixed point algebra of 
a certain automorphism group $G$ of the extended chiral algebra \cite{FS},
where $G$ is the character group for the simple current group $\Gamma$
used in the extension
and hence isomorphic to $\Gamma$; 
$G = \Gamma^* \cong \Gamma$.
For the case of $su(2)_4 \subset su(3)_1$,
$G$ is nothing but the charge-conjugation automorphism group $G_c$
of $su(3)_1$, since $su(2)_4$ is the fixed point algebra
of the charge-conjugation $\omega_c$.
The eigenvalue of $\omega_c$ is $\pm 1$, since $\omega_c$ has order 2.
The vacuum representation of $su(3)_1$ is therefore
decomposed into two parts according to the eigenvalue of $\omega_c$.
Since $\omega_c$ fixes $su(2)_4 \subset su(3)_1$,
this decomposition of $(0,0)$ coincides with the decomposition
\eqref{eq:su3su2} into representations of $su(2)_4$. 
The representation $(0)$ is $su(2)_4$ itself and corresponds to $\omega_c = 1$.
The representation $(4)$ contains the highest root of $su(3)$,
whose sign is reversed under the exchange of two simple roots
of $su(3)$, \textit{i.e.} the charge-conjugation $\omega_c$.
The representation $(4)$ therefore corresponds to $\omega_c = -1$,
\beq
  \omega_c : \, (0) \mapsto (0) \, , \quad (4) \mapsto -(4) \, .
\eeq
From this action, 
one can express the Ishibashi states of $su(3)_1$ in terms of
those of $su(2)_4$ as follows,
\bes
  \dket{(0,0)} = \frac{1}{\sqrt{2}}
  (\dket{0} + \dket{4}) \, , \quad
  \dket{(0,0); \omega_c} = \frac{1}{\sqrt{2}}
  (\dket{0} - \dket{4}) \, ,
\ees
where we have used the definition \eqref{eq:twistIshibashi}
of twisted Ishibashi states.
Conversely, one obtains 
\bes
  \dket{0} &= \frac{1}{\sqrt{2}}
  (\dket{(0,0)} + \dket{(0,0); \omega_c})
  = \dket{((0,0); +)} \, , \\
  \dket{4} &= \frac{1}{\sqrt{2}}
  (\dket{(0,0)} - \dket{(0,0); \omega_c})
  = \dket{((0,0); -)} \, ,
\ees
where $\pm$ stands for the irreducible representations of
$G_c = \{1, \omega_c \} \cong \Z_2$. 
The remaining two representations of $su(3)_1$ have the trivial
stabilizer $\Sc((1,0)) = \Sc((0,1)) = \{1\}$
and we have only one basis for each of them
\beq
  \dket{2_1} = \dket{(1,0)} \, , \quad
  \dket{2_2} = \dket{(0,1)} \, .
\eeq
We have distinguished two bulk fields with representation $(2)$
in \eqref{eq:su2D} by putting the subscript. 

In this way,
the basis \eqref{eq:Ishibashinew} of twisted Ishibashi states
in $su(3)_1$ can be regarded
as the (untwisted) Ishibashi states of $su(2)_4$. 
This implies that the regular states \eqref{eq:regular}
of the generalized fusion algebra $\F(su(3)_1;G_c)$
yield a complete set of boundary states in \eqref{eq:su2D},
whose coefficient matrix $\Psi$ is given by the generalized
$S$-matrix \eqref{eq:Shat}
\beq
  \Psi = \Shat = \frac{1}{\sqrt{6}}
  \begin{pmatrix}
    1 & 1 & \sqrt{2} & \sqrt{2} \\
    1 & 1 & \sqrt{2} \kappa & \sqrt{2} \bar{\kappa} \\
    1 & 1 & \sqrt{2} \bar{\kappa} & \sqrt{2} \kappa \\
    \sqrt{3} & -\sqrt{3} & 0 & 0 
  \end{pmatrix} \quad (\kappa = e^{\frac{2 \pi i}{3}} ) \, . 
\eeq
Here 
the columns are ordered as 
$\{((0,0);+), ((0,0);-), (1,0), (0,1)\}$.
The first three rows correspond to three representations 
$\{(0,0),(1,0),(0,1)\}$ of
$su(3)_1$, respectively,
while the fourth row stands for the representation of
the twisted chiral algebra $A_2^{(2)}$ at level 1. 
The boundary states of \eqref{eq:su2D} that preserve $su(2)_4$ 
are therefore labeled by $L \in \Ihat$ for the generalized
fusion algebra $\F(su(3)_1; G_c)$. 

The structure constants \eqref{eq:Gfusion} of the graph fusion algebra
associated with $\Psi$ is defined by summing over the column indices
of $\Psi$.
Since $\Psi = \Shat$ in the present case, 
this is nothing but the generalized Verlinde formula
\eqref{eq:gen_Verlinde2} if we choose $(0,0) \in \Ihat$ as 
the node $0$ in \eqref{eq:Gfusion}.
Namely, the structure constants of the graph fusion algebra
are given by the generalized fusion coefficients 
of $\F(su(3)_1; G_c)$,
\beq
  G_{LM}{}^N = \fusionhat{L}{M}{N} \, .
\eeq
The graph fusion algebra associated with \eqref{eq:su2D} therefore
coincides with the generalized fusion algebra $\F(su(3)_1; G_c)$.
Since the generalized fusion algebra contains the ordinary
fusion algebra as a subalgebra,
this identification of the graph fusion algebra
with the generalized fusion
algebra naturally explains the observation that the graph fusion algebra
contains the fusion algebra of the extended theory as a subalgebra.
\medskip

This argument for the relation of graph fusion algebras with
the generalized fusion algebras of the extended theory
is readily extended to 
simple current extensions other than $su(2)_4 \subset su(3)_1$.
For instance, the case of $su(3)_3 \subset so(8)_1$ in \eqref{eq:D6}
can be treated
in completely the same way.

First, we give 
the branching rule for the representations \eqref{eq:so8I}
of $so(8)_1$,
\beq
  O = (0,0) \oplus (3,0) \oplus (0,3) \, , \quad
  V = (1,0) \, , \quad
  S = (1,0) \, , \quad
  C = (1,0) \, . \quad
  \label{eq:so8branching}
\eeq
As is shown in Appendix~\ref{sec:su3so8}, 
the unextended chiral algebra $su(3)_3$ is characterized
as the fixed point algebra of an automorphism group 
$\langle \pi' \rangle \cong \Z_3$ of $so(8)_1$.
Since $\pi'$ has order 3, the eigenvalue of $\pi'$ takes the form
$\kappa^n \, (\kappa = e^{\frac{2\pi i}{3}}\, ; n=0,1,2)$.
The decomposition \eqref{eq:so8branching} of 
the vacuum representation $O$ of $so(8)_1$
into those of $su(3)_3$ corresponds to the decomposition
according to the eigenvalue of $\pi'$, since $\pi'$ fixes $su(3)_3$.
From the explicit realization of $su(3)_3 \subset so(8)_1$ given
in Appendix~\ref{sec:su3so8}, one can show that
$\pi'$ acts on the vacuum representation $O$ of $so(8)_1$ as follows
\beq
  \pi' : 
  (0,0) \mapsto (0,0) \, , \,\,
  (3,0) \mapsto \bar{\kappa} (3,0) \, , \,\,
  (0,3) \mapsto \kappa (0,3) \, . \,\,
\eeq
The twisted Ishibashi states of $so(8)_1$ are then expressed 
in terms of the (untwisted) Ishibashi states of $su(3)_3$,
\bes
  \dket{O} &= \frac{1}{\sqrt{3}}
  (\dket{(0,0)} + \dket{(3,0)} + \dket{(0,3)} ) \, , \\
  \dket{O; \pi'} &= \frac{1}{\sqrt{3}}
  (\dket{(0,0)} + \bar{\kappa} \dket{(3,0)} + \kappa \dket{(0,3)} ) \, , \\
  \dket{O; {\pi'}^2} &= \frac{1}{\sqrt{3}}
  (\dket{(0,0)} + \kappa \dket{(3,0)} + \bar{\kappa} \dket{(0,3)} ) \, , 
\ees
where we have used again the definition \eqref{eq:twistIshibashi}
of twisted Ishibashi states.
From this expression, one obtains
\bes
  \dket{(0,0)} &= \frac{1}{\sqrt{3}}
  (\dket{O} + \dket{O;\pi'} + \dket{O;{\pi'}^2}) 
  = \dket{(O;\rho'_0)} \, , \\
  \dket{(3,0)} &= \frac{1}{\sqrt{3}}
  (\dket{O} + \kappa \dket{O;\pi'} + \bar{\kappa} \dket{O;{\pi'}^2}) 
  = \dket{(O;\rho'_1)} \, , \\  
  \dket{(0,3)} &= \frac{1}{\sqrt{3}}
  (\dket{O} + \bar{\kappa} \dket{O;\pi'} + \kappa \dket{O;{\pi'}^2}) 
  = \dket{(O;\rho'_2)} \, ,
  \label{eq:su3so8dict1}
\ees
where $\rho'_n \, (n=0,1,2)$
are the irreducible representations of $\langle \pi' \rangle$ 
defined as $\rho'_n(\pi') = \kappa^{2n}$.
The other three representations of $so(8)_1$ have the trivial stabilizer
$\{1\} \subset \langle \pi' \rangle$ and
we can relate the corresponding Ishibashi states 
with those of  $su(3)_3$ as follows,
\beq
  \dket{(1,1)_1} = \dket{V} \, , \quad
  \dket{(1,1)_2} = \dket{S} \, , \quad 
  \dket{(1,1)_3} = \dket{C} \, .
  \label{eq:su3so8dict2}
\eeq
We have distinguished three bulk fields with representation $(1,1)$ 
in \eqref{eq:D6} again by putting the subscript.
The six untwisted Ishibashi states of $su(3)_3$ are thus identified
with the six twisted Ishibashi states of $so(8)_1$ for
the automorphism group $\langle \pi' \rangle \cong \Z_3$.
Hence the boundary states of $su(3)_3$ associated with 
the modular invariant \eqref{eq:D6} are identified with
the twisted boundary states of $so(8)_1$ for 
$\langle \pi' \rangle \cong \Z_3$.

Since $\pi'$ is composed of the triality automorphism $\pi$
and an inner automorphism of $so(8)_1$, 
the twisted chiral algebra for $\pi'$ is also isomorphic to
the twisted affine Lie algebra $D_4^{(3)}$ at level 1 and
we have only one
representation for $\pi'$.
The same argument holds for ${\pi'}^2$.
The generalized fusion algebra $\F(so(8)_1; \langle \pi' \rangle)$
therefore consists of $4 + 1 + 1 = 6$ representations.
From the definition \eqref{eq:Shat},
one obtains its generalized $S$-matrix in the form
\beq
  \Shat = \frac{1}{2\sqrt{3}}
  \begin{pmatrix} 
  1 & 1 & 1 & \sqrt{3} & \sqrt{3} & \sqrt{3} \\
  1 & 1 & 1 & \sqrt{3} & -\sqrt{3} & -\sqrt{3} \\
  1 & 1 & 1 & -\sqrt{3} & \sqrt{3} & -\sqrt{3} \\
  1 & 1 & 1 & -\sqrt{3} & -\sqrt{3} & \sqrt{3} \\
  2 & 2\kappa & 2\bar{\kappa} & 0 & 0 & 0 \\
  2 & 2\bar{\kappa} & 2\kappa & 0 & 0 & 0 
  \end{pmatrix} \, .
  \label{eq:su33psi}
\eeq
Here the columns are ordered as
$\{(O;\rho'_0), (O;\rho'_1), (O;\rho'_2), (V), (S), (C) \}$.
The first four rows express respectively the untwisted representations
$O, V, S$ and $C$,
while the last two rows correspond to representations
in $\I^{{\pi'}^2}$ and $\I^{\pi'}$.
This matrix gives the boundary state coefficients for
the regular states of $\F(so(8)_1; \langle \pi' \rangle)$.
Using the correspondence of Ishibashi states given in
\eqref{eq:su3so8dict1} and \eqref{eq:su3so8dict2},
we can regard this as the boundary state coefficient matrix
$\Psi_{D^{(6)}}$ for the untwisted 
boundary states associated with \eqref{eq:D6}.
The graph fusion algebra
corresponding to $\Psi_{D^{(6)}}$ 
is therefore identified with the generalized fusion algebra
$\F(so(8)_1; \langle \pi' \rangle)$.
Since $\F(so(8)_1; \langle \pi' \rangle)$ 
contains the ordinary fusion algebra $\F(so(8)_1)$,
which is the group algebra of $\Z_2 \times \Z_2$,
the graph fusion algebra of $\Psi_{D^{(6)}}$ also contains
it as a subalgebra.
\medskip

In the modular invariant \eqref{eq:D6}, 
in addition to the untwisted boundary states,
we have the twisted
boundary states for the charge-conjugation
automorphism group $G_c = \{1, \omega_c\}$ of $su(3)_3$.
As is shown in Appendix~\ref{sec:su3so8}, 
the automorphism group $G_c$ has a lift 
$\tilde{G}_c = \langle \pi',  \sigma' \rangle$ 
to $so(8)_1$, which is isomorphic to the symmetric group $S_3$.
Using this fact, we can construct the $\omega_c$-twisted states
in \eqref{eq:D6} starting from the $\tilde{G}_c$-twisted
states of $so(8)_1$.

Since $\sigma' \in \tilde{G}_c$ is composed of the chirality flip and
an inner automorphism of $so(8)_1$, the twisted chiral algebra
for $\sigma'$ is isomorphic to $D_4^{(2)}$ at level 1
and the generalized fusion algebra $\F(so(8)_1; \tilde{G}_c)$
has exactly the same form as $\F(so(8)_1; S_3)$ presented in 
Section~\ref{sec:so8}.
Consequently, one obtains $12$ twisted boundary states
for $\tilde{G}_c$, 
of which six states for 
$\langle \pi' \rangle = \{1, \pi', {\pi'}^2\}$
correspond to the untwisted states of $su(3)_3$ while
the remaining six for the coset  
$\sigma' \langle \pi' \rangle = \{\sigma', \pi'\sigma', {\pi'}^2 \sigma' \}$
are regarded as the $\omega_c$-twisted ones, since
the restriction of $\sigma' \langle \pi' \rangle$ to
$su(3)_3$ yields $\omega_c$. 
For example, two $\sigma'$-twisted states of $so(8)_1$ take the form
\beq
  \frac{1}{\sqrt{2}} (\dket{O;\sigma'} \pm \dket{V;\sigma'}) \, .
  \label{eq:so8sigmastates}
\eeq
The $\sigma'$-twisted Ishibashi states of $so(8)_1$
can be expressed in terms of the $\omega_c$-twisted Ishibashi
states of $su(3)_3$,
\bea
  \dket{O;\sigma'} = R(\sigma') \dket{O} 
  &= \frac{1}{\sqrt{3}} R(\omega_c)
  (\dket{(0,0)} + \dket{(3,0)} 
   + \dket{(0,3)} ) \notag \\
  &= \frac{1}{\sqrt{3}}
  (\dket{(0,0);\omega_c} + \dket{(0,3);\omega_c} 
   + \dket{(3,0);\omega_c} ) \, , \\
  \dket{V; \sigma'} = R(\sigma') \dket{V} 
  &= R(\omega_c) \dket{(1,1)_1} = \dket{(1,1)_1; \omega_c} \, .
\eea
Substituting these expressions into \eqref{eq:so8sigmastates},
we obtain two $\omega_c$-twisted states of $su(3)_3$,
\beq
  \frac{1}{\sqrt{6}} (\dket{(0,0);\omega_c} + \dket{(3,0);\omega_c} 
   + \dket{(0,3);\omega_c}
   \pm \sqrt{3} \dket{(1,1)_1; \omega_c}) \, .
\eeq
The case of $\pi' \sigma'$ and ${\pi'}^2 \sigma'$ can be treated 
in the same way and we obtain four more $\omega_c$-twisted states,
\bea
  \lefteqn{%
  \frac{1}{\sqrt{2}} (\dket{O;\pi' \sigma'} \pm \dket{C;\pi' \sigma'})}
  \qquad & \notag \\
  &=   \frac{1}{\sqrt{6}} (\dket{(0,0);\omega_c} 
   + \bar{\kappa} \dket{(3,0);\omega_c} 
   + \kappa \dket{(0,3);\omega_c}
   \pm \sqrt{3} \dket{(1,1)_3; \omega_c}) \, , \\
  \lefteqn{%
  \frac{1}{\sqrt{2}} 
  (\dket{O;{\pi'}^2 \sigma'} \pm \dket{S;{\pi'}^2 \sigma'})}
  \qquad & \notag \\
  &=   \frac{1}{\sqrt{6}} (\dket{(0,0);\omega_c} 
   + \kappa \dket{(3,0);\omega_c} 
   + \bar{\kappa} \dket{(0,3);\omega_c}
   \pm \sqrt{3} \dket{(1,1)_2; \omega_c}) \, .
\eea
The $\omega_c$-twisted boundary states obtained in this way
form a complete set of twisted states,
since the number of the twisted boundary states
is the same as the number $\abs{\E(\omega_c)}$
of the twisted Ishibashi states available in \eqref{eq:D6}.
The resulting boundary states have the following coefficient matrix
\beq
  \Psi_{D^{(6)*}} = \frac{1}{\sqrt{6}}
  \begin{pmatrix}
  1 & 1 & 1 & \sqrt{3} & 0 & 0 \\
  1 & 1 & 1 & -\sqrt{3} & 0 & 0 \\
  1 & \kappa & \bar{\kappa} & 0 & \sqrt{3} & 0 \\
  1 & \kappa & \bar{\kappa} & 0 & -\sqrt{3} & 0 \\
  1 & \bar{\kappa} & \kappa & 0 & 0 & \sqrt{3} \\
  1 & \bar{\kappa} & \kappa & 0 & 0 & -\sqrt{3}
  \end{pmatrix} \, ,
  \label{eq:su33sctwistpsi}
\eeq
where the columns are ordered as 
$\{(0,0),(3,0),(0,3),(1,1)_1,(1,1)_2,(1,1)_3\}$.
This matrix $\Psi_{D^{(6)*}}$ realizes the $D^{(6)*}$ NIM-rep of
the ordinary fusion algebra $\F(su(3)_3)$.
By appropriately mixing three columns of $\Psi_{D^{(6)*}}$
corresponding to $(1,1)_n \,(n=1,2,3)$, 
one can make the first row of $\Psi_{D^{(6)*}}$ 
consist of only non-vanishing entries.
The graph fusion algebra for $D^{(6)*}$ is then constructed
by taking the first row as the node $0$.
It would be interesting to relate this algebra with some
generalized fusion algebra, possibly $\F(so(8)_1; \tilde{G}_c)$.

\section{Summary and Discussion}

In this paper, we have studied a consistency condition 
of boundary states satisfying the boundary condition
twisted by an automorphism group $G$ of the chiral algebra $\A$
and clarified the relation of twisted boundary states with
the generalized fusion algebra. 
We have shown that the fusion coefficients of the generalized fusion
algebra is expressed by a formula analogous to the Verlinde formula
even for non-abelian cases
and thereby determined irreducible representations of the generalized
fusion algebra, which generalize quantum dimensions
of the ordinary fusion algebras.
For a non-abelian $G$,
some irreducible representations have dimension greater than 1
reflecting the fact that the generalized fusion algebra is
non-commutative.
Based on these results, 
we have shown that a consistent set of twisted boundary states
forms a NIM-rep of the generalized fusion algebra.
As a check of our argument,
we have considered twisted boundary states in several models,
which includes non-diagonal modular invariants as well as
the case of non-abelian automorphisms.
We have seen that several NIM-reps of the ordinary fusion algebra
are organized into a single NIM-rep of the generalized fusion algebra.
In particular, the $D$ and $E$ type NIM-reps of 
$su(3)_k\,(k=3,5)$ are paired
with their counterpart $D^*$ and $E^*$, respectively, to yield
a NIM-rep of the generalized fusion algebra for the charge-conjugation
automorphism group of $su(3)_k$. 
Finally, we have given an argument that the graph fusion algebra
associated with a simple current extension can be regarded
as the generalized fusion algebra of the extended chiral algebra,
which naturally explains the observation that the graph fusion algebra
contains the fusion algebra of the extended theory as a subalgebra.

Having obtained these results, 
a natural problem is the classification of NIM-reps of the generalized
fusion algebra,
which generalizes the corresponding problem \cite{BPPZ,Gannon}
for the case of ordinary fusion algebras. 
These two problems are related with each other,
since the generalized fusion algebra contains the ordinary fusion
algebra as a subalgebra and a NIM-rep of the former is decomposed
into NIM-reps of the latter.
More generally, if the automorphism group $G$ of the chiral algebra
$\A$ has a subgroup $H \subsetneq G$, 
the generalized fusion algebra $\F(\A;G)$ has a subalgebra $\F(\A;H)$,
and a NIM-rep of $\F(\A;G)$ is decomposed into NIM-reps of
$\F(\A;H)$.
On the other hand,
as we have pointed out in Section~\ref{sec:u1},
there exist some unphysical NIM-reps of 
the ordinary fusion algebra $\F(u(1)_6)$
that are not originated from those of the generalized fusion algebra
$\F(u(1)_6; G_c)$.
This example shows that not all of the NIM-reps 
for $\F(\A;H)$ can be obtained from those of $\F(\A;G)$.
It is interesting to find other examples with this property
and understand its significance. In particular, any relation
with the notion of physical NIM-reps \cite{Gannon}
would be desirable.

The case of block diagonal modular invariants is of particular interest.
In that case, the chiral algebra $\A$ has an extension $\A_\mathrm{ext}$
and one can consider a lift of 
the automorphism group $G$ of $\A$ to $\A_\mathrm{ext}$.
If a lift $G_\mathrm{ext}$ of $G$ exists,~\footnote{%
This is actually an assumption.
There is in general an obstruction to lifting 
the automorphism group $G$ of $\A$ to 
the extended chiral algebra $\A_\mathrm{ext}$ \cite{FSW}.
}
one can construct $G_\mathrm{ext}$-twisted boundary states 
of the extended theory, which can also be regarded as $G$-twisted states
of the unextended theory.
This suggests that there is some relation between
two generalized fusion algebras $\F(\A; G)$ and
$\F(\A_\mathrm{ext}; G_\mathrm{ext})$.
Indeed, for the case of simple current extensions,
NIM-reps of $\F(\A_\mathrm{ext}; G_\mathrm{ext})$
can be considered as NIM-reps of $\F(\A; G)$,
as we have seen in Section~\ref{sec:graph}.
Probably, this correspondence of NIM-reps is a common feature of
the cases in which the unextended chiral algebra $\A$ is 
characterized as the fixed point algebra
of a certain automorphism group $H_\mathrm{ext} \subset G_\mathrm{ext}$
of $\A_\mathrm{ext}$.
This is exactly the setting of the Galois theory for 
vertex operator algebras (see \cite{FSW} and references therein)
and the mutual relation of 
generalized fusion algebras in such cases
might be described also by the Galois theory.
For exceptional invariants, however, the unextended chiral algebra
$\A$ can not be obtained as the fixed point algebra 
for any automorphism group of $\A_\mathrm{ext}$ in general,
and one would need another tool, such as the operator-algebraic methods
\cite{Xu,BEK,FS2},
for a deep understanding of
the mutual relation of generalized fusion algebras associated with
an exceptional invariant.
\bigskip

\newpage
\appendix
\section{Algebra embedding $su(3)_3 \subset so(8)_1$}
\label{sec:su3so8}

In this appendix, we give an explicit realization of
the algebra embedding $su(3)_3 \subset so(8)_1$
and show that the charge-conjugation automorphism group
$G_c = \{1, \omega_c \} \cong \Z_2$ of $su(3)_3$ has a lift 
$\tilde{G}_c$ to
$so(8)_1$, which is isomorphic to the symmetric group $S_3$.

The affine Lie algebra $so(8)_1$ can be expressed by 
four pairs of complex fermions 
$\psi_i^\pm \,(i=1,2,3,4)$.
The Cartan element $\tilde{H}_i$ and
the simple root $\tilde{E}^{\alpha_i}$
of $so(8)$ have the form
\beq
  \tilde{H}_i = \psi_i^+ \psi_i^- \, , \quad
  \tilde{E}^{\alpha_1} = \psi_1^+ \psi_2^- \, , \quad
  \tilde{E}^{\alpha_2} = \psi_2^+ \psi_3^- \, , \quad
  \tilde{E}^{\alpha_3} = \psi_3^+ \psi_4^- \, , \quad
  \tilde{E}^{\alpha_4} = \psi_3^+ \psi_4^+ \, .
\eeq
The other roots are written as $\psi_i^\pm \psi_j^\pm$ or
$\psi_i^\pm \psi_j^\mp$ with $i \neq j$.
The subalgebra $su(3) \subset so(8)$ with embedding index $3$
can be realized as follows,
\bes
  H_1 &= \frac{1}{\sqrt{2}}
  (-\tilde{H}_1 - 2 \tilde{H}_2 +  \tilde{H}_3) \, , \quad
  H_2 = \sqrt{\frac{3}{2}} (\tilde{H}_1 + \tilde{H}_3) \, , \\
  E^1 &= 
  \psi_1^- \psi_3^+ + \kappa\, \psi_4^- \psi_2^- 
  + \bar{\kappa}\, \psi_4^+ \psi_2^- \, ,\\
  E^2 &= 
  \psi_2^+ \psi_3^+ + \kappa\, \psi_1^+ \psi_4^+ 
  + \bar{\kappa}\, \psi_1^+ \psi_4^- \, , \\
  E^\theta &= 
  \tilde{E}^{\alpha_1} + \tilde{E}^{\alpha_3} + \tilde{E}^{\alpha_4}
  = \psi_1^+ \psi_2^- + \psi_3^+ \psi_4^- + \psi_3^+ \psi_4^+ \, , 
  \label{eq:su3so8}
\ees
where $H_i \,(i=1,2)$ and
$E^i (i=1,2)$ 
stand for the Cartan elements and the simple roots of $su(3)$,
respectively.
$E^\theta = [E^1, E^2]$ is the highest root
of $su(3)$ and $\kappa = e^{\frac{2\pi i}{3}}$.
The $28$ elements of $so(8)$ can be decomposed into irreducible representations
of this $su(3)$ as 
$\mathbf{28} = \mathbf{8} \oplus \mathbf{10} \oplus \overline{\mathbf{10}}$, 
where $\mathbf{10}$ ($\overline{\mathbf{10}}$) is
the highest weight representation with the Dynkin label
$(3,0)$ ($(0,3)$)
and contains $\psi_2^- \psi_3^+$ ($\psi_1^+ \psi_3^+$)
as the highest weight state.

One can construct an automorphism of $so(8)$ that fixes  
the $su(3)$ subalgebra \eqref{eq:su3so8} from the triality
of $so(8)$. 
Let $\pi$ be the triality automorphism of $so(8)$,
which generates an automorphism group 
$\langle \pi \rangle = \{1, \pi, \pi^2 \} \cong \Z_3$. 
The action of $\pi$ on the simple roots of $so(8)$ reads
\beq
  \pi : 
  \tilde{E}^{\alpha_1} \mapsto \tilde{E}^{\alpha_3} \mapsto 
  \tilde{E}^{\alpha_4} \mapsto \tilde{E}^{\alpha_1} \, , \,\,
  \tilde{E}^{\alpha_2} \mapsto \tilde{E}^{\alpha_2} \, ,
\eeq
which in turn implies the following action on the Cartan elements,
\bes
  \pi : 
  \tilde{H}_1 &\mapsto \frac{1}{2}
  (\tilde{H}_1 + \tilde{H}_2 + \tilde{H}_3 - \tilde{H}_4) \, , \\
  \tilde{H}_2 &\mapsto \frac{1}{2}
  (\tilde{H}_1 + \tilde{H}_2 - \tilde{H}_3 + \tilde{H}_4) \, , \\
  \tilde{H}_3 &\mapsto \frac{1}{2}
  (\tilde{H}_1 - \tilde{H}_2 + \tilde{H}_3 + \tilde{H}_4) \, , \\
  \tilde{H}_4 &\mapsto \frac{1}{2}
  (\tilde{H}_1 - \tilde{H}_2 - \tilde{H}_3 - \tilde{H}_4) \, .
\ees
The action on the other elements of $so(8)$ can be determined
by the commutation relations and one obtains
\bes
  \pi : 
  \psi_1^+ \psi_3^- &\mapsto 
  \psi_4^- \psi_2^+ \mapsto 
  \psi_4^+ \psi_2^+ \mapsto \psi_1^+ \psi_3^- \, , \,\,
  \psi_2^+ \psi_3^+ \mapsto 
  \psi_1^+ \psi_4^+ \mapsto 
  \psi_1^+ \psi_4^- \mapsto \psi_2^+ \psi_3^+ \, , \\  
  \psi_1^+ \psi_3^+ &\mapsto \psi_1^+ \psi_3^+ \, , \,\,
  \psi_1^+ \psi_2^+ \mapsto \psi_1^+ \psi_2^+ \, .
\ees
Here we give only the action on the positive roots;
the action on the negative roots can be obtained
by taking the hermitian conjugation of this equation.
Since the order of $\pi$ is 3, its eigenvalue is of the form
$\kappa^n \, (n=0,1,2)$,
and $so(8)$ is decomposed into
the eigenspaces of $\pi$ as 
$\mathbf{28} = \mathbf{14} \oplus \mathbf{7} \oplus \mathbf{7}$.
The fixed point algebra of $\pi$ has dimension 14 and
coincides with $G_2 \subset so(8)$ with embedding index 1.

$\pi$ acts on $su(3)$ in \eqref{eq:su3so8} as follows,
\beq
  \pi : 
  E^1 \mapsto \kappa\, E^1 \, , \,\,
  E^2 \mapsto \bar{\kappa}\, E^2 \, , \,\,
  E^\theta \mapsto E^\theta \, , \,\,
  H_i \mapsto H_i \, .
\eeq
Hence $\pi$ almost fixes $su(3)$. 
One can construct an automorphism $\pi'$ fixing $su(3)$ 
in the form $\pi' = \mathrm{Ad}_h \pi$, where
$\mathrm{Ad}_h$ is an inner automorphism 
$J \mapsto \mathrm{Ad}_h (J) = h J h^{-1}$
with the element $h = e^{\frac{2 \pi i}{3}(H_1 + H_2)} \in SO(8)$.
From the action of $\mathrm{Ad}_h$ on $\psi_i^\pm$,
\beq
  \mathrm{Ad}_h : 
  \psi_1^\pm \mapsto \kappa^{\pm 1} \psi_1^\pm \, , \,\,
  \psi_2^\pm \mapsto \kappa^{\pm 1} \psi_2^\pm \, , \,\,
  \psi_3^\pm \mapsto \psi_3^\pm \, , \,\,
  \psi_4^\pm \mapsto \psi_4^\pm \, ,
\eeq
one can readily check that $\pi' = \mathrm{Ad}_h \pi$ 
actually fixes $su(3)$. 
Since $\mathrm{Ad}_h$ commutes with $\pi$ and $(\mathrm{Ad}_h)^3 = 1$,
the automorphism $\pi'$ has also order 3 and hence generates
an automorphism group 
$\langle \pi' \rangle = \{1,\pi',\pi'^2\} \cong \Z_3$. 
The elements of $so(8)$ are again decomposed according to
the eigenvalue $\kappa^n \,(n=0,1,2)$ of $\pi'$ as
$\mathbf{28} = \mathbf{8} \oplus \overline{\mathbf{10}} \oplus 
\mathbf{10}$, which coincides with 
the decomposition into irreducible representations of $su(3)$
mentioned above.
In this way,
the $su(3)$ subalgebra \eqref{eq:su3so8} is
characterized as the fixed point algebra of
the automorphism group $\langle \pi' \rangle \cong \Z_3$ of $so(8)$.

The fact that $su(3)$ in \eqref{eq:su3so8} is characterized
as the fixed point algebra of $\pi'$ implies
that the identity automorphism of $su(3)$ has a lift 
$\langle \pi' \rangle \cong \Z_3$ to $so(8)$. 
One can show that the charge-conjugation automorphism $\omega_c$
of $su(3)$ also has a lift to $so(8)$.
Consider the following automorphism $\sigma'$ of $so(8)$,
\beq
  \sigma' : 
  \psi_1^\pm \mapsto \psi_2^\mp \, , \,\,
  \psi_2^\pm \mapsto \psi_1^\mp \, , \,\,
  \psi_3^\pm \mapsto \psi_3^\pm \, , \,\,
  \psi_4^\pm \mapsto -\psi_4^\mp \, ,
\eeq
which has order 2. 
This is an outer automorphism of $so(8)$, since 
$\sigma'$ acts on eight real fermions as an element of
$O(8)$ with determinant $-1$.
This automorphism $\sigma'$ acts on $su(3)$ in \eqref{eq:su3so8}
as the exchange of two simple roots
\beq
  \sigma' : E^1 \mapsto E^2 \, , \,\,
  E^\theta \mapsto - E^\theta \, .
\eeq
In particular, $\sigma'$ keeps $su(3) \subset so(8)$ invariant
and can be restricted to $su(3)$.
Clearly, the restriction of $\sigma'$ to $su(3)$ is
the charge-conjugation $\omega_c$ of $su(3)$.
In other words, $\sigma'$ is a lift of $\omega_c$ to $so(8)$. 
Together with $\pi'$, $\sigma'$ forms a lift $\tilde{G}_c$ of 
$G_c = \{1, \omega_c \}$.
By explicit calculations,
one can show that $\sigma' \pi' \sigma'^{-1} = \pi'^{-1}$,
which means that $\tilde{G}_c = \langle \pi', \sigma' \rangle$
is isomorphic to the symmetric group $S_3$; 
the charge-conjugation automorphism group $G_c$ of $su(3) \subset so(8)$
has a lift $\tilde{G}_c$ to $so(8)$, which is isomorphic to $S_3$.
The restriction $G_c$ of $\tilde{G}_c$ 
can be identified with the quotient group 
of $\tilde{G}_c$ by the stabilizer of $su(3) \subset so(8)$ as
$\tilde{G}_c / \langle \pi' \rangle \cong S_3 / \Z_3 \cong \Z_2$. 

\section{Twisted boundary states for $su(3)_k$}
\label{sec:su3detail}

In this appendix, we describe in detail the construction of
twisted boundary states for the chiral algebra 
$su(3)_k \, (k=3,5)$ and the charge-conjugation automorphism
group $G_c = \{1,\omega_c\}$, and check that the resulting boundary
states realize a NIM-rep of the generalized fusion algebra
$\F(su(3)_k; G_c)$.
\medskip

Let us first calculate the explicit form of the generalized fusion algebra 
${\cal F}(su(3)_k;G_c)$ for $k=3$ and $5$. 
The set $\I$ of the irreducible representations of $su(3)_k$ reads
\beq
  \I = P_+^k(su(3))=\{\lambda=(\lambda_1,\lambda_2)|
   \lambda_1+\lambda_2 \le k\}.
\label{eq:su3kccspec}
\eeq
The modular transformation matrix of the characters of these 
representations is given 
by the Kac-Peterson formula~\cite{Kac}.
The charge conjugation $\omega_c$ acts on ${\cal I}$ as 
\beq
   \omega_c~:~(\lambda_1,\lambda_2) \mapsto (\lambda_2,\lambda_1).
\eeq
The corresponding twisted chiral algebra is 
the twisted affine Lie algebra $A_2^{(2)}$.
Since its horizontal subalgebra is $so(3)$, 
the irreducible representations of $A_2^{(2)}$ are labeled
by a single Dynkin label which we denote by $\tilde{\lambda}$, 
\beq
   \I^{\omega_c} = P_+^{k}(A_2^{(2)}) = \{\tilde{\lambda}~|~2 
   \tilde{\lambda} \le k\}.
\label{eq:su3kcctwiststate}
\eeq  
The set $\I(\omega_c)$ of representations fixed by $\omega_c$ reads
\beq
   \I({\omega_c}) = \{\mu=(\mu_1,\mu_1)~|~2\mu_1 \le k\}.
\label{eq:su3kcctwistspec}
\eeq
The modular transformation matrix $S^{\omega_c}$ of the twisted 
representation 
$\tilde{\lambda} \in \I^{\omega_c}$ takes the form~\cite{Kac}
\beq
   S^{\omega_c}_{\tilde{\lambda} \mu} = \frac{2}{\sqrt{k+3}}
                                \,\mbox{sin}\Big(\frac{2\pi}{k+3}
                                  (\tilde{\lambda}+1)(\mu_1+1)\Big).
\label{eq:su3kcctwistS}
\eeq

By using these data, one can construct untwisted and $\omega_c$-twisted 
boundary states compatible with the charge conjugation modular invariant
of $su(3)_k$. 
The total number $|\hat{{\cal I}}|$ of boundary states is
\bes
|\hat{{\cal I}}|= |{\cal I}| + |{\cal I}^{\omega_c}| = 
\left\{
\begin{array}{ll}
10 +2 =12  &~~ \mbox{for} \quad k=3,\\
21 +3 =24  &~~ \mbox{for} \quad k=5.
\end{array}
\right.
\label{eq:su3kregNIM}
\ees
As we have shown in Section \ref{sec:Gfusion},
the generalized fusion algebra ${\cal F}(su(3)_k;G_c)$, or equivalently
the regular NIM-reps of ${\cal F}(su(3)_k;G_c)$,
is obtained from the overlaps of these boundary states. 
We give only the part of $\F(su(3)_k; G_c)$ concerning
the twisted representations.
The part containing only the untwisted representations,
\textit{i.e.}, the ordinary fusion algebra $\F(su(3)_k)$,
is found in, \textit{e.g.}, \cite{CFTbook}.
\medskip \\
\underline{$k=3$}   
\bes
   (\lambda_1,\lambda_2) \times (0) &= (\lambda_1,\lambda_2) \times (1) 
                        = \left\{
                         \begin{array}{ll}
                         (0)          & \, (\lambda_1,\lambda_2) \in \{(0,0),(3,0),(0,3)\} , \\
                         (0) + (1)    & \, (\lambda_1,\lambda_2) \in \{(1,0),(0,1),(2,0),(0,2),(2,1),(1,2)\} ,
                         \end{array} 
                         \right.  \\
   (1,1) \times (0) &=  (0) + 2\,(1) ,  \\
   (1,1) \times (1) &= 2\,(0) + (1),  \\                        
   (0) \times (0) &= (1) \times (1) 
                   = \sum_{(\lambda_1, \lambda_2) \in \I} 
                     (\lambda_1,\lambda_2) ,\\
   (0) \times (1) &= (1,0)+(0,1)+(2,0)+(0,2)+(2,1)+(1,2)+2\,(1,1).   
\label{eq:su33twistfusion}
\ees
\underline{$k=5$}   
\bes
 & (\lambda_1,\lambda_2) \times (\tilde{\lambda})  = (\tilde{\lambda}) 
    \hspace{4.0cm} (\lambda_1,\lambda_2) \in \{(0,0),(5,0),(0,5)\}, \quad
                   \tilde{\lambda} \in {\cal I}^{\omega_c}=\{0,1,2\} ,\\ 
 & (\lambda_1,\lambda_2) \times \left\{
                   \begin{array}{l}
                   (0) = (0) + (1)  \\
                   (1) = (0) + (1) + (2)  \\
                   (2) = (1) + (2) 
                   \end{array}
                   \right.      \hspace{1.2cm} (\lambda_1,\lambda_2) \in \{(1,0),(0,1),(4,0),(0,4),(4,1),(1,4)\},  \\
 & (\lambda_1,\lambda_2) \times \left\{
                   \begin{array}{l}
                   (0) = (0) + (1) + (2) \\
                   (1) = (0) + 2\,(1) + (2)  \\
                   (2) = (0) + (1) + (2) 
                   \end{array}
                   \right.      \quad \quad~ (\lambda_1,\lambda_2) \in \{(2,0),(0,2),(3,0),(0,3),(3,2),(2,3)\}, \\
 & (\lambda_1,\lambda_2) \times \left\{
                   \begin{array}{l}
                   (0) = (0) + 2\,(1) + (2)  \\
                   (1) = 2\,(0) + 2\,(1) + 2\,(2)  \\
                   (2) = (0) + 2\,(1) + (2) 
                   \end{array}
                   \right.      \quad  (\lambda_1,\lambda_2)
                    \in \{(1,1),(3,1),(1,3)\},  \\
 & (\lambda_1,\lambda_2) \times \left\{
                   \begin{array}{l}
                   (0) = (0) + 2\,(1) + 2\,(2)  \\
                   (1) = 2\,(0) + 3\,(1) + 2\,(2)  \\
                   (2) = 2\,(0) + 2\,(1) + (2) 
                   \end{array}
                   \right.      \quad  (\lambda_1,\lambda_2)
                    \in \{(2,1),(1,2),(2,2)\},  \\
 & (0)\times (0) = (2) \times (2) = 
   \sum_{(\lambda_1,\lambda_2) \in \I} (\lambda_1,\lambda_2) , \\
 & (1)\times (1) = (0,0)+(5,0)+(0,5)+(1,0)+(0,1)+(4,0)+(0,4)+(4,1)+(1,4)  \\ 
 &   \hspace{1.6cm} +2\Bigl((2,0)+(0,2)+(3,0)+(0,3)+(3,2)+(2,3)+(1,1)+(3,1)+(1,3)\Bigr) \\
 &   \hspace{1.6cm} +3\Bigl((2,1)+(1,2)+(2,2)\Bigr),                \\
 & (0) \times (1) = (1) \times (2) 
                  = (1,0)+(0,1)+(4,0)+(0,4)+(4,1)+(1,4) \\
 &   \hspace{3.7cm} +(2,0)+(0,2)+(3,0)+(0,3)+(3,2)+(2,3)  \\
 &   \hspace{3.7cm} +2\Bigl((1,1)+(3,1)+(1,3)+(2,1)+(1,2)+(2,2)\Bigr),  \\
 & (0)\times(2) = (2,0)+(0,2)+(3,0)+(0,3)+(3,2)+(2,3)+(1,1)+(3,1)+(1,3) \\
 &  \hspace{1.6cm}  +2\Bigl((2,1)+(1,2)+(2,2)\Bigr).    
\label{eq:su35twistfusion}
\ees

\subsection{$su(3)_3$}
\label{sec:Appsu33}

The modular invariant 
\beq
    Z_{D^{(6)}} = |\chi_{(0,0)}+\chi_{(3,0)}+\chi_{(0,3)}|^2
                               +3 |\chi_{(1,1)}|^2
\label{eq:su33scinv}
\eeq
is a simple current extension of $su(3)_3$ by $(3,0)$ and 
the corresponding extended chiral algebra is $so(8)_1$.
As we have shown in Section~\ref{sec:graph}, 
the twisted boundary states for the automorphism group $G_c$
are obtained from those for $\tilde{G}_c$, a lift of $G_c$ to $so(8)_1$.
The resulting boundary state coefficients are given in
\eqref{eq:su33psi} and \eqref{eq:su33sctwistpsi}.
The set $\Vhat$ of all the twisted boundary states
consists of 12 elements
\beq
  \abs{\Vhat} = \abs{\V} + \abs{\V^{\omega_c}} = 6+6 =12.
\eeq
By calculating the overlaps of these $12$  
boundary states, we obtain a set of $12 \times 12$ matrices 
$\{\hat{n}_{N}\,|N \in \Ihat\}$ defined in \eqref{eq:BA}. 
These matrices turn out to be non-negative integer valued. The explicit form 
is as follows,
\bes
  & \hat{n}_{(0,0)} = \hat{n}_{(3,0)} = \hat{n}_{(0,3)} 
  = \begin{pmatrix}
     I & O  \\
     O & I
    \end{pmatrix} , \\
  & \hat{n}_{(1,0)} = \hat{n}_{(0,2)} = \hat{n}_{(2,1)} 
  = \hat{n}_{(0,1)}^T = \hat{n}_{(2,0)}^T = \hat{n}_{(1,2)}^T 
  = \begin{pmatrix}
     \varrho & O  \\
     O & \varrho'
    \end{pmatrix} ,  \\
  & \hat{n}_{(1,1)}
  = \begin{pmatrix}
     \varrho_{(1,1)} & O  \\
     O & \varrho'_{(1,1)}
    \end{pmatrix} ,    \\
  & \hat{n}_{(0)}
  = \begin{pmatrix}
     O & \varrho_0^T  \\
     \varrho_0 & O
    \end{pmatrix} ,  \\
  & \hat{n}_{(1)}
  = \begin{pmatrix}
     O & \varrho_1^T  \\
     \varrho_1 & O
    \end{pmatrix} ,  \\  
\ees
where the first six rows and columns stand for the states of
$\Psi_{D^{(6)}}$ in \eqref{eq:su33psi}
and the last six for those of $\Psi_{D^{(6)*}}$ in
\eqref{eq:su33sctwistpsi}.
$O$ and $I$ are the 6 $\times$ 6 zero and the unit matrices, respectively,
and the other matrices are defined as
\bes
  \varrho
 &= \begin{pmatrix}
    0 & 0 & 0 & 0 & 1 & 0  \\
    0 & 0 & 0 & 0 & 1 & 0  \\
    0 & 0 & 0 & 0 & 1 & 0  \\
    0 & 0 & 0 & 0 & 1 & 0  \\   
    0 & 0 & 0 & 0 & 0 & 2  \\
    1 & 1 & 1 & 1 & 0 & 0 
    \end{pmatrix} , ~~~~~
  \varrho'
 = \begin{pmatrix}
    0 & 0 & 0 & 0 & 1 & 1 \\
    0 & 0 & 0 & 0 & 1 & 1 \\
    1 & 1 & 0 & 0 & 0 & 0 \\
    1 & 1 & 0 & 0 & 0 & 0 \\
    0 & 0 & 1 & 1 & 0 & 0 \\
    0 & 0 & 1 & 1 & 0 & 0 
    \end{pmatrix} ,  \\
 \varrho_{(1,1)}
 &= \begin{pmatrix}
    0 & 1 & 1 & 1 & 0 & 0  \\
    1 & 0 & 1 & 1 & 0 & 0  \\
    1 & 1 & 0 & 1 & 0 & 0  \\
    1 & 1 & 1 & 0 & 0 & 0  \\   
    0 & 0 & 0 & 0 & 3 & 0  \\
    0 & 0 & 0 & 0 & 0 & 3 
    \end{pmatrix} , ~
  \varrho'_{(1,1)}
 = \begin{pmatrix}
    1 & 2 & 0 & 0 & 0 & 0  \\
    2 & 1 & 0 & 0 & 0 & 0  \\
    0 & 0 & 1 & 2 & 0 & 0  \\
    0 & 0 & 2 & 1 & 0 & 0  \\   
    0 & 0 & 0 & 0 & 1 & 2  \\
    0 & 0 & 0 & 0 & 2 & 1 
    \end{pmatrix} ,  \\   
 \varrho_0
 &= \begin{pmatrix}
    1 & 1 & 0 & 0 & 1 & 1  \\
    0 & 0 & 1 & 1 & 1 & 1  \\
    1 & 0 & 1 & 0 & 1 & 1  \\
    0 & 1 & 0 & 1 & 1 & 1  \\   
    1 & 0 & 0 & 1 & 1 & 1  \\
    0 & 1 & 1 & 0 & 1 & 1 
    \end{pmatrix} , ~~~~~ 
 \varrho_1
 = \begin{pmatrix}
    0 & 0 & 1 & 1 & 1 & 1  \\
    1 & 1 & 0 & 0 & 1 & 1  \\
    0 & 1 & 0 & 1 & 1 & 1  \\
    1 & 0 & 1 & 0 & 1 & 1  \\   
    0 & 1 & 1 & 0 & 1 & 1  \\
    1 & 0 & 0 & 1 & 1 & 1 
    \end{pmatrix} .   
\ees
One can check that these matrices $\hat{n}$ form a NIM-rep
of $\F(su(3)_3; G_c)$
defined in \eqref{eq:su33twistfusion}.
For example, $\hat{n}$ satisfies the following relation
\bes
 & \hat{n}_{(1,1)}\,\hat{n}_{(0)} = \hat{n}_{(0)} + 2\,\hat{n}_{(1)},\quad
  \hat{n}_{(1,1)}\,\hat{n}_{(1)} = 2\,\hat{n}_{(0)} +  \hat{n}_{(1)}, \\
 & \hat{n}_{(0)}\,\hat{n}_{(1)} = \hat{n}_{(1,0)}+\hat{n}_{(0,1)}+\hat{n}_{(2,0)}                                 +\hat{n}_{(0,2)}+\hat{n}_{(2,1)}+\hat{n}_{(1,2)}
                                 +2\,\hat{n}_{(1,1)} ,
   \quad \mbox{etc.} 
\ees
In this way,
two NIM-reps, $D^{(6)}$ and $D^{(6)*}$, associated with
the simple current modular invariant \eqref{eq:su33scinv}
are combined to yield a NIM-rep 
of the generalized fusion algebra $\F(su(3)_3; G_c)$.

\subsection{$su(3)_5$}

We next turn to the construction of twisted boundary states
associated with a non-trivial modular invariant of $su(3)_5$
and check that the resulting boundary states realize
a NIM-rep of the 
generalized fusion algebra $\F(su(3)_5;G_c)$.

\subsubsection{The case of $Z_{E^{(8)}}$}

We first consider the exceptional modular invariant of $su(3)_5$,
\bes
   Z_{E^{(8)}} & = |\chi_{(0,0)}+\chi_{(2,2)}|^2 
                 + |\chi_{(0,2)}+\chi_{(3,2)}|^2 
                 + |\chi_{(1,2)}+\chi_{(5,0)}|^2 \\
       & +|\chi_{(3,0)}+\chi_{(0,3)}|^2 
         +|\chi_{(2,1)}+\chi_{(0,5)}|^2 + |\chi_{(2,0)}+\chi_{(2,3)}|^2 \, .
\label{eq:su35E8}
\ees
This invariant is originated from 
the charge conjugation modular invariant 
of $su(6)_1$ through the conformal embedding $su(3)_5 \subset su(6)_1$,
in which the representations of $su(6)_1$ branch to those of
$su(3)_5$ as follows,
\bes
   \Lambda_0 & \mapsto (0,0) \oplus (2,2), \\  
   \Lambda_1 & \mapsto (0,2) \oplus (3,2), \\ 
   \Lambda_2 & \mapsto (1,2) \oplus (5,0), \\ 
   \Lambda_3 & \mapsto (3,0) \oplus (0,3), \\ 
   \Lambda_4 & \mapsto (2,1) \oplus (0,5), \\ 
   \Lambda_5 & \mapsto (2,0) \oplus (2,3). 
\label{eq:su35su61branch}
\ees
In the modular invariant (\ref{eq:su35E8}), 
there are twelve untwisted Ishibashi states available,~\footnote{%
For simplicity, we omit in this appendix
the label for the anti-holomorphic representation of Ishibashi states.
}
\beq
   \E  = \{(0,0),(5,0),
                 (0,5),(2,2),
                 (1,2),(2,1),   
              (3,0),(2,3),
              (0,2),(0,3),
              (2,0),(3,2)\}.
\label{eq:su35espec}
\eeq
Correspondingly, we have twelve untwisted boundary states.
Six of them are identified with the untwisted states of $su(6)_1$
and the other six states are generated by using the fusion 
with the representations of $su(3)_5$.
The resulting boundary state coefficient matrix $\Psi_{E^{(8)}}$
takes the following form
\beq
      \Psi_{E^{(8)}} = \frac{1}{2}
        \begin{pmatrix}
        a_1 K  &  a_2 \,K   &    \,K   &    \,K  \\
        a_1 K  &  a_2 \,K   &  - \,K   &  - \,K  \\
        a_2 K & -a_1\, K & i \,K & -i \,K \\
        a_2 K & -a_1\, K & -i \,K&   i \,K  
        \end{pmatrix} \, , 
\label{eq:su35untwisted}
\eeq 
where 
$a_1 = \sqrt{2} \sin \frac{\pi}{8} = \sqrt{\frac{2-\sqrt{2}}{2}}$,
$a_2 = \sqrt{2} \cos \frac{\pi}{8} = \sqrt{\frac{2+\sqrt{2}}{2}}$
and the matrix $K$ is defined as
\beq
  K = \frac{1}{\sqrt{3}} 
           \begin{pmatrix}
           1    &   1    &    1    \\
           1    &  e^{2\pi i/3}  &   e^{-2\pi i/3}   \\
           1    &  e^{-2\pi i/3}  &   e^{2\pi i/3} 
           \end{pmatrix} \, .
\label{eq:a1a2}           
\eeq
Here the columns of $\Psi_{E^{(8)}}$ 
are ordered as in (\ref{eq:su35espec}).
The first six rows correspond to the 
states of the $su(6)_1$ theory.
One can check that these untwisted states yield the $E^{(8)}$ NIM-rep
of the ordinary fusion algebra $\F(su(3)_5)$. 

The labels of the available $\omega_c$-twisted Ishibashi states 
are obtained from \eqref{eq:su35E8}, 
\beq
 \E(\omega_c) = \{(0,0),(2,2),
                        (0,3),(3,0)\}.
\label{eq:su35etwistspec}
\eeq
As is seen from the branching rule \eqref{eq:su35su61branch}, 
the charge conjugation $\omega_c$ can be
lifted to the charge conjugation $\tilde{\omega_c}$ of $su(6)_1$,
which acts on the fundamental weights of $su(6)_1$ as follows,
\beq
    \tilde{\omega_c}~:~ 
    \Lambda_0 \mapsto \Lambda_0 \, , \quad
    \Lambda_i \leftrightarrow \Lambda_{6-i} 
    \quad(i = 1,2,3,4,5).
\eeq
The $\tilde{\omega}_c$-twisted boundary states in the $su(6)_1$ theory 
take the form
\beq
    \ket{\pm} = \frac{1}{\sqrt{2}}
    (\dket{\Lambda_0;\tilde{\omega}_c} 
    \pm \dket{\Lambda_3;\tilde{\omega}_c}).
\label{eq:su61twisted}
\eeq
One can express these states in terms of the $\omega$-twisted 
Ishibashi states of $su(3)_5$ in the same way as 
the $D^{(6)*}$ states considered in Section~\ref{sec:graph}.
First,
from the branching rule \eqref{eq:su35su61branch} and 
the modular transformation matrix of $su(3)_5$, 
the Ishibashi state for the vacuum representation of $su(6)_1$
can be decomposed as follows, 
\beq
  \dket{\Lambda_0} = \frac{1}{\sqrt{2}}
  (a_1 \dket{(0,0)} + a_2 \dket{(2,2)}) \, , 
\label{eq:su35euntwistedphase}
\eeq
where $a_1, a_2$ are the constants used in \eqref{eq:su35untwisted}.
The $\tilde{\omega}_c$-twisted Ishibashi state
$\dket{\Lambda_0; \tilde{\omega}_c}$ is then obtained by the action of
$\tilde{\omega}_c$,
\beq
  \dket{\Lambda_0; \tilde{\omega}_c} 
  = R(\tilde{\omega}_c) \dket{\Lambda_0} 
  = \frac{1}{\sqrt{2}}
  (a_1 \dket{(0,0); \omega_c} - a_2 \dket{(2,2); \omega_c}) \, .
\label{eq:su35etwistedphase}
\eeq
Here we use the fact that the $\tilde{\omega}_c$ acts on
$(2,2)$ as $-1$, which can be shown, for example, by
the calculation of $\bra{\Lambda_0}\tilde{q}^{\frac{1}{2}H_c}\ket{+}$.
In this way, we obtain two $\omega_c$-twisted states.
The remaining two states are generated by the fusion of $su(3)_5$ and
we obtain four $\omega_c$-twisted states, 
whose boundary state coefficient matrix takes the form
\bes
   &  \Psi_{E^{(8)*}} = \frac{1}{2}
                      \begin{pmatrix}
                      a_1   &    -a_2    &    1   &    1  \\
                      a_1   &    -a_2    &   -1   &   -1  \\
                      a_2 &  a_1  &   i  &  -i  \\
                      a_2 &  a_1  &  -i  &   i  
                      \end{pmatrix} \, .
\label{eq:su35etwistmatrix}
\ees
Here the columns are ordered 
as in (\ref{eq:su35etwistspec}) and the first (second) row corresponds
to $\ket{+}$ ($\ket{-}$).
These states form the $E^{(8)*}$ NIM-rep of $\F(su(3)_5)$.

The number of all the twisted boundary states is thus 
\beq
|\hat{{\cal V}}| = |{\cal V}| + |{\cal V}^{\omega_c}| = 12+4=16 \, .
\eeq
From the explicit form of the boundary state coefficients,
one can calculate the corresponding overlap matrices $\hat{n}$,
which are $16$-dimensional.
The result is as follows,
\bes
& \hat{n}_{(0,0)}=I_{16\times 16}, \\
& \hat{n}_{(1,0)}   
 =  \left(
    \begin{array}{cccc|cc}
    O      &     O     &  \varsigma  &      O     &           &              \\
    O      &     O     &     O       &  \varsigma &           &    \OO       \\
    O      & \varsigma & \varsigma   &  \varsigma &           &              \\
 \varsigma &     O     & \varsigma   &  \varsigma &           &              \\ \hline      
          &            &             &            &      O'     &    I'         \\
          &    \OO     &             &            & \varsigma' & I'+\varsigma' \\
    \end{array}
    \right) ,  \\ 
&  \hat{n}_{(2,0)}  
 =  \left(
    \begin{array}{cccc|cc}
    O      &\varsigma^t &\varsigma^t  &      O     &            &               \\
\varsigma^t&     O      &     O       &\varsigma^t &            &    \OO          \\
    O      & \varsigma^t& \varsigma^t & 2\,\varsigma^t &            &             \\
\varsigma^t&     O      &2\,\varsigma^t   & \varsigma^t&            &             \\ \hline      
           &            &             &            & \varsigma' &    I'         \\
           &    \OO     &             &            & \varsigma' & I'+2\,\varsigma' \\
    \end{array}
    \right) ,  
 \hat{n}_{(1,1)} 
 =  \left(
    \begin{array}{cccc|cc}
    O      &     O      &     I       &      I     &            &               \\
    O      &     O      &     I       &      I     &            &   \OO         \\
    I      &     I      &     2I      &      2I    &            &               \\
    I      &     I      &     2I      &      2I    &            &               \\ \hline      
           &            &             &            &      O'    & I'+ \varsigma'\\
           &   \OO      &             &            &I'+\varsigma' & 2(I'+\varsigma') \\
    \end{array}
    \right) ,  \\
& \hat{n}_{(3,0)} 
 =  \left(
    \begin{array}{cccc|cc}
    O      &     I      &     O       &      I     &            &               \\
    I      &     O      &     I       &      O     &            &  \OO          \\
    I      &     O      &     I       &      2I     &            &               \\
    O      &     I      &     2I       &      I     &            &               \\ \hline      
           &            &             &            & \varsigma' &    \varsigma'  \\
           &   \OO      &             &            &      I'    & I'+2\,\varsigma' \\
    \end{array}
    \right) ,  
 \hat{n}_{(2,1)}  
 =  \left(
    \begin{array}{cccc|cc}
 \varsigma &      O      & \varsigma  & \varsigma     &            &               \\
    O      &\varsigma    & \varsigma  & \varsigma     &            &   \OO          \\
 \varsigma &\varsigma    &3\,\varsigma&2\,\varsigma   &            &                \\
 \varsigma &\varsigma    &2\,\varsigma&3\,\varsigma   &            &               \\ \hline      
           &            &             &            &      I'      &I'+\varsigma'       \\
           &    \OO     &             &            &I'+\varsigma' & 3I'+2\,\varsigma' \\
    \end{array}
    \right) ,    \\ 
& \hat{n}_{(4,0)} 
 =  \left(
    \begin{array}{cccc|cc}
    O      &      O      &     O      & \varsigma     &            &               \\
    O      &      O      & \varsigma  &      O        &            &   \OO          \\
 \varsigma &      O      & \varsigma  & \varsigma     &            &               \\
    O      &\varsigma    & \varsigma  & \varsigma     &            &               \\ \hline      
           &            &             &            &      O'      & \varsigma'       \\
           &    \OO     &             &            &      I'     & I'+\varsigma' \\
    \end{array}
    \right) , 
  \hat{n}_{(3,1)} 
 =  \left(
    \begin{array}{cccc|cc}
    O      &      O      & \varsigma^t     & \varsigma^t   &            &               \\
    O      &      O      & \varsigma^t     & \varsigma^t   &            &   \OO          \\
\varsigma^t& \varsigma^t & 2\,\varsigma^t  &2\,\varsigma^t &            &               \\
\varsigma^t& \varsigma^t & 2\,\varsigma^t  &2\,\varsigma^t &            &               \\ \hline      
           &            &             &            &      O'       & I'+\varsigma'       \\
           &   \OO      &             &            &I'+\varsigma' &2(I'+\varsigma') \\
    \end{array}
    \right) ,    \\
& \hat{n}_{(2,2)} 
 =  \left(
    \begin{array}{cccc|cc}
    I      &     O      &     I       &      I     &            &               \\
    O      &     I      &     I       &      I     &            &   \OO          \\
    I      &     I      &     3I      &      2I    &            &                \\
    I      &     I      &     2I      &      3I    &            &                \\ \hline      
           &            &             &            &      I'     & I'+\varsigma' \\
           &   \OO      &             &            &I'+\varsigma'& 3I'+2\,\varsigma' \\
    \end{array}
    \right) ,  
 \hat{n}_{(5,0)} 
 =  \left(
    \begin{array}{cccc|cc}
\varsigma^t&      O      &     O      &      O        &            &               \\
    O      & \varsigma^t &     O      &      O        &            &   \OO          \\
    O      &      O      &\varsigma^t &      O        &            &               \\
    O      &      O      &     O      & \varsigma^t   &            &               \\ \hline      
           &            &             &            &      I'      &     O'       \\
           &    \OO     &             &            &      O'       &     I'      \\
    \end{array}
    \right) ,   \\
& \hat{n}_{(4,1)} 
 =  \left(
    \begin{array}{cccc|cc}
    O      &     O      &     I       &      O     &            &               \\
    O      &     O      &     O       &      I     &            &   \OO          \\
    O      &     I      &     I       &      I     &            &               \\
    I      &     O      &     I       &      I     &            &               \\ \hline      
           &            &             &            &     O'     &    I'         \\
           &   \OO      &             &            & \varsigma' & I'+\varsigma' \\
    \end{array}
    \right) ,  
 \hat{n}_{(3,2)} 
 =  \left(
    \begin{array}{cccc|cc}
    O      &\varsigma   &\varsigma    &      O     &            &               \\
\varsigma  &     O      &     O       &\varsigma   &            &    \OO          \\
    O      & \varsigma  & \varsigma   & 2\,\varsigma   &            &              \\
\varsigma  &     O      &2\,\varsigma & \varsigma      &            &              \\ \hline      
           &            &             &            & \varsigma' &    I'         \\
           &   \OO      &             &            & \varsigma' & I'+2\,\varsigma' \\
    \end{array}
    \right) ,  \\
& \hat{n}_{(0)} = \hat{n}_{(2)}   
 =  \left(
    \begin{array}{cc}
    \OO & A^t  \\
    A & \OO
    \end{array}
    \right) , ~
   \hat{n}_{(1)} 
 =  \left(
    \begin{array}{cc}
    \OO & B^t  \\
    B & \OO
    \end{array}
    \right) .
\ees
The matrices not presented above are obtained by
$\hat{n}_{(\lambda_2,\lambda_1)}
=\hat{n}_{(\lambda_1,\lambda_2)}^T$.
The first 12 rows and columns of these matrices $\hat{n}$ are 
ordered as in the rows of $\Psi_{E^{(8)}}$ in \eqref{eq:su35untwisted} and 
the last four are ordered as in $\Psi_{E^{(8)*}}$ defined in 
\eqref{eq:su35etwistmatrix}.
$O$ and $O'$ ($I$ and $I'$) are respectively $3\times 3$ and $2\times 2$ 
zero (unit) matrices while the other matrices are defined as 
\bes
  \varsigma 
 =  \begin{pmatrix}
    0 & 1 & 0 \\
    0 & 0 & 1 \\
    1 & 0 & 0      
    \end{pmatrix} , \,\,
  \varsigma' 
  = \begin{pmatrix}
    0 & 1   \\
    1 & 0   
    \end{pmatrix}, \,\,
  A=\begin{pmatrix}
    0_6 & 1_6   \\
    0_6 & 1_6   \\
    1_6 & 2_6   \\
    1_6 & 2_6
    \end{pmatrix}, \,\,
 B=\begin{pmatrix}
    1_6 & 1_6 \\
    1_6 & 1_6 \\
    1_6 & 3_6 \\
    1_6 & 3_6 
    \end{pmatrix},
\ees
where $m_6 \, (m=0,1)$ is 
a row vector with all entries being $m$, $m_6 = (m,m,m,m,m,m)$.
One can explicitly check that these non-negative integer valued matrices $\{\hat{n}_N\}$ satisfy the generalized fusion algebra
$\F(su(3)_5; G_c)$ in \eqref{eq:su35twistfusion}, for example, 
\bes
 & \hat{n}_{(2,2)}\,\hat{n}_{(0)}  
 = \hat{n}_{(0)} + 2\, \hat{n}_{(1)}+2\, \hat{n}_{(2)}, \\
 & \hat{n}_{(2,2)}\,\hat{n}_{(1)} 
 = 2\,\hat{n}_{(0)} + 3\, \hat{n}_{(1)} +2\,\hat{n}_{(2)}, \\
 & \hat{n}_{(2,2)}\,\hat{n}_{(2)}  
 = 2\,\hat{n}_{(0)} + 2\,\hat{n}_{(1)}+ \hat{n}_{(2)}, \\
 & \hat{n}_{(0)}\,\hat{n}_{(2)}
 = \hat{n}_{(2,0)}+\hat{n}_{(0,2)}+\hat{n}_{(3,0)}+\hat{n}_{(0,3)}
  +\hat{n}_{(3,2)}+\hat{n}_{(2,3)}+\hat{n}_{(1,1)}+\hat{n}_{(3,1)}+\hat{n}_{(1,3)} \\
 &\hspace{1.5cm} +2\bigl(\hat{n}_{(2,1)}+\hat{n}_{(1,2)}+\hat{n}_{(2,2)}\bigr)  
                                      \quad \quad\mbox{etc.} 
\ees

\subsubsection{The case of $Z_{D^{(8)}}$}

We next consider twisted boundary states
associated with the simple current invariant 
\bes
  Z_{D^{(8)}}=& |\chi_{(0,0)}|^2 + |\chi_{(1,1)}|^2+|\chi_{(2,2)}|^2 + |\chi_{(3,0)}|^2
               +|\chi_{(0,3)}|^2 + |\chi_{(4,1)}|^2+|\chi_{(1,4)}|^2   \\
              &+(\chi_{(5,0)}\overline{\chi_{(0,5)}} + \chi_{(3,1)}\overline{\chi_{(1,3)}} 
               +\chi_{(1,2)}\overline{\chi_{(2,1)}} \\
              &+\chi_{(2,3)}\overline{\chi_{(0,2)}} 
               +\chi_{(2,0)}\overline{\chi_{(3,2)}} + \chi_{(0,4)}\overline{\chi_{(1,0)}} 
              +\chi_{(0,1)}\overline{\chi_{(4,0)}} + \mathrm{c.c}) \, .
\label{eq:d8inv}
\ees
Since this invariant is the $\Z_3$-orbifold
of $su(3)_5$, one can obtain the boundary states for \eqref{eq:d8inv}
by averaging the $\Z_3$-orbit of boundary states for the charge-conjugation
invariant.
For the untwisted states, this procedure yields the following 
coefficient matrix,
\beq
  \Psi_{D^{(8)}} = \frac{1}{2\sqrt{2}}
         \begin{pmatrix}
         1-\frac{1}{\sqrt{2}} & \sqrt{2}  & 1+\frac{1}{\sqrt{2}}  
       & 1                    & 1         &  \frac{1}{\sqrt{2}}    &   \frac{1}{\sqrt{2}}    \\
         \frac{1}{\sqrt{2}}  & \sqrt{2}  & -\frac{1}{\sqrt{2}} 
       & i                    & -i        & i-\frac{1}{\sqrt{2}}   & -i-\frac{1}{\sqrt{2}}   \\
         1                   & 0         & 1                     
       & -1+i                 & -1-i      & -i                     & i                       \\
          \sqrt{2}            & 0         &   -\sqrt{2}        
       & 0                    & 0         &  \sqrt{2}              &    \sqrt{2}             \\
         1                    & 0         & 1                     
       & -1-i                 & -1+i      & i                      & -i                      \\
          \frac{1}{\sqrt{2}}  & \sqrt{2}  & -\frac{1}{\sqrt{2}} 
       &-i                    & i         &-i-\frac{1}{\sqrt{2}}   & i-\frac{1}{\sqrt{2}}    \\      
         1+\frac{1}{\sqrt{2}} & -\sqrt{2} & 1-\frac{1}{\sqrt{2}}  
       & 1                    & 1         & -\frac{1}{\sqrt{2}}    &  -\frac{1}{\sqrt{2}}    
       \end{pmatrix} \, , 
\label{eq:su35D8Psi}
\eeq
where the columns are ordered as 
\beq
  \E = \{(0,0),(1,1),(2,2),(3,0),(0,3),(4,1),(1,4) \}.
\label{eq:su35D8spec}
\eeq
Since all of three $\omega_c$-twisted states for the charge-conjugation
invariant are the fixed point of the $\Z_3$ action,
we obtain $3 \times 3 = 9$ twisted states for \eqref{eq:d8inv}
by an appropriate resolution of the fixed points.
The resulting coefficient matrix takes the following form
\beq
  \Psi_{D^{(8)*}} = \frac{1}{\sqrt{3}}
                    \begin{pmatrix}
                    \tau       & \tau       & \tau      \\
                    \tau       & e^{2 \pi i/3}\, \tau   & e^{-2\pi i/3}\, \tau   \\
                    \tau       & e^{-2 \pi i/3}\, \tau  & e^{2\pi i/3} \,\tau   \\
                    \end{pmatrix}
                    \quad  \mbox{with} \quad
   \tau = \frac{1}{2}
          \begin{pmatrix}
          1        &   \sqrt{2} &  1         \\
          \sqrt{2} &      0     &  -\sqrt{2} \\
          1        &  -\sqrt{2} &  1                
          \end{pmatrix} \, .
\label{eq:su35D8Psicc}
\eeq 
where the columns are ordered as 
\beq
  \E(\omega_c) = \{(0,0),(1,1),(2,2),(5,0),(3,1),(1,2),
 (0,5),(1,3),(2,1)\}.
\label{eq:su35D8ccspec}  
\eeq
In this way, we obtain 
\beq
 |\hat{{\cal V}}| = |{\cal V}| + |{\cal V}^{\omega_c}| = 7+9=16
\eeq
boundary states associated with the modular invariant \eqref{eq:d8inv}.
By calculating the overlap of these states,
we obtain a set of $16 \times 16$ matrices
\bes
 &  \hat{n}_{(0,0)} = I_{16\times 16},  \\
 &  \hat{n}_{(1,0)} = \left(
                     \begin{array}{c|c}
                     A_{D^{(8)}} & \OO      \\ \hline
                     \OO           & A_{D^{(8)\ast}}
                     \end{array} \right) \, , 
\ees
where $A_{D^{(8)}}$ and $A_{D^{(8)\ast}}$ are the adjacency matrix for 
the graphs of type $D^{(8)}$ and $D^{(8)\ast}$, respectively,
\beq
   A_{D^{(8)}} = \begin{pmatrix}
                 0 & 1 & 0 & 0 & 0 & 0 & 0  \\
                 0 & 0 & 1 & 0 & 0 & 1 & 0  \\
                 0 & 0 & 0 & 1 & 1 & 0 & 0  \\
                 0 & 1 & 0 & 0 & 1 & 0 & 1  \\
                 0 & 0 & 0 & 0 & 0 & 1 & 1  \\
                 1 & 0 & 0 & 1 & 0 & 0 & 0  \\
                 0 & 0 & 1 & 1 & 0 & 0 & 1
                 \end{pmatrix}, \,\,
   A_{D^(8)\ast}= \begin{pmatrix}
                 O & O & \alpha   \\
                 \alpha & O & O   \\
                 O & \alpha & O
                 \end{pmatrix}, \,\,
   \alpha =  \begin{pmatrix}
                 1 & 1 & 0   \\
                 1 & 1 & 1   \\
                 0 & 1 & 1
                 \end{pmatrix} .             
\eeq
The remaining matrices for the untwisted representations
take the following form,
\bes
&  \hat{n}_{(2,0)} = \left(
                     \begin{array}{c|ccc}
                     A_{(2,0)} &   &\OO &  \\ \hline
                       & O & \alpha_{(2,0)}  & O  \\
                     \OO & O & O & \alpha_{(2,0)}  \\
                       & \alpha_{(2,0)} & O & O  
                     \end{array}
                     \right),~ 
  \hat{n}_{(1,1)} = \left(
                     \begin{array}{c|ccc}
                     A_{(1,1)} &  & \OO &  \\ \hline
                       & \alpha_{(1,1)}  & O & O \\
                     \OO & O & \alpha_{(1,1)} & O \\
                       & O & O & \alpha_{(1,1)}   
                     \end{array}
                     \right) ,   \\                 
&  \hat{n}_{(3,0)} = \left(
                     \begin{array}{c|ccc}
                     A_{(2,0)}^t &  &\OO &  \\ \hline
                      & \alpha_{(2,0)}  & O & O \\
                     \OO & O & \alpha_{(2,0)} & O \\
                      & O & O & \alpha_{(2,0)}   
                     \end{array}
                     \right),~                 
  \hat{n}_{(2,1)} = \left(
                     \begin{array}{c|ccc}
                     A_{(2,2)} &   &\OO &  \\ \hline
                      & O & O & \alpha_{(2,2)}   \\
                     \OO & \alpha_{(2,2)} & O & O   \\
                      & O & \alpha_{(2,2)} & O   
                     \end{array}
                     \right) ,  \\
&  \hat{n}_{(4,0)} = \left(
                     \begin{array}{c|ccc}
                     A_{D^{(8)}}^t &  &\OO &  \\ \hline
                      & O & O & \alpha  \\
                     \OO & \alpha & O & O  \\
                      & O & \alpha & O  
                     \end{array}
                     \right),~ 
  \hat{n}_{(3,1)} = \left(
                     \begin{array}{c|ccc}
                     A_{(1,1)} &  &\OO &  \\ \hline
                      & O & \alpha_{(1,1)}  & O  \\
                     \OO & O & O & \alpha_{(1,1)}  \\
                      & \alpha_{(1,1)} & O & O  
                     \end{array}
                     \right) ,\\
&   \hat{n}_{(2,2)} = \left(
                     \begin{array}{c|ccc}
                     A_{(2,2)} &  &\OO &  \\ \hline
                      & \alpha_{(2,2)} & O & O  \\
                     \OO & O & \alpha_{(2,2)} & O  \\
                      & O & O & \alpha_{(2,2)}  
                     \end{array}
                     \right),~                              
  \hat{n}_{(5,0)} = \left(
                     \begin{array}{c|ccc}
                     I_{7\times 7} &  & \OO & \\ \hline
                      & O & 1_{3\times3}  & O  \\
                     \OO & O & O & 1_{3\times3}  \\
                      & 1_{3\times 3} & O & O  
                     \end{array}
                     \right),    \\
&  \hat{n}_{(4,1)} = \left(
                     \begin{array}{c|ccc}
                     A_{D^{(8)}} &  & \OO &  \\ \hline
                      & \alpha & O & O \\
                     \OO & O & \alpha & O  \\
                      & O & O & \alpha   
                     \end{array}
                     \right) ,~
  \hat{n}_{(3,2)} = \left(
                     \begin{array}{c|ccc}
                     A_{(2,0)} &  &\OO &  \\ \hline
                       & O & O & \alpha_{(2,0)}  \\
                     \OO & \alpha_{(2,0)} & O & O  \\
                       & O & \alpha_{(2,0)} & O   
                     \end{array}
                     \right) ,
\ees
where 
\beq
   A_{(2,0)}={\small 
               \begin{pmatrix}
                      0 &\bsp 0 &\bsp 1 &\bsp 0 &\bsp 0 &\bsp 0 &\bsp 0   \\
                      0 &\bsp 0 &\bsp 0 &\bsp 1 &\bsp 1 &\bsp 0 &\bsp 0  \\  
                      0 &\bsp 0 &\bsp 0 &\bsp 0 &\bsp 1 &\bsp 1 &\bsp 1   \\
                      0 &\bsp 0 &\bsp 1 &\bsp 1 &\bsp 0 &\bsp 1 &\bsp 1   \\       
                      1 &\bsp 0 &\bsp 0 &\bsp 1 &\bsp 0 &\bsp 0 &\bsp 1   \\
                      0 &\bsp 1 &\bsp 0 &\bsp 0 &\bsp 0 &\bsp 0 &\bsp 1   \\
                      0 &\bsp 1 &\bsp 1 &\bsp 1 &\bsp 1 &\bsp 0 &\bsp 1   \\
               \end{pmatrix} } ,
   A_{(1,1)} = {\small 
               \begin{pmatrix}
                     0 &\bsp 0 &\bsp 0 &\bsp 1 &\bsp 0 &\bsp 0 &\bsp 0   \\
                     0 &\bsp 1 &\bsp 0 &\bsp 0 &\bsp 1 &\bsp 0 &\bsp 1  \\  
                     0 &\bsp 0 &\bsp 1 &\bsp 1 &\bsp 0 &\bsp 1 &\bsp 1   \\
                     1 &\bsp 0 &\bsp 1 &\bsp 2 &\bsp 1 &\bsp 0 &\bsp 1   \\       
                     0 &\bsp 1 &\bsp 0 &\bsp 1 &\bsp 1 &\bsp 0 &\bsp 1   \\
                     0 &\bsp 0 &\bsp 1 &\bsp 0 &\bsp 0 &\bsp 1 &\bsp 1   \\
                     0 &\bsp 1 &\bsp 1 &\bsp 1 &\bsp 1 &\bsp 1 &\bsp 2   \\
               \end{pmatrix} } ,
   A_{(2,2)} = {\small 
               \begin{pmatrix}
                     0 &\bsp 0 &\bsp 0 &\bsp 0 &\bsp 0 &\bsp 0 &\bsp 1   \\
                     0 &\bsp 0 &\bsp 1 &\bsp 1 &\bsp 0 &\bsp 0 &\bsp 1  \\  
                     0 &\bsp 1 &\bsp 1 &\bsp 1 &\bsp 1 &\bsp 0 &\bsp 1   \\
                     0 &\bsp 1 &\bsp 1 &\bsp 1 &\bsp 1 &\bsp 1 &\bsp 2   \\       
                     0 &\bsp 0 &\bsp 1 &\bsp 1 &\bsp 1 &\bsp 1 &\bsp 1   \\
                     0 &\bsp 0 &\bsp 0 &\bsp 1 &\bsp 1 &\bsp 0 &\bsp 1   \\
                     1 &\bsp 1 &\bsp 1 &\bsp 2 &\bsp 1 &\bsp 1 &\bsp 2   \\
               \end{pmatrix} } ,  
\eeq
\beq
  \alpha_{(2,0)} = 
                 \begin{pmatrix}
                 1 & 1 & 1   \\
                 1 & 2 & 1   \\
                 1 & 1 & 1
                 \end{pmatrix},  \,\,     
   \alpha_{(1,1)} = 
                 \begin{pmatrix}
                 1 & 2 & 1   \\
                 2 & 2 & 2   \\
                 1 & 2 & 1
                 \end{pmatrix},  \,\,
   \alpha_{(2,2)} = 
                 \begin{pmatrix}
                 1 & 2 & 2   \\
                 2 & 3 & 2   \\
                 2 & 2 & 1
                 \end{pmatrix}.                         
\eeq
For the twisted representations, we obtain
\beq
\hat{n}_{(0)} = \left(
          \begin{array}{c|ccc}
          \OO & B^t_{0} & B^t_0 & B^t_0 \\ \hline
          B_0 &      &       &       \\  
          B_0 &      &  \OO  &       \\          
          B_0 &      &       &       
          \end{array}
          \right), 
\hat{n}_{(1)} = \left(
          \begin{array}{c|ccc}
         \OO  & B^t_{1} & B^t_1 & B^t_1 \\ \hline
          B_1 &       &       &       \\  
          B_1 &       &  \OO  &       \\          
          B_1 &       &       &       
          \end{array}        
          \right),
\hat{n}_{(2)} = \left(
         \begin{array}{c|ccc}
         \OO  & B^t_{2} & B^t_2 & B^t_2 \\ \hline
          B_2 &       &       &       \\  
          B_2 &       &  \OO  &       \\          
          B_2 &       &       &       
          \end{array}           
          \right),
\eeq
where
\beq
B_0 = {\small
      \begin{pmatrix}
      1 &\bsp 1 &\bsp 1 &\bsp 1 &\bsp 1 &\bsp 1 &\bsp 1  \\
      0 &\bsp 1 &\bsp 1 &\bsp 2 &\bsp 1 &\bsp 1 &\bsp 2  \\
      0 &\bsp 0 &\bsp 1 &\bsp 1 &\bsp 1 &\bsp 0 &\bsp 2  
      \end{pmatrix} }, \,\,
B_1 = {\small 
      \begin{pmatrix}
      0 &\bsp 1 &\bsp 1 &\bsp 2 &\bsp 1 &\bsp 1 &\bsp 2  \\
      1 &\bsp 1 &\bsp 2 &\bsp 2 &\bsp 2 &\bsp 1 &\bsp 3  \\
      0 &\bsp 1 &\bsp 1 &\bsp 2 &\bsp 1 &\bsp 1 &\bsp 2      
      \end{pmatrix}},  \,\,        
B_2 = {\small 
      \begin{pmatrix}
      0 &\bsp 0 &\bsp 1 &\bsp 1 &\bsp 1 &\bsp 0 &\bsp 2  \\
      0 &\bsp 1 &\bsp 1 &\bsp 2 &\bsp 1 &\bsp 1 &\bsp 2  \\
      1 &\bsp 1 &\bsp 1 &\bsp 1 &\bsp 1 &\bsp 1 &\bsp 1   
      \end{pmatrix}}.                              
\eeq
We can check that
these matrices form a NIM-rep of 
the generalized fusion algebra ${\cal F}(su(3)_5;G_c)$
given in \eqref{eq:su35twistfusion}.
\medskip

In summary, we have checked that the twisted boundary states
of $su(3)_5$ realize a NIM-rep of the generalized fusion algebra
$\F(su(3)_5; G_c)$.
In other words,
we have constructed three NIM-reps of ${\cal F}(su(3)_5;G_c)$.
Besides the regular one, which has dimension 24,
we have obtained a 12-dimensional 
one associated to the exceptional modular invariant $Z_{E^{(8)}}$
and a 16-dimensional one associated to the simple current invariant
$Z_{D^{(8)}}$.

\section{Twisted boundary states for $su(3)_1^{\oplus 3}$}
\label{sec:E6detail}

In this appendix, we show that the $S_3$-twisted boundary states
associated with the modular invariant \eqref{eq:su3E6main}
of $su(3)_1^{\oplus 3}$ yield a NIM-rep of the generalized 
fusion algebra $\F(su(3)_1^{\oplus 3};S_3)$. 

\subsection{Generalized fusion algebra $\F(su(3)_1^{\oplus 3}; S_3)$}

For $su(3)_1$, there are three integrable representations 
corresponding to three fundamental weights $\Lambda_0, \Lambda_1$
and $\Lambda_2$. In the following, we denote them as 
$(0) = (\Lambda_0), (1) = (\Lambda_1)$ and $(2) = (\Lambda_2)$.
In this notation, the modular transformation matrix of $su(3)_1$
can be written as
\beq
  S^{su(3)_1}_{mn} = \frac{1}{\sqrt{3}} e^{\frac{2 \pi i}{3} mn} \quad
  (m, n = 0, 1, 2)
  \, .
\eeq
Since this expression is invariant under the shift $m \rightarrow m + 3$
(and also $n \rightarrow n + 3$),
we can consider that the representations of $su(3)_1$ are labeled
by an integer modulo 3, namely, 
$(m + 3) = (m)$. 
Actually, the fusion algebra of $su(3)_1$ is the group algebra 
of $\Z_3$,
\beq
  (m) \times (n) = (m+n) \, . 
\eeq
In this notation,
the action of the charge-conjugation is expressed as
\beq
  (m)^* = (-m) \, .
\eeq
The representations of $su(3)_1^{\oplus 3}$ are then labeled 
by three integers $(n_1, n_2, n_3)$ and
the set $\I$ of integrable representations reads
\beq
  \I = \{ (n_1, n_2, n_3) \, | \, n_i = 0, 1, 2 \} \, .
\eeq
Hence there are $3^3 = 27$ representations for $su(3)_1^{\oplus 3}$.
The modular transformation matrix $S$ is simply the tensor product
of three $S^{su(3)_1}$,
\beq
  S_{(m_1,m_2,m_3)(n_1,n_2,n_3)} = 
  S^{su(3)_1}_{m_1 n_1} S^{su(3)_1}_{m_2 n_2} S^{su(3)_1}_{m_3 n_3} =
  \frac{1}{3 \sqrt{3}} e^{\frac{2 \pi i}{3} (m_1 n_1 + m_2 n_2 + m_3 n_3)}
  \, .
\eeq
The fusion algebra of $su(3)_1^{\oplus 3}$ is the group algebra of
$\Z_3 \times \Z_3 \times \Z_3$,
\beq
  (m_1, m_2, m_3) \times (n_1, n_2, n_3)
  = (m_1 + n_1, m_2 + n_2, m_3 + n_3) \, .
\eeq

The automorphism group $S_3$ is generated by two elements
$\pi$ and $\sigma$, which act on the elements of $\I$ as
\bes
  \pi    &: (n_1, n_2, n_3) \mapsto (n_3, n_1, n_2) \, , \\
  \sigma &: (n_1, n_2, n_3) \mapsto (n_1, n_3, n_2) \, .
  \label{eq:pisigma}
\ees
In terms of $\pi$ and $\sigma$, the elements of 
$S_3$ can be expressed as follows, 
\beq
  S_3 = \{1, \pi, \pi^2, \sigma, \pi \sigma, \pi^2 \sigma \} \, . 
\eeq
The fixed point of $\pi \in S_3$ is of the form $(n,n,n)$,
\beq
  \I(\pi) = \{ (n,n,n) \, | \, n = 0,1,2 \} \, .
\eeq
Therefore we have three twining characters for $\pi$,
which can be expressed by those of $su(3)_1$ as follows
\beq
  \chi_{(n,n,n)}^{\pi}(\tilde{q}) = 
  \trace_{(n,n,n)} \pi \, \tilde{q}^{L_0 - \frac{c}{24}} =
  \chi^{su(3)_1}_{n}(\tilde{q}^3) \, .
\eeq
As is shown in eq.\eqref{eq:Somega},
the $\pi$-twisted representations are obtained by the modular transformation
of the twining characters $\chi^\pi$,
\beq
  \chi^{su(3)_1}_{m}(q^{\frac{1}{3}}) =
  \sum_{n = 0,1,2} S^{su(3)_1}_{mn} \chi^{su(3)_1}_{n}(\tilde{q}^3)
  = \sum_{n =0,1,2} S^{su(3)_1}_{mn} \chi_{(n,n,n)}^{\pi}(\tilde{q}) \, .
\eeq
Since the left-hand side is labeled by an integer $m \,(\textrm{mod}\, 3)$,
we denote by $(m)_\pi$
the $\pi$-twisted representation of $su(3)_1^{\oplus 3}$, 
\beq
  \I^\pi = \{(m)_\pi \, | \, m = 0,1,2 \} \, .
\eeq
From the above equation,
the characters and the modular transformation matrix for $(m)_\pi$ read
\beq
  \chi_{(m)_\pi}(q) = \chi^{su(3)_1}_{m}(q^\frac{1}{3}) \, , \quad
  S^\pi_{(m)_\pi (n,n,n)} = S^{su(3)_1}_{mn} \, . 
\eeq
For $\sigma \in S_3$, 
the fixed points and their twining characters have the form
\bea
  \I(\sigma) &= \{ (n_1, n_2, n_2) \, | \, n_1, n_2 = 0,1,2 \} \, , \\
  \chi_{(n_1,n_2,n_2)}^{\sigma}(\tilde{q}) &= 
  \trace_{(n_1,n_2,n_2)} \sigma \, \tilde{q}^{L_0 - \frac{c}{24}} =
  \chi^{su(3)_1}_{n_1}(\tilde{q}) \chi^{su(3)_1}_{n_2}(\tilde{q}^2) \, .  
\eea
We can therefore label 
the $\sigma$-twisted representations by a pair of integers,
\beq
  \I^\sigma = \{(m_1,m_2)_\sigma \, | \, m_1, m_2 = 0,1,2 \} \, ,
\eeq
for which 
the characters and the modular transformation matrix read
\beq
  \chi_{(m_1, m_2)_\sigma}(q) 
  = \chi^{su(3)_1}_{m_1}(q)\, \chi^{su(3)_1}_{m_2}(q^\frac{1}{2}) \, , \quad
  S^\sigma_{(m_1, m_2)_\sigma (n_1,n_2,n_2)} 
  = S^{su(3)_1}_{m_1 n_1}\, S^{su(3)_1}_{m_2 n_2} \, . 
\eeq
The remaining cases are treated in the same way and we give only the results
below,
\bea
  \I^{\pi^2} &= \{(m)_{\pi^2} \, | \, m = 0,1,2 \}\, ,  &
  S^{\pi^2}_{(m)_{\pi^2} (n,n,n)} &= S^{su(3)_1}_{mn} \, , \\
  \I^{\pi\sigma}   &= \{(m_1,m_2)_{\pi\sigma} \, | \, m_1, m_2 = 0,1,2 \}
  \, , & 
  S^{\pi\sigma}_{(m_1, m_2)_{\pi\sigma} (n_1,n_1,n_2)} 
  &= S^{su(3)_1}_{m_1 n_1}\, S^{su(3)_1}_{m_2 n_2} \, , \\
  \I^{\pi^2\sigma} &= \{(m_1,m_2)_{\pi^2\sigma} \, | \, m_1, m_2 = 0,1,2 \}
  \, , &
  S^{\pi^2\sigma}_{(m_1, m_2)_{\pi^2\sigma} (n_1,n_2,n_1)} 
  &= S^{su(3)_1}_{m_1 n_1}\, S^{su(3)_1}_{m_2 n_2} \, .
\eea
The set $\Ihat$ of all the representations 
eventually consists of 60 elements,
\beq
  \abs{\Ihat} = \abs{\I} + \abs{\I^\pi} + \abs{\I^{\pi^2}}
  + \abs{\I^\sigma} + \abs{\I^{\pi \sigma}} + \abs{\I^{\pi^2 \sigma}}
  = 3^3 + 3 \times 2 + 3^2 \times 3 = 60 \, .
\eeq
From the formula \eqref{eq:Sbar},
the conjugation acts on the twisted representations as follows,
\bes
  (m_1, m_2, m_3)^* &= (-m_1, -m_2, -m_3) \, , \\
  (m)_\pi^* &= (-m)_{\pi^2} \, , \\
  (m_1, m_2)_\omega^* &= (-m_1, -m_2)_\omega \quad
  (\omega \in \{\sigma, \pi \sigma, \pi^2 \sigma\}) \, .
\ees
\begin{table}[t]
{\footnotesize
\bes
  &\renewcommand{\arraystretch}{1.5}
  \begin{array}{c | ccc}
    & (m_1, m_2, m_3) 
    & (m)_\pi
    & (m)_{\pi^2}\\
    \hline
    (l_1, l_2, l_3)
    & (l_1 + m_1, l_2 + m_2, l_3 + m_3)
    & (l_1 + l_2 + l_3 + m)_\pi 
    & (l_1 + l_2 + l_3 + m)_{\pi^2} \\
    (l)_\pi 
    & (l + m_1 + m_2 + m_3)_\pi
    & 3 (l + m)_{\pi^2}
    & \sum_{n_1 + n_2 + n_3 = l + m} (n_1, n_2, n_3) \\
    (l)_{\pi^2} 
    & (l + m_1 + m_2 + m_3)_{\pi^2}
    & \sum_{n_1 + n_2 + n_3 = l + m} (n_1, n_2, n_3)   
    & 3 (l + m)_{\pi} \\
    (l_1, l_2)_\sigma 
    & (l_1 + m_1, l_2 + m_2 + m_3)_\sigma
    & \sum_{n_1 + n_2 = l_1 + l_2 + m} (n_1, n_2)_{\pi^2 \sigma}
    & \sum_{n_1 + n_2 = l_1 + l_2 + m} (n_1, n_2)_{\pi \sigma} \\
    (l_1, l_2)_{\pi\sigma}
    & (l_1 + m_1 + m_2, l_2 + m_3)_{\pi\sigma}
    & \sum_{n_1 + n_2 = l_1 + l_2 + m} (n_1, n_2)_{\sigma}
    & \sum_{n_1 + n_2 = l_1 + l_2 + m} (n_1, n_2)_{\pi^2 \sigma} \\    
    (l_1, l_2)_{\pi^2\sigma}
    & (l_1 + m_1 + m_3, l_2 + m_2)_{\pi^2\sigma}
    & \sum_{n_1 + n_2 = l_1 + l_2 + m} (n_1, n_2)_{\pi\sigma}
    & \sum_{n_1 + n_2 = l_1 + l_2 + m} (n_1, n_2)_{\sigma}    
  \end{array} \\[5\jot]
  &\renewcommand{\arraystretch}{1.5}
  \begin{array}{c | ccc}
    & (m_1, m_2)_\sigma
    & (m_1, m_2)_{\pi\sigma}
    & (m_1, m_2)_{\pi^2\sigma} \\
    \hline
    (l_1, l_2, l_3)
    & (l_1 + m_1, l_2 + l_3 + m_2)_\sigma 
    & (l_1 + l_2 + m_1, l_3 + m_2)_{\pi \sigma} 
    & (l_1 + l_3 + m_1, l_2 + m_2)_{\pi^2 \sigma} \\
    (l)_\pi
    & \sum_{n_1 + n_2 = l + m_1 + m_2} (n_1, n_2)_{\pi \sigma}
    & \sum_{n_1 + n_2 = l + m_1 + m_2} (n_1, n_2)_{\pi^2 \sigma}
    & \sum_{n_1 + n_2 = l + m_1 + m_2} (n_1, n_2)_{\sigma} \\
    (l)_{\pi^2}
    & \sum_{n_1 + n_2 = l + m_1 + m_2} (n_1, n_2)_{\pi^2 \sigma}
    & \sum_{n_1 + n_2 = l + m_1 + m_2} (n_1, n_2)_{\sigma}
    & \sum_{n_1 + n_2 = l + m_1 + m_2} (n_1, n_2)_{\pi \sigma} \\    
    (l_1, l_2)_\sigma 
    & \sum_{n_2 + n_3 = l_2 + m_2} (l_1 + m_1, n_2, n_3)
    & (l_1 + l_2 + m_1 + m_2)_{\pi^2}
    & (l_1 + l_2 + m_1 + m_2)_{\pi} \\
    (l_1, l_2)_{\pi\sigma}
    & (l_1 + l_2 + m_1 + m_2)_{\pi}
    & \sum_{n_1 + n_2 = l_1 + m_1} (n_1, n_2, l_2 + m_2)
    & (l_1 + l_2 + m_1 + m_2)_{\pi^2} \\
    (l_1, l_2)_{\pi^2\sigma}
    & (l_1 + l_2 + m_1 + m_2)_{\pi^2}
    & (l_1 + l_2 + m_1 + m_2)_{\pi} 
    & \sum_{n_1 + n_3 = l_1 + m_1} (n_1, l_2 + m_2, n_3)
  \end{array} \notag
\ees
}
\caption{Multiplication table of the generalized fusion algebra
$\F(su(3)_1^{\oplus 3}; S_3)$.
The subscripts stand for the automorphism type of twisted representations.
$l_i, m_i$ and $n_i$ take values in $\Z_3 = \{0, 1, 2\}$. 
Since $\sigma \pi = \pi^2 \sigma \neq \pi \sigma$, this algebra
is non-commutative. 
}
\label{tab:S3}
\end{table}

Having obtained twisted representations and their modular transformation
matrices, it is straightforward to calculate the generalized fusion 
coefficients of $\F(su(3)_1^{\oplus 3}; S_3)$
using the formula \eqref{eq:gen_Verlinde}. 
For example, the coefficient 
$\fusionhat{(l_1,l_2)_{\sigma}}{(m)_{\pi}}{(n_1,n_2)_{\pi^2 \sigma}}$
can be obtained as follows,
\bes
  \fusionhat{(l_1,l_2)_{\sigma}}{(m)_{\pi}}{(n_1,n_2)_{\pi^2 \sigma}}
  &=   \sum_{\lambda \in
  \I(\sigma) \cap \I(\pi)}
  \frac{%
    S^{\sigma}_{(l_1,l_2)_{\sigma}\, \lambda} \,
    S^{\pi}_{(m)_{\pi}\, \lambda} \,
    \overline{S^{\pi^2 \sigma}_{(n_1,n_2)_{\pi^2 \sigma}\, \lambda}}
    }{S_{(0,0,0) \lambda}} \\
  &=   \sum_{p = 0,1,2}
  \frac{%
    S^{\sigma}_{(l_1,l_2)_{\sigma} (p,p,p)} \,
    S^{\pi}_{(m)_{\pi} (p,p,p)} \,
    \overline{S^{\pi^2 \sigma}_{(n_1,n_2)_{\pi^2 \sigma}\, (p,p,p)}}
    }{S_{(0,0,0) (p,p,p)}}  \\
  &=  \sum_{p=0,1,2}
  \frac{%
    S^{su(3)_1}_{l_1 p} S^{su(3)_1}_{l_2 p} 
    S^{su(3)_1}_{m p}
    \overline{S^{su(3)_1}_{n_1 p} S^{su(3)_1}_{n_2 p}}
    }{(S^{su(3)_1}_{0 p})^3} \\
  &= \frac{1}{3} \sum_{p=0,1,2} 
     e^{\frac{2 \pi i}{3} p (l_1 + l_2 + m - n_1 - n_2)} \\
  &= \delta^{(3)}_{l_1 + l_2 + m, \, n_1 + n_2} \, ,
\ees
where $\delta^{(3)}$ is the Kronecker delta for $\Z_3$. 
The other cases can be calculated in the same manner;
we give the result in Table~\ref{tab:S3}.

\subsection{Non-trivial NIM-rep of $\F(su(3)_1^{\oplus 3}; S_3)$}

\begin{figure}
\begin{center}
\includegraphics[height=.17\textheight]{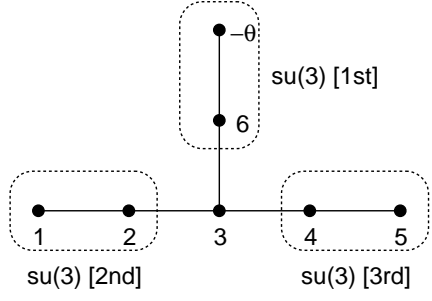}
\end{center}
\caption{Extended Dynkin diagram of $E_6$.
$\theta$ stands for the highest root of $E_6$. 
Each box expresses the simple roots of $su(3) \subset E_6$.
}
\label{fig:E6}
\end{figure}

For the chiral algebra $su(3)_1^{\oplus 3}$, 
the following block diagonal invariant is available, 
\bes
  Z_{E_6} &= 
  \abs{\chi_{(0,0,0)} + \chi_{(1,1,1)} + \chi_{(2,2,2)}}^2 \\
  & \qquad
  + \abs{\chi_{(0,2,1)} + \chi_{(1,0,2)} + \chi_{(2,1,0)}}^2
  + \abs{\chi_{(0,1,2)} + \chi_{(2,0,1)} + \chi_{(1,2,0)}}^2 \, ,
  \label{eq:su3E6}
\ees
which originates from the conformal embedding 
$su(3)_1^{\oplus 3} \subset E_{6, 1}$ (see Fig.~\ref{fig:E6}).~\footnote{%
This invariant is also considered to be a simple current extension
by $(1,1,1) \in \I$.
}
For $E_{6,1}$, there are three integrable representations,
$(\tilde{\Lambda}_0), (\tilde{\Lambda}_1)$ and $(\tilde{\Lambda}_5)$,
\footnote{%
We distinguish the fundamental weights of $E_{6,1}$ from
those of $su(3)_1$ by putting a tilde.
}
and the charge-conjugation invariant reads
\beq
  Z = \abs{\chi_{\tilde{\Lambda}_0}}^2 
  +   \abs{\chi_{\tilde{\Lambda}_1}}^2
  +   \abs{\chi_{\tilde{\Lambda}_5}}^2 \, .
\eeq
From this together with the branching rule
\bes
  (\tilde{\Lambda}_0) &= (0,0,0) \oplus (1,1,1) \oplus (2,2,2) \, , \\
  (\tilde{\Lambda}_1) &= (0,2,1) \oplus (1,0,2) \oplus (2,1,0) \, , \\
  (\tilde{\Lambda}_5) &= (0,1,2) \oplus (1,2,0) \oplus (2,0,1) \, ,
  \label{eq:E6branching}
\ees
one obtains the invariant \eqref{eq:su3E6}. 

There are three boundary states preserving $E_{6,1}$
corresponding to three integrable representations,~\footnote{%
For simplicity, we omit in this appendix
the label for the anti-holomorphic representation of Ishibashi states
and denote by $\dket{\lambda;\omega}$ 
instead of $\dket{(\lambda,\mu^*);\omega}$.
}
\beq
  \ket{\tilde{\Lambda}_i} = 
  \sum_{j=0,1,5} S^{E_6}_{\tilde{\Lambda}_i \tilde{\Lambda}_j}
  \dket{\tilde{\Lambda}_j)} \quad
  (i = 0, 1, 5) \, .
  \label{eq:E6regular}
\eeq
$S^{E_6}$ is the modular transformation matrix of $E_{6,1}$
\beq
  S^{E_6} = \frac{1}{\sqrt{3}}
  \begin{pmatrix} 
  1 & 1 & 1 \\ 
  1 & e^{-\frac{2\pi i}{3}} & e^{\frac{2\pi i}{3}} \\
  1 & e^{\frac{2\pi i}{3}}  & e^{-\frac{2\pi i}{3}} 
  \end{pmatrix} ,
  \label{eq:E6S}
\eeq
where the rows and the columns are ordered as
$\{(\tilde{\Lambda}_0), (\tilde{\Lambda}_1), (\tilde{\Lambda}_5)\}$.
From the branching rule \eqref{eq:E6branching}, one can
express the Ishibashi states of $E_{6,1}$ in terms of
those of $su(3)_1^{\oplus 3}$, 
\bes
  \dket{\tilde{\Lambda}_0} &= \frac{1}{\sqrt{3}} 
  (\dket{(0,0,0)} + \dket{(1,1,1)} + \dket{(2,2,2)}) \, , \\
  \dket{\tilde{\Lambda}_1} &= \frac{1}{\sqrt{3}} 
  (\dket{(0,2,1)} + \dket{(1,0,2)} + \dket{(2,1,0)}) \, , \\
  \dket{\tilde{\Lambda}_5} &= \frac{1}{\sqrt{3}} 
  (\dket{(0,1,2)} + \dket{(1,2,0)} + \dket{(2,0,1)}) \, .
  \label{eq:E6Ishibashibranching}
\ees
Substituting this into \eqref{eq:E6regular},
one obtains three boundary states preserving $su(3)_1^{\oplus 3}$.
The remaining $6$ ($=9-3$) states can be constructed by the fusion 
with $(1,0,0) \in \I$.
Since $(1,0,0)^3 = (3,0,0) = (0,0,0)$, we obtain three boundary states
for each state preserving $E_{6,1}$. 
We therefore label the resulting states as follows
\bes
  \V = \{
  & (0,0) = (\tilde{\Lambda}_0), (0,1), (0,2), \\
  & (1,0) = (\tilde{\Lambda}_1), (1,1), (1,2), 
  (2,0) = (\tilde{\Lambda}_5), (2,1), (2,2)
  \} \, .
  \label{eq:E6V}
\ees
The boundary state coefficient takes the form
\beq
  \Psi = \frac{1}{\sqrt{3}} 
  \begin{pmatrix} 
  K & K & K \\ 
  K & e^{-\frac{2\pi i}{3}} K & e^{\frac{2\pi i}{3}} K \\
  K & e^{\frac{2\pi i}{3}} K & e^{-\frac{2\pi i}{3}} K
  \end{pmatrix} = \frac{1}{\sqrt{3}}
  \begin{pmatrix} 
  1 & 1 & 1 \\ 
  1 & e^{-\frac{2\pi i}{3}} & e^{\frac{2\pi i}{3}} \\
  1 & e^{\frac{2\pi i}{3}}  & e^{-\frac{2\pi i}{3}} 
  \end{pmatrix} \otimes K \, , 
  \label{eq:E6untwisted}
\eeq
where $K$ is a $3 \times 3$ unitary matrix
\beq
  K = \frac{1}{\sqrt{3}}
  \begin{pmatrix} 
  1 & 1 & 1 \\ 
  1 & e^{\frac{2\pi i}{3}} & e^{-\frac{2\pi i}{3}} \\
  1 & e^{-\frac{2\pi i}{3}}  & e^{\frac{2\pi i}{3}} 
  \end{pmatrix} \, .
  \label{eq:K}
\eeq
The row of $\Psi$ is ordered as \eqref{eq:E6V} while
the column is ordered as
\beq
  \E = \{(0,0,0),(1,1,1),(2,2,2),(0,2,1),(1,0,2),(2,1,0),
         (0,1,2),(1,2,0),(2,0,1) \} \, .
\eeq
The overlap matrices $\hat{n}$ can be calculated using
the formula \eqref{eq:nhat}, and the result is
\beq
  (\hat{n}_{(m_1,m_2,m_3)})_{(\alpha,a)}{}^{(\beta,b)} =
  (P^{m_2 + 2 m_3})_\alpha{}^\beta \otimes 
  (P^{m_1 + m_2 + m_3})_a{}^b \, ,
  \label{eq:su3E6nhat}
\eeq
where $(\alpha,a), (\beta, b) \in \V$ and
$P$ is a $3 \times 3$ permutation matrix
\beq
  P = \begin{pmatrix} 
  0 & 1 & 0 \\
  0 & 0 & 1 \\
  1 & 0 & 0 \end{pmatrix} \, .
  \label{eq:P}
\eeq
These matrices $\{\hat{n}_{(m_1,m_2,m_3)} | (m_1,m_2,m_3) \in \I \}$
satisfy the ordinary fusion algebra $\F(su(3)_1^{\oplus 3})$
\beq
  \hat{n}_{(l_1,l_2,l_3)} \, \hat{n}_{(m_1,m_2,m_3)}
  = P^{l_2 + 2 l_3 + m_2 + 2 m_3} \otimes
    P^{l_1 + l_2 + l_3 + m_1 + m_2 + m_3}
  = \hat{n}_{(l_1 + m_1, l_2 + m_2, l_3 + m_3)} \, , 
\eeq
which means that the untwisted boundary states
\eqref{eq:E6untwisted} form a NIM-rep of $\F(su(3)_1^{\oplus 3})$.

\begin{figure}
\begin{center}
\includegraphics[height=.17\textheight]{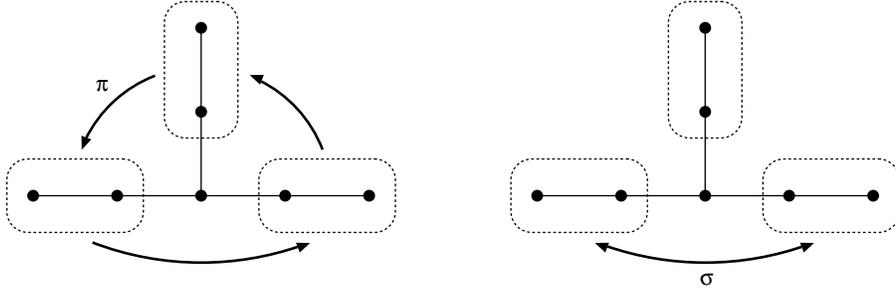}
\end{center}
\caption{
The automorphism group $S_3$ of $su(3)^{\oplus 3}$
has a lift to $E_6$.
}
\label{fig:E6auto}
\end{figure}

We turn to the construction of twisted boundary states in 
the invariant \eqref{eq:su3E6}.
The automorphism group $S_3$ of $su(3)_1^{\oplus 3}$ has
a lift to $E_{6,1}$ (see Fig.~\ref{fig:E6auto}). 
The automorphism group $\mathrm{Aut}(E_6)$
of $E_6$ has a normal subgroup $\mathrm{Aut}_0(E_6)$
consisting of all the inner automorphisms.
The quotient group $\mathrm{Aut}(E_6)/\mathrm{Aut}_0(E_6)$
has two elements:
one is the identity and corresponds to $\mathrm{Aut}_0(E_6)$ while
the other comes from the outer automorphisms
that contains the charge conjugation
$\tilde{\omega}_c$ of $E_6$,
\beq
  \tilde{\omega}_c : 
  (\tilde{\Lambda}_0) \mapsto (\tilde{\Lambda}_0) \, , \quad
  (\tilde{\Lambda}_1) \leftrightarrow (\tilde{\Lambda}_5) \, .
\eeq
As is seen from the definition \eqref{eq:pisigma} of $\pi \in S_3$
and the branching rule \eqref{eq:E6branching},
$\pi$ does not change any representation of $E_6$.
Hence the lift of $\pi$ to $E_6$ is an inner automorphism
of $E_6$.
On the other hand, $\sigma \in S_3$ exchanges 
$(\tilde{\Lambda}_1)$ with $(\tilde{\Lambda}_5)$
and its lift is an outer automorphism of $E_6$. 

Let $\tilde{\pi}$ be the lift of $\pi$ to $E_6$.
Since $\tilde{\pi}$ is inner, the corresponding twisted
boundary states of $E_6$ are expressed by the same
boundary state coefficient as the untwisted ones, 
namely the modular transformation matrix \eqref{eq:E6S}. 
Therefore, applying the fusion $(1,0,0) \in \I$ to 
the $\tilde{\pi}$-twisted states, we obtain exactly the same
boundary state coefficient for the $\pi$-twisted states
as the untwisted ones,
\beq
  \Psi^\pi = \Psi \, .
\eeq
Accordingly the labels of the $\pi$-twisted boundary states
has the same structure as $\V$, 
\beq
  \V^\pi = \{ (\alpha, a)_\pi \,|\, \alpha = 0,1,2; \, a = 0,1,2  \} \, ,
  \quad
  \E(\pi) = \E \, .
\eeq
We can construct the $\pi^2$-twisted states in the same way
as the case of $\pi$ and obtain the result
\beq
  \Psi^{\pi^2} = \Psi \, , \quad
  \V^{\pi^2} = 
  \{ (\alpha, a)_{\pi^2} \,|\, \alpha = 0,1,2; \, a = 0,1,2  \} \, ,
  \quad
  \E(\pi^2) = \E \, . 
\eeq

The lift $\tilde{\sigma}$ of $\sigma \in S_3$ to $E_6$ is outer.
Hence we have to start from a non-trivial boundary state
coefficient in $E_6$ instead of the modular transformation matrix. 
Since $\tilde{\sigma}$ fixes only $(\tilde{\Lambda}_0)$
among the integrable representations of $E_{6,1}$, 
there is only one $\tilde{\sigma}$-twisted boundary state
\footnote{%
One can regard the boundary state coefficient of
$\ket{(0)_\sigma}$ as
the modular transformation `matrix' of 
the twisted chiral algebra $E_6^{(2)}$ at level $1$.
}
\beq
  \ket{(0)_\sigma} = \dket{\tilde{\Lambda}_0; \tilde{\sigma}} 
  = \frac{1}{\sqrt{3}}
   (\dket{(0,0,0); \sigma} + \dket{(1,1,1); \sigma} 
  + \dket{(2,2,2); \sigma})  \, .
\eeq
Applying the fusion with $(1,0,0) \in \I$ 
yields the remaining two states. The result is as follows,
\beq
  \Psi^\sigma = K \, , \quad
  \V^\sigma = \{(a)_\sigma \, | \, a = 0,1,2 \} \, , \quad
  \E(\sigma) = \{(0,0,0),(1,1,1),(2,2,2) \} \, ,
\eeq
where $K$ is the matrix defined in \eqref{eq:K}. 
The case of $\pi \sigma$ and $\pi^2 \sigma$ can be treated 
in the same way as $\sigma$ and yields the result
\bea
  \Psi^{\pi \sigma} &= K \, , & 
  \V^{\pi\sigma} &= \{(a)_{\pi\sigma} \, | \, a = 0,1,2 \} \, ,& 
  \E(\pi\sigma) &= \E(\sigma) \, , \\
  \Psi^{\pi^2 \sigma} &= K \, , & 
  \V^{\pi^2\sigma} &= \{(a)_{\pi^2\sigma} \, | \, a = 0,1,2 \} \, ,& 
  \E(\pi^2\sigma) &= \E(\sigma) \, .
\eea
In this way, we obtain $36$ boundary states in the block diagonal
invariant \eqref{eq:su3E6},
\beq
  \abs{\Vhat} = \abs{\V} + \abs{\V^\pi} + \abs{\V^{\pi^2}}
  + \abs{\V^\sigma} + \abs{\V^{\pi \sigma}} + \abs{\V^{\pi^2 \sigma}}
  = 9 \times 3 + 3 \times 3 = 36 \, .
\eeq

From the boundary state coefficients obtained above,
we can calculate the overlap matrices $\hat{n}$ by the formula
\eqref{eq:nhat}.
In expressing $\hat{n}$, which is a $36 \times 36$ matrix,
it is convenient to factorize $\Vhat$ in the manner similar to
\eqref{eq:su3E6nhat},
\beq
  \Vhat = \{(0),(1),(2),(0)_\pi,(1)_\pi,(2)_\pi,
            (0)_{\pi^2},(1)_{\pi^2},(2)_{\pi^2},
            (0)_\sigma,(0)_{\pi\sigma},(0)_{\pi^2 \sigma} \} \otimes
          \{(a) \, | \, a = 0,1,2 \} \, ,
\eeq
which is related to the original notation as, \textit{e.g.},
$(\alpha)_\pi \otimes (a) = (\alpha,a)_\pi \in \V^\pi$, 
$(0)_\sigma \otimes (a) = (a)_\sigma \in \V^\sigma$. 
In this basis, one can show that 
the matrices $\hat{n}$ take the following form,
\bea
  \hat{n}_{(m_1,m_2,m_3)} &=
  \begin{pmatrix} 
  P^{m_2 + 2 m_3} & O & O & O \\ 
  O & P^{m_2 + 2 m_3} & O & O \\
  O & O & P^{m_2 + 2 m_3} & O \\
  O & O & O & I \end{pmatrix} \otimes P^{m_1 + m_2 + m_3} \, , \notag \\
  \hat{n}_{(m)_\pi} &=
  \begin{pmatrix} 
  O & T & O & O \\ 
  O & O & T & O \\
  T & O & O & O \\
  O & O & O & 3 P^2 \end{pmatrix} \otimes P^{m} \, , \quad
  \hat{n}_{(m)_{\pi^2}} =
  \begin{pmatrix} 
  O & O & T & O \\ 
  T & O & O & O \\
  O & T & O & O \\
  O & O & O & 3 P \end{pmatrix} \otimes P^{m} \, ,   \\
  \hat{n}_{(m_1,m_2)_\sigma} &=
  \begin{pmatrix} 
  O & O & O & E_1 \\ 
  O & O & O & E_2 \\
  O & O & O & E_3 \\
  E_1^T & E_2^T & E_3^T & O \end{pmatrix} \otimes P^{m_1 + m_2} \, , \,\,
  \hat{n}_{(m_1,m_2)_{\pi\sigma}} =
  \begin{pmatrix} 
  O & O & O & E_2 \\ 
  O & O & O & E_3 \\
  O & O & O & E_1 \\
  E_2^T & E_3^T & E_1^T & O \end{pmatrix} \otimes P^{m_1 + m_2} \, , \notag \\
  \hat{n}_{(m_1,m_2)_{\pi^2 \sigma}} &=
  \begin{pmatrix} 
  O & O & O & E_3 \\ 
  O & O & O & E_1 \\
  O & O & O & E_2 \\
  E_3^T & E_1^T & E_2^T & O \end{pmatrix} \otimes P^{m_1 + m_2} \, , \notag
\eea
where $P$ is a permutation matrix of \eqref{eq:P} while
$O$ and $I$ are the zero and the unit matrices, respectively.
The matrices $T, E_1, E_2$ and $E_3$ are defined as follows,
\beq
  T = \begin{pmatrix} 1 & 1 & 1 \\ 1 & 1 & 1 \\ 1 & 1 & 1 \end{pmatrix} \, ,
  \quad
  E_1 = \begin{pmatrix} 1 & 0 & 0 \\ 1 & 0 & 0 \\ 1 & 0 & 0 \end{pmatrix} \, ,
  \quad
  E_2 = \begin{pmatrix} 0 & 1 & 0 \\ 0 & 1 & 0 \\ 0 & 1 & 0 \end{pmatrix} \, ,
  \quad
  E_3 = \begin{pmatrix} 0 & 0 & 1 \\ 0 & 0 & 1 \\ 0 & 0 & 1 \end{pmatrix} \, .  
\eeq
We have checked that these $60$ matrices 
$\{\hat{n}_N | N \in \Ihat \}$
satisfy the generalized fusion
algebra $\F(su(3)_1^{\oplus 3}; S_3)$. 
For example, 
$(0,0)_\sigma \times (0)_\pi 
= (0,0)_{\pi^2 \sigma} + (1,2)_{\pi^2 \sigma} + (2,1)_{\pi^2 \sigma}$
is satisfied as follows
\bes
  \hat{n}_{(0,0)_\sigma}\, \hat{n}_{(0)_\pi}
  &= \begin{pmatrix} 
  O & O & O & E_1 \\ 
  O & O & O & E_2 \\
  O & O & O & E_3 \\
  E_1^T & E_2^T & E_3^T & O \end{pmatrix} 
  \begin{pmatrix} 
  O & T & O & O \\ 
  O & O & T & O \\
  T & O & O & O \\
  O & O & O & 3 P^2 \end{pmatrix} \otimes I \\
  &= \begin{pmatrix} 
  O & O & O & 3 E_3 \\ 
  O & O & O & 3 E_1 \\
  O & O & O & 3 E_2 \\
  3 E_3^T & 3 E_1^T & 3 E_2^T & O \end{pmatrix} \otimes I 
  = \hat{n}_{(0,0)_{\pi^2 \sigma}} 
  + \hat{n}_{(1,2)_{\pi^2 \sigma}}
  + \hat{n}_{(2,1)_{\pi^2 \sigma}} \, . 
\ees
The matrices $\{\hat{n}_N \}$ therefore form a $36$-dimensional
NIM-rep of $\F(su(3)_1^{\oplus 3}; S_3)$.
Together with the regular NIM-rep, which is $60$-dimensional,
we have obtained two NIM-reps of $\F(su(3)_1^{\oplus 3}; S_3)$
corresponding to two modular invariants of $su(3)_1^{\oplus 3}$.


\end{document}